\documentclass[12pt]{article}
\pdfoutput=1 %pdflatex
\usepackage{amsmath,amssymb,amsfonts,amsthm}
\usepackage{float,graphicx, multirow, setspace}
\usepackage{subfiles}
\usepackage{placeins}
\usepackage{url}
\usepackage{xr, xr-hyper}
\usepackage{hyperref}
\usepackage{algpseudocode}
\usepackage{algorithm}
\usepackage{subfig}

\newcommand{\E}{{\mathbb E}}
\renewcommand{\Pr}{{\mathbb P}}
\renewcommand{\P}{{\mathbf P}}
\newcommand{\D}{{\mathbf D}}
\newcommand{\OO}{{\mathbf O}}
\newcommand{\XX}{{\mathbf X}}
\newcommand{\N}{\mathcal{N}}
\newcommand{\bP}{\mathcal{P}}
\newcommand{\rmse}{\text{rmse}}
\newcommand{\xRA}{\text{xRA}}
\newcommand{\RA}{\text{RA}}
\newcommand{\FIP}{\text{FIP}}
\newcommand{\ifFIP}{\text{ifFIP}}
\newcommand{\WAR}{\text{WAR}}
\newcommand{\GWAR}{\text{GWAR}}
\newcommand{\FWAR}{\text{FWAR}}
\newcommand{\FWARr}{\text{FWAR (RA/9)}}
\newcommand{\FWARf}{\text{FWAR (FIP)}}
\newcommand{\wrep}{w_{\text{rep}}}
\newcommand{\xgb}{\text{XGBoost}}
\newcommand{\var}{\text{Var}}
\newcommand{\NL}{\text{NL}}
\newcommand{\AL}{\text{AL}}
\newcommand{\Poisson}{\text{Poisson}}
\newcommand{\Skellam}{\text{Skellam}}

\usepackage{fullpage}
\usepackage{parskip}
\usepackage{mathptmx}
\usepackage[dvipsnames]{xcolor}
\usepackage[margin=1in]{geometry}
\usepackage[round]{natbib}
\numberwithin{equation}{section}
\definecolor{SkyBlue}{RGB}{14, 118, 188}
\definecolor{BrightRed}{RGB}{223,82, 78}
\hypersetup{pdfborder = {0 0 0.5 [3 3]}, colorlinks = true, linkcolor = BrightRed, citecolor = SkyBlue, filecolor = BrightRed}
\title{Introducing Grid WAR: Rethinking WAR for Starting Pitchers}
\author{Ryan S. Brill\thanks{Graduate Group in Applied Mathematics and Computational Science, University of Pennsylvania. Correspondence to: ryguy123@sas.upenn.edu} \ and Abraham J. Wyner\thanks{Dept.~of Statistics and Data Science, The Wharton School, University of Pennsylvania}}

\begin{document}
\maketitle

\begin{abstract}
The baseball statistic ``Wins Above Replacement'' (WAR) has emerged as one of the most popular evaluation metrics. But it is not readily observed and tabulated; WAR is an estimate of a parameter in a vaguely defined model with all its attendant assumptions.  
Industry-standard models of WAR for starting pitchers from FanGraphs and Baseball Reference all assume that season-long averages are  sufficient statistics for a  pitcher's performance. This provides an invalid mathematical foundation for many reasons, especially because WAR should not be linear with respect to any counting statistic.  
To repair this defect, as well as many others, we devise a new measure, Grid WAR, which accurately estimates a starting pitcher's WAR on a per-game basis. The convexity of Grid WAR diminishes the impact of ``blow-up'' games and up-weights exceptional games, raising the valuation of pitchers like Sandy Koufax, Whitey Ford, and Catfish Hunter who exhibit fundamental game-by-game variance. Although Grid WAR is designed to accurately measure historical performance, it has predictive value insofar as a pitcher's Grid WAR is better than Fangraphs' FIP WAR at predicting future performance. Finally, at \url{https://gridwar.xyz} we host a Shiny app which displays the Grid WAR results of each MLB game since 1952, including career, season, and game level results, which updates automatically every morning.

% The baseball statistic ``Wins Above Replacement'' ($\WAR$) has emerged as one of the most popular evaluation metrics. 
% But it is not readily observed and tabulated; $\WAR$ is an estimate of a parameter in a vaguely defined model with all its attendant assumptions.  
% Industry-standard models of $\WAR$ for starting pitchers from FanGraphs and Baseball Reference all assume that season-long averages are  sufficient statistics for a  pitcher's performance. 
% This provides an invalid mathematical foundation for many reasons, especially because $\WAR$ should not be linear with respect to any counting statistic.  
% To repair this defect, as well as many others, we devise a new measure, Grid $\WAR$, which accurately estimates a starting pitcher's $\WAR$ on a per-game basis. 
% The convexity of Grid $\WAR$ diminishes the impact of ``blow-up'' games and up-weights exceptional games, raising the valuation of pitchers like Sandy Koufax, Whitey Ford, and Catfish Hunter who exhibit fundamental game-by-game variance.
% Although Grid $\WAR$ is designed to accurately measure historical performance, it has predictive value insofar as a pitcher's Grid $\WAR$ is better than Fangraphs' $\FIP$ $\WAR$ at predicting future performance.
% Finally, at \url{https://gridwar.xyz} we host a Shiny app which displays the Grid $\WAR$ results of each MLB game since 1952, including career, season, and game level results, which updates automatically every morning.
\end{abstract}

% %%%%%%%%%%%%%%%%%%%%%%%%%%%%%
% \documentclass[12pt]{article}
% \input{../header}
% \usepackage{fullpage, parskip}
% \onehalfspacing
% %%%%%%%%%%%%%%%%%%%%%%%%%%%%%

% \begin{document}

%%%%%%%%%%%%%%%%%%%%%%%%%%%%%%%%%%%%%%%%%%%%%%%%%%%%%%%%%%%%%%%%%%%%%
%%%%%%%%%%%%%%%%%%%%%%%%%%%%%%%%%%%%%%%%%%%%%%%%%%%%%%%%%%%%%%%%%%%%%
\section{Introduction}\label{sec:intro}

%%%%%%%%%%%%%%%%%%%%%%%%%%%%%%%%%%%%%%%%%%%%%%%%%%%%%%%%%%%%%%%%%%%%%%%%%%%
\subsection{Why calculate $\WAR$?}

Player valuation is one of the fundamental goals of sports analytics.
In team sports, the goal in particular is to quantify the contributions of individual players towards the collective performance of their team. 
In baseball player roles are clearly defined and events are discrete (see Appendix~\ref{app:rules_of_baseball} for a review of the rules of baseball). 
This has led to the development of separate player valuation measures for each aspect of the game: hitting, pitching, fielding, and baserunning \citep{BaumerJensenMatthews+2015+69+84}. 
Historically fundamental measures for evaluating hitters include batting average (BA), on-base percentage (OBP), and slugging percentage (SLG).
Classical measures of pitching include earned run average (ERA) and walks and hits per inning pitched (WHIP).
For a more thorough review of other player performance measures in baseball, we refer to the reader to \citet{hiddenBaseball}, \citet{moneyball}, \citet{curveBall}, \citet{numbersGame}, \citet{theBook}, \citet{saberRev}, and \citet{BaumerJensenMatthews+2015+69+84}.

The benefit of separately measuring different aspects of baseball is that it isolates different aspects of player ability. 
The drawback is that it doesn't provide a comprehensive measure of overall player performance.
This makes it difficult to compare the value of players across positions.
For instance, is a starting pitcher with a $2.50$ ERA more valuable than a shortstop with a $2.80$ batting average?
As the fundamental result of a baseball game is a win or loss, the ideal unified measure of a player's performance is his contribution to the number of games his team wins. 
On this view, an analyst's task is to apportion this total number of wins to each player.

To this end, \textit{wins above replacement (WAR)} estimates an individual baseball player’s contribution on the scale of team wins relative to a replacement level player.
% Win contribution is estimated relative to a replacement level player, rather than a league average player, because average players themselves are valuable.
Win contribution is not estimated relative to a league average player because average players themselves are valuable.
A better baseline is a \textit{replacement level} player who is freely available to add to your team in the absence of the player being evaluated (e.g. a minor league free agent).
% We term such a player \textit{replacement level}.
With this baseline, Sandy Koufax having 11.5 wins above replacement in 1966 means that losing Koufax to injury is associated, on average, with his team dropping about 11.5 games in the standings over an entire season.
% So, it is not reasonable to assume a team can replace a given player with another average player.
% Rather, a team can replace a given player with a replacement level player who is freely available to add to your team in the absence of the player being evaluated (e.g. a minor league free agent).

\subsection{Standard $\WAR$ calculations}

Traditional measures of player performance like ERA and others listed previously are counting statistics.
In other words, they are statistics which tabulate what factually happened during a season.
For instance, in 1966 Sandy Koufax had 62 earned runs in 323 innings, resulting in a 1.73 ERA.
$\WAR$, on the other hand, is not a readily observed counting statistic.
In other words, how many of the 1966 Dodgers' 95 wins were due to Koufax's contributions is not a known or observable quantity.
Rather, $\WAR$ must be \textit{estimated} from historical data.

% There are many different ways to estimate a baseball player's $\WAR$. 
% It is difficult to disentangle the impact of each individual baseball player within a game.
Estimating a player's wins above replacement is an inherently difficult task because the ultimate win/loss result of a game is the culmination of a complex interaction of event outcomes involving players of varying roles.
% Due to the wide variation in the performance and roles across players, this is a difficult task.
% \adi{can you expand on why it is hard to do WAR right?}
% Accordingly, there are many different ways to estimate a baseball player's $\WAR$. 
Accordingly, there have been many different attempts to estimate a baseball player's $\WAR$. 
The fundamental idea behind each of these estimation procedures is to capture the contribution of a player's observed performance isolated from the factors outside of his control.
The way in which we measure  observed performance is a crucial component of estimating his contribution to winning games.

% In this paper, we focus on estimating $\WAR$ for starting pitchers.
To estimate $\WAR$, a baseball analyst first chooses a base metric of performance and then maps this base metric to wins.
Different choices of base metric yield substantially different estimates for $\WAR$.
For instance, FanGraphs builds separate $\WAR$ values for pitchers from two counting stats: fielding-independent pitching  ($\FIP$) and average runs allowed per nine innings ($\RA/9$).
$\FIP$ is a weighted average of a pitcher's isolated pitching metrics\footnote{
    FanGraphs builds $\WAR$ from $\ifFIP$ ($\FIP$ with infield flies), %\citep{war_FG}, 
    $$\ifFIP := \frac{13\cdot HR + 3\cdot(BB+HBP) - 2\cdot(K+IFFB)}{IP} + ifFIP constant,$$
    which is built from home runs ($HR$), walks ($BB$), hit by pitches ($HBP$), strikeouts ($K$), infield fly balls ($IFFB$), and innings pitched ($IP$).
} (e.g., home runs, walks, and strikeouts) \citep{war_FG}.
% Baseball Reference builds $\WAR$ values for pitchers from expected runs allowed ($\xRA$), which  estimates runs allowed from observed outcomes by removing sequencing randomness \citep{war_BR}. 
Baseball Reference builds $\WAR$ for pitchers from expected runs allowed ($\xRA$), which assigns to each potential at-bat outcome\footnote{
    The seven potential outcomes of an at-bat are out, walk, hit-by-pitch, single, double, triple, and home run.
} an expected run value \citep{war_BR}. 
For instance, the expected runs allowed off a single reflects the average number of runs that follow a single in a half-inning.

The difference between Runs Allowed, $\FIP$, and $\xRA$, and hence the associated $\WAR$ values for players, can be substantial. 
$\FIP$, unlike runs allowed and $\xRA$, ignores at-bat outcomes involving fielders in order to fully disentangle pitching performance from fielding.
Specifically, $\FIP$ excludes balls-in-play (singles, doubles, triples, ground outs, fly outs, etc.) because fielders' actions play a role in the outcome of these plays.
Furthermore, 
$\FIP$ and $\xRA$, unlike runs allowed, do not depend on the sequencing of events in an inning.
For example, consider an inning where a pitcher strikes out three, while allowing a home run, two walks, and a single. 
Depending on the sequence of the events, the pitcher could be charged with one to four runs.
The pitcher's $\FIP$ and $\xRA$ for this inning, however, are the same regardless of the sequence.
For a more thorough review of the differences in their methodologies, we refer the reader to the supplementary materials of \citet{BaumerJensenMatthews+2015+69+84}.

After choosing a base measure of player performance, a baseball analyst then decides how to aggregate player performance over the course of a season.
Current implementations of $\WAR$ from FanGraphs and Baseball Reference average pitcher performance over the entire season (e.g., $\RA$ per nine innings, $\FIP$ per inning, and $\xRA$ per out).

%%%%%%%%%%%%%%%%%%%%%%%%%%%%%%%%%%%%%%%%%%%%%%%%%%%%%%%%%%%%%%%%%%%%%%%%%%%%%
\subsection{Problems with standard $\WAR$ calculations for starting pitchers}

% We argue that, for 
For starting pitchers in particular, there are two problems with standard $\WAR$ calculations.
First, a base measure of pitcher performance should account for sequencing context; $\FIP$ and $\xRA$ do not.
Second, pitcher performance should not be a simple average across innings and games.

\textbf{Problem 1: excluding sequencing context.} 
Standard $\WAR$ calculations are functions of pitcher performance, traditionally one of $\FIP$, $\xRA$, or runs allowed.
If the pitcher cannot affect the sequence of events in an inning, then it makes sense to measure his performance using $\FIP$ or $\xRA$.
If sequencing variation were due to chance alone, it would not be controlled by the pitcher and should reasonably be excluded.
For starting pitchers, however,  sequencing variability has other causes. Blake Snell in 2023 is a prime example:  he has the highest walk rate since 2000\footnote{
    At this time of this writing, Snell has about five walks per nine innings and a walk rate of $13.4\%$, the $14^{th}$ highest walk rate since 1912. 
}
yet has the best ERA in baseball (2.33) and an exceptional 1.2 ERA through the last 23 games at the time of this writing.  
High-strikeout pitchers like Snell can give up walks strategically without giving up runs; he can increase his effort to strike out a batter when necessary. All this indicates that he exerts some control over the sequencing of events.\footnote{ 
    Great pitchers are much better in high leverage situations, where the game is on the line. See for example \url{https://www.baseball-reference.com/players/split.fcgi?id=koufasa01&year=1966&t=p}.
}
Thus, we believe starting pitchers bear responsibility for their sequence of outcomes and that their performance in situations of varying leverage should be taken into account. 
For batters, there is enough evidence of this to generate a substantial debate -- see for example, Bill James' comparison of Altuve and Judge\footnote{
    \url{https://www.billjamesonline.com/judge_and_altuve/}
} -- but for pitchers the argument against it is nonsense. 
Thus, we should not use $\FIP$ or $\xRA$, which remove sequencing context, as a base measure of pitcher performance to estimate a starting pitcher's $\WAR$.  
Rather, in this paper we use runs allowed.\footnote{
    Critics of runs allowed as a base measure of pitcher performance argue it is confounded with the fielding, but we argue in Section~\ref{sec:Discussion} that the impact of fielding on $\WAR$ is small (smaller than ballpark, which itself is small).
}

\textbf{Problem 2: averaging across games.} 
After selecting a base measure of pitcher performance, standard $\WAR$ calculations from FanGraphs and Baseball Reference then average pitcher performance over the entire season (e.g., $\RA$ per nine innings, $\FIP$ per inning, and $\xRA$ per out).
If the pitcher doesn't affect his variability across games, then it makes sense to average his performance across games. 
If variation across games were due to chance alone, it would not be controlled by the pitcher and should reasonably be excluded.
For starting pitchers, however, we believe that game-by-game variability has causes other than chance.
Sandy Koufax in 1966 is a prime example: he had an astounding 17 complete games in which he allowed at most one run\footnote{
    In 1996 Koufax had eight complete game shutouts and nine one-run complete games.
} 
yet had three terrible ``blow-up'' games.
The variability of pitchers like Koufax appears in noticeable patterns, indicating that his non-stationarity across games is a fundamental trait. Our method will account for the possibility that the version of Sandy Koufax that starts a game and gets tagged for six runs in two innings is fundamentally different from the Koufax who strikes out the first six batters (colloquially known as the ``the left arm of God''). 

% Since a starting pitcher's game-by-game variability is not entirely due to chance, averaging pitcher performance across games is \textit{wrong}, specifically because \textit{not all runs have the same value} within each game.
% To understand this,
% % Formally, 
% think of a starting pitcher's $\WAR$ in a single game as a function $R \mapsto \WAR(R)$ where $R$ is the number of runs allowed in that game. 
% We expect $\WAR$ to be a decreasing function of $R$ because allowing more runs in a game should correspond to fewer wins above replacement. 
% Additionally, we expect $\WAR$ to be a \textit{convex} function in $R$ (i.e. its second derivative is positive.) 
% As $R$ increases, we expect the relative impact of allowing an extra run, given by $\WAR(R+1) - \WAR(R)$, to decrease. 
% For instance, allowing two runs instead of one should have a much steeper drop off in $\WAR$ than allowing eight runs instead of seven.\footnote{
%     Because \textit{you can only lose a game once}.
% }

Since a starting pitcher's game-by-game variability is not entirely due to chance, averaging pitcher performance across games is \textit{wrong}, specifically because \textit{not all runs have the same value} within each game.
To understand this, think of a starting pitcher's $\WAR$ in a single game as a function $R \mapsto \WAR(R)$ where $R$ is the number of runs allowed in that game.\footnote{
    As we will discuss in Section~\ref{sec:def_grid_war}, think of game $\WAR$ as measuring a context-neutral version of win probability added derived only from a pitcher's performance, invariant to factors outside of his control (e.g., his team's batting).
    This is very different from the usual win-probability-added calculation, which is completely dependent on the starting pitcher's team's offense.
} 
We expect $\WAR$ to be a decreasing function of $R$ because allowing more runs in a game should correspond to fewer wins above replacement. 
Additionally, we expect $\WAR$ to be a \textit{convex} function in $R$ (i.e. its second derivative is positive.) 
As $R$ increases, we expect the relative impact of allowing an extra run, given by $\WAR(R+1) - \WAR(R)$, to decrease. 
For instance, allowing two runs instead of one should have a much steeper drop off in $\WAR$ than allowing eight runs instead of seven.\footnote{
    Because \textit{you can only lose a game once}.
}
Therefore, by Jensen's inequality,
\begin{equation}
\WAR(\E[R]) \leq \E[\WAR(R)].
\label{eqn:wjensen1}
\end{equation}
Traditional methods for computing $\WAR$ are reminiscent of the left side of Equation \eqref{eqn:wjensen1}: first average a pitcher's performance, then compute his $\WAR$ from the resulting average scaled by the number of innings pitched. 
Averaging weighs each run allowed equally, causing a pitcher's ``blow-up'' games to unfairly dilute the value of high quality performances.
Because winning a baseball game is defined by the runs allowed during a game, $\WAR$ should look like the right side of Equation \eqref{eqn:wjensen1} -- compute the $\WAR$ of each of a pitcher's individual games, and then aggregate. 
Crucially, these quantities are not the same.

%%%%%%%%%%%%%%%%%%%
\begin{table}[htb!]
\centering
\begin{tabular}{rccccccc} \hline
\text{game} & 1 & 2 & 3 & 4 & 5 & 6 & \text{total} \\ \hline
\text{earned runs} & 0  & 1 & 2 & 1 & 1 & 10 & 15  \\
\text{innings pitched} & 9  & 6 & 7 & 8 & 7 & 4 & 41 \\ \hline
\end{tabular}
\caption{Max Scherzer's performance over six games prior to the 2014 All Star break, consisting of five excellent games and one bad ``blow-up'' game.}
\label{Tab:Scherzer}
\end{table}
%%%%%%%%%%%%%%%%%%%

For concreteness, consider Max Scherzer's six game stretch from June 12, 2014 through the 2014 All Star game, shown in Table \ref{Tab:Scherzer} \citep{Scherzer}. 
We re-arrange the order of these games to aid our explanation.
He was excellent in games one through five and had a 1.2 ERA (five runs in 37 innings). 
This corresponds to about 2 $\WAR$ according to standard metrics (the Detroit Tigers did in fact win all 5). 
In game six, Scherzer was rocked for ten runs in four innings, exiting in the fifth inning with runners on second and third and no outs. 
His ERA over the six game stretch ballooned to 3.3 (15 runs allowed in 41 innings), reducing his total $\WAR$ to about $1/2$ according to standard metrics.
This is a complete absurdity: because a game can't be lost more than once, accumulated ``real'' $\WAR$ cannot drop from 2 to 1/2 with the addition of one game!
The correct assessment would charge Scherzer with the maximum possible damage, about $-0.40$ wins. So  Scherzer's ``real'' $\WAR$  over the six games should be about 1.5, which is three times higher than the standard calculation.  By evaluating Scherzer's performances using only the average, standard $\WAR$ significantly devalues his contributions during this six game stretch. %,  because no game can be ``lost'' more than once. 
The correct approach would be to calculate $\WAR$ per game and sum them  up.

Here is another revealing albeit hypothetical example. 
Suppose a pitcher tosses two nearly flawless eight inning starts, allowing one run in each start, followed by a terrible two-inning blow-up where he gives up eight runs. 
His averaged performance over the three games is a thoroughly mediocre five runs per nine innings, which translates to a $\WAR$ of about 0.0 when calculated using standard metrics.  In contrast, it is clear that over the three starts his team will win, with near certainty, two out of three, which translates to a ``real'' $\WAR$ of about $0.8$ in total. 
Our hypothetical pitcher, who is great in two out of three starts and terrible in every third, would  accumulate more than eight $\WAR$ over a full season, constituting an all-time great season worthy of a Cy Young award.\footnote{
    Just two pitchers since 2016 have eclipsed eight Grid $\WAR$.
} 
 In contrast, standard WAR metrics would suggest he be designated for assignment.  
What drives the difference?  A  poor performance can greatly affect the average, effectively allowing a single game to be ``lost'' more than once. 
Specifically, standard metrics allow the one blow-up game to count for two losses, resulting in 0 $\WAR$.
The example is somewhat extreme, but not that rare.\footnote{
    Yankees' starting pitcher Domingo Germán is a perfect real life example.
    In two June 2023, he threw a perfect game one game after being rocked for ten runs allowed in three innings.
    Across an entire season, a pitcher like Germán who alternates having great and terrible games would accumulate more than $3$ real $\WAR$ and $-4.5$ standard $\WAR$.
}

%%%%%%%%%%%%%%%%%%%%%%%%%%%%%%%%%%%%%%%%%%%%%%%%%%%%%%%%%%%%%%%%%%%%%%%%%%%
\subsection{Paper organization}

These examples illustrate that calculating $\WAR$ as a function of pitcher performance averaged across games is wrong.
Hence in Section~\ref{sec:def_grid_war} we devise Grid $\WAR$, which estimates a starting pitcher's $\WAR$ in each of his games.
Grid $\WAR$ estimates the completely context-neutral win probability added above replacement at the point when a pitcher exits the game.
Then in Section~\ref{sec:Results} we discuss our results.
We find that averaging pitcher performance across games tends to, in general, undervalue mediocre and highly variable pitchers.
This is because the convexity of $\GWAR$ diminishes the impact of games in which a pitcher allows many runs, and these pitchers have more of those games. 
We also find that Grid $\WAR$ has predictive value: past Grid $\WAR$ is more predictive than past FanGraphs $\WAR$ of future Grid $\WAR$.
This suggests that some pitchers’ game-by-game variance in performance is not just ``bad luck'' but a measurable characteristic.
Finally, in Section~\ref{sec:Discussion} we conclude with a discussion.
Notably, we compare the careers of starting pitchers via Grid $\WAR$ across baseball history.
Although Grid $\WAR$ values many pitchers' careers similarly as other metrics, it substantially changes our view of some pitchers who exhibit fundamental game-by-game variance.
Sandy Koufax is a prime example: although his 1966 season is just the $20^{th}$ best season of all time according to FanGraphs $\WAR$, it is the best season of all time according to Grid $\WAR$.
Other methods incorrectly overweight his three outlying blow-up games.

%%%%%%%%%%%%%%%%%%%%%%%%%%%%%%%%%%%%%%%%%%%%%%%%%%%%%%%%%%%%%%%%%%%%%
%%%%%%%%%%%%%%%%%%%%%%%%%%%%%%%%%%%%%%%%%%%%%%%%%%%%%%%%%%%%%%%%%%%%%
\section{Defining Grid $\WAR$ for starting pitchers}\label{sec:def_grid_war}

Our task is to estimate a starting pitcher's $\WAR$ for an individual game, which we call Grid $\WAR$ ($\GWAR$). 
The idea is to estimate a context-neutral version of win probability added derived only from a pitcher's performance, invariant to factors outside of his control such as his team's batting.\footnote{
    Note that computing a player's context neutral value added in each individual game was first introduced as ``Support Neutral Win Loss'' \citep{wolvertonSNWL,wolvertonSNWLbp2,wolvertonSNWLbp1}.
    In our study, we use different methods, explore the statistical aspects of Grid $\WAR$ in much greater depth, and conduct a much more extensive comparison of Grid $\WAR$ to other $\WAR$ metrics which average pitcher performance over the entire season (specifically, FanGraphs $\WAR$).
} 
This is very different from the usual win-probability-added calculation, which is completely dependent on the starting pitcher's team's offense.
In Section~\ref{sec:gwar_formulation} we detail our mathematical formulation of Grid $\WAR$.
We give a brief overview of our data in Section~\ref{sec:our_data}, and then
we discuss how we estimate the grid functions $f$ and $g$ (Sections \ref{sec:estimate_f} and \ref{sec:estimate_g}), the constant $\wrep$ (Section~\ref{sec:estimate_wrep}), and the park effects $\alpha$ (Section~\ref{sec:parkFxFinal}) which allow us to compute a starting pitcher's Grid $\WAR$ for a baseball game.

%%%%%%%%%%%%%%%%%%%%%%%%%%%%%%%%%%%%%%%%%%%%%%%%%%%%%%%%%%%%%%%%%%%%%
\subsection{Grid $\WAR$ formulation}\label{sec:gwar_formulation}

In baseball, each team's starting pitcher is the first pitcher in the game for that team. 
A starter usually exits midway through a game according to the discretion of the field manager (the equivalent of a head coach in baseball).
Generally a starter pitches for a significant portion of the game and sometimes pitches the entire game (known as a complete game).
Often times, a starter is removed between innings.\footnote{
    Recall that in a baseball game, the two teams switch back and forth between batting and fielding; the batting team's turn to bat is over once the fielding team records three outs. 
    One turn batting for each team constitutes an inning. 
    A game is usually composed of nine innings, and the team with the greater number of runs at the end of the game wins \citep{rules_of_baseball}. 
}
We first define a starter's Grid $\WAR$ for a game in which he exits at the end of an inning. 
% First, we define a starting pitcher's Grid $\WAR$ for a game in which he exits at the end of an inning. 
To do so, we define the function $f=f(I,R)$ which, assuming both teams have league-average offenses, computes the probability a team wins a game after giving up $R$ runs through $I$ innings (the values of $f(I,R)$ for integer values of $I$ and $R$ can be displayed in a simple grid). In short, 
$f$ is a context-neutral version of win probability, as it depends only on the starter's performance. 

Note that $f$ also depends on the league ($\AL$ vs. $\NL$), season, and ballpark.
For example, games in which the home team is in the National League ($\NL$) prior to 2022 did not feature designated hitters, whereas American League ($\AL$) games did, leading to different run environments. 
Additionally, baseballs have different compositions across seasons, leading to different proportions of home runs and base hits, and hence different run environments.
Finally, it is easier to score runs at some ballparks that at others.
For instance, Coors Field at high altitude in Denver features many more home runs that other parks.
Consequently, $f = f(I,R)$ is implicitly a function of league, season, and ballpark.

To compute a wins \textit{above replacement} metric, we need to compare this context-neutral win-contribution to that of a potential replacement-level pitcher. 
A replacement-level player is freely available to add to your team in the absence of the player being evaluated (e.g. a minor league free agent).
We use a constant $\wrep$ which denotes the probability a team wins a game with a replacement-level starting pitcher, assuming both teams have league-average offenses. We expect $\wrep < 0.5$ since replacement-level pitchers are worse than league-average pitchers. 
Then, we define a starter's Grid $\WAR$ during a game in which he gives up $R$ runs through $I$ complete innings as 
\begin{equation}
f(I, R) - \wrep.
\label{eqn:war_f}
\end{equation}
We call our metric Grid $\WAR$ because the function $f=f(I,R)$ is defined on the 2D grid $\{1,...,9\} \times \{0,...,R_{max}=10\}$. 

Next, we define a starting pitcher's Grid $\WAR$ for a game in which he exits midway through an inning. 
In this case, a starter exits an inning having thrown a certain number of outs $O \in \{0,1,2\}$ and having potentially left baserunners on first, second, or third base, encoded by the base-state
$$S \in \{000,100,010,001,110,101,011,111\}.$$
We define an auxiliary function $g=g(r|S,O)$ which, assuming both teams have league-average offenses, computes the probability that, starting midway through an inning with $O$ outs and base-state $S$,
a team scores exactly $r$ runs through the end of the inning. 
Given $g$, we define a starter's Grid $\WAR$ during a game in which he gives up $R$ runs and leaves midway through inning $I$ with $O$ outs and base-state $S$ as the expected Grid $\WAR$ at the end of the inning,
\begin{equation}
\sum_{r \geq 0} g(r|S,O) f(I,r+R) - \wrep.
\label{eqn:war_g}
\end{equation}
Finally, we define a starting pitcher's Grid $\WAR$ for an entire season as the sum of the Grid $\WAR$ of his individual games.

%%%%%%%%%%%%%%%%%%%%%%%%%%%%%%%%%%%%%%%%%%%%%%%%%%%%%%%%%%%%%%%%%%%%%%
\subsection{Our Data}\label{sec:our_data}

In the remainder of Section~\ref{sec:def_grid_war}, we discuss how we estimate the grid functions $f$ and $g$, the constant $\wrep$, and the park effects $\alpha$ which allow us to compute a starting pitcher's Grid $\WAR$ for a baseball game (Equations \ref{eqn:war_f} and \eqref{eqn:war_g}). 
In our analysis. we use play-by-play data from Retrosheet.
% We use play-by-play data from Retrosheet in our analysis.
We scraped every plate appearance from 1990 to 2020 from the Retrosheet database.
For each plate appearance, we record the pitcher, batter, home team, away team, league, park, inning, runs allowed, base state, and outs count.
% We provide a link to our final dataset in Appendix~\ref{sec:data_and_code}.
Our final dataset is publicly available online.\footnote{
    \url{https://upenn.app.box.com/v/retrosheet-pa-1990-2000}
}
In our study, we restrict our analysis to every plate appearance from 2010 to 2019 featuring a starting pitcher. 
Additionally, we scraped FanGraphs $\RA/9$ $\WAR$ (abbreviated henceforth as $\FWARr$) and FanGraphs $\FIP$ $\WAR$ (abbreviated henceforth as $\FWARf$) using the \texttt{baseballr} package \citep{petti_gilani_2021}.
% All computations in our analysis are performed in \textsf{R}, and our code is publicly avaiable online (see Appendix~\ref{sec:data_and_code}).
All computations in our analysis are performed in \textsf{R}, and our code is publicly available online.\footnote{
    \url{https://github.com/snoopryan123/grid_war}
}

%%%%%%%%%%%%%%%%%%%%%%%%%%%%%%%%%%%%%%%%%%%%%%%%%%%%%%%%%%%%%%%%%%%%%%
\subsection{Estimating the grid function $f$}\label{sec:estimate_f}

Now, we estimate the grid function $f=f(I,R)$ which, assuming both teams have average offenses\footnote{
    Technically, we assume both teams have offenses that are \textit{randomly drawn} from our dataset, rather than league-average offenses, since we don't explicitly adjust for offensive quality.
}, computes the probability a team wins a game after giving up $R$ runs through $I$ complete innings. 
We call $f$ a grid because the values of $f(I,R)$ for integer values of $I$ and $R$ can be displayed in a simple 2D grid.
To account for different run environments across different seasons, leagues ($\NL$ vs. $\AL$), and ballparks, it is imperative to estimate a different grid for each league-season-ballpark.
Thus we estimate $f$ using a parametric mathematical model rather than a statistical or machine learning model fit from historical data.
Here, we give an overview as to why; see Appendix~\ref{app:estimate_f} for a more detailed discussion.

The naive solution to estimating $f$ is the empirical grid: across all combinations of $I$ and $R$, simply take the observed proportion of times a starter's team won the game.
% Due to a lack of data, the empirical grid fit on all half-innings from any given league-season is terrible.
Due to a lack of data, the empirical grid fit on all half-innings from any given league-season massively overfits.
In particular, it fails to be monotonic in $I$ or $R$ even though it should be.\footnote{
    For instance, $f$ should be monotonic decreasing in $R$ because as a pitcher allows more runs through a fixed number of innings, his team is less likely to win the game. 
}
% Thus any ballpark-adjusted empirical grid would also be terrible.
Thus any ballpark-adjusted empirical grid would also massively overfit.
To force the grid to be monotonic, we try $\xgb$ with monotonic constraints. 
While the fitted $f$ is monotonic and also convex in $R$ as expected, it still overfits, especially towards the tails (e.g., for $R$ large).
% Put bluntly, refitting the grid using a statistical model for each league-season is awful and clumsy. 
Refitting the grid using a statistical model for each league-season is clearly not optimal.

As there is not enough data to use machine learning to fit a separate grid for each league-season without overfitting, we turn to a parametric mathematical model.
Specifically, we use an Empirical Bayes Poisson model from which we explicitly compute context-neutral win probability. %, rather than a logistic regression, $\xgb$, or empirical distribution fit from observational data.
We find that our Poisson model is a powerful approximation technique, yielding reasonable grids for our application which vary across each league, season, and ballpark without overfitting.
Indeed, by distilling the information of our dataset into just a few parameters, our Poisson model creates a strong model from limited data.

% Due to a lack of data within each individual league-season, we estimate $f$ using a parametric mathematical model rather than a statistical or machine learning model fit from historical data.
% In particular, we use an Empirical Bayes Poisson model, from which we explicitly compute context-neutral win probability, rather than a logistic regression, $\xgb$, or empirical distribution fit from observational data.
% We detail why our mathematical model is superior to statistical models in Appendix~\ref{app:estimate_f}.

Because the runs allowed in a half-inning is a natural number, we begin our parametric modeling process by supposing that the runs allowed in a half-inning is a $\Poisson$ random variable.
In particular, denoting the runs scored by the pitcher's team's batters in inning $i$ by $X_i$ and the runs scored by the opposing team in inning $i$ by $Y_i$ (for innings after the pitcher exits the game), we model
\begin{equation}
    X_i \overset{i.i.d.}{\sim} \Poisson(\lambda_X) \quad \text{and} \quad Y_i \overset{i.i.d.}{\sim} \Poisson(\lambda_Y).
\end{equation}
The two teams have their own runs per inning parameters $\lambda_X$ and $\lambda_Y$ because a baseball season involves teams of varying strength playing against each other.
%%%%%%%%%%%%%%
The assumption that runs scored in an inning is independent across innings given a team's strength, while technically false due to non-stationarity across innings,\footnote{
    For instance, there are differences in batter quality across innings.
    Usually better batters are clumped towards the top of the batting lineup and worse batters appear towards the bottom, so some innings feature more good batters while other innings feature more bad batters.
    Additionally, there are differences in pitcher quality across innings.
    Middle relievers, generally appearing in innings five through seven, are usually worse than starting and closing pitchers. 
}
is justified by \textit{working}.
In particular, it yields a grid which looks like a smoothed, non-overfit version of the empirical and $\xgb$ grids, and so we deem it a reasonable enough assumption.
In other words, we view the independence and Poisson assumptions as tools for creating flexible, non-overfit grids which vary across league, season, and ballpark.

Given the team strength parameters, the probability that a pitcher wins the game after allowing $R$ runs through $I$ innings, assuming win probability in overtime is $1/2$, is 
\begin{align}
\label{eqn:Apoisson_model}
    & f(I,R|\lambda_X,\lambda_Y) := \Pr\bigg(\sum_{i=1}^{9} X_i > R + \sum_{i=I+1}^{9} Y_i\bigg) + \frac{1}{2}\cdot\Pr\bigg(\sum_{i=1}^{9} X_i = R + \sum_{i=I+1}^{9} Y_i\bigg).
\end{align}
If $I = 9$, this is equal to
\begin{align}
    & \Pr\big(\Poisson(9\lambda_X) > R\big) + \frac{1}{2}\cdot\Pr\big(\Poisson(9\lambda_X) = R\big).
\end{align}
If $I < 9$, it is equal to
\begin{align}
    & \Pr\big(\Skellam(9\lambda_X, (9-I-1)\lambda_Y) > R\big) + \frac{1}{2}\cdot\Pr\big(\Skellam(9\lambda_X, (9-I-1)\lambda_Y) = R\big),
\end{align}
noting that the $\Skellam$ distribution arises as a difference of two independent $\Poisson$ random variables.
To capture variability in team strength across each of the 30 MLB teams, we impose a positive normal prior,
\begin{equation}
\label{eqn:Apoisson_model_2prior_tuned}
    \lambda_X, \lambda_Y \sim \N_{+}(\lambda, k\cdot\sigma^2_\lambda).
\end{equation}
We estimate the prior hyperparameters $\lambda$ and $\sigma_\lambda^2$ separately for each league-season by computing each team's mean and variance of the runs allowed in each half inning, respectively, and then averaging over all teams.
The initial estimated values of ${\sigma}^2_\lambda$ are too large (e.g., the prior is overdispersed), so we include a tuning parameter $k$ designed to tune the dispersion across team strengths to match observed data.
In particular, we use $k = 0.28$, which minimizes the log-loss between the observed win/loss column and predictions from the induced grid. 
In particular, the induced grid is given by the posterior mean grid, which we estimate using Monte Carlo integration with $B=100$ samples,
\begin{align}
\label{eqn:Apoisson_model_2post}
    f(I,R|\lambda, \sigma^2_\lambda, k) := \frac{1}{B} \sum_{b=1}^{B} f(I,R|\lambda_X^{(b)}, \lambda_Y^{(b)}),
    % f(I,R|\widehat{\lambda}, \widehat{\sigma}^2_\lambda, k) \approx \frac{1}{B} \sum_{b=1}^{B} f(I,R|\lambda_X^{(b)}, \lambda_Y^{(b)}),
\end{align}
where $\lambda_X^{(b)}$ and $\lambda_Y^{(b)}$ are i.i.d. samples from the prior distribution~\eqref{eqn:Apoisson_model_2prior_tuned}.

%%%%%%%%%%%%%%%%%%%%%%%%%%%%%%%%%%%%%%%%%%%%%%%%%%
% \subsection{Park adjustment}\label{sec:park_adjustment}
Additionally, recall that $f=f(I,R)$ is implicitly a function of ballpark.
To adjust for ballpark, we first define the \textit{park effect} $\alpha$ of a ballpark as the expected runs allowed in one half-inning at that park above that of an average park, if an average offense faces an average defense. 
With our parametric model, the ballpark adjustment is easy: we simply incorporate the park effect into our Poisson parameters.
With a statistical model, the ballpark adjustment is more difficult and prone to overfitting, providing yet another justification of our mathematical model.
As $\lambda$ represents the mean runs allowed in a half-inning for a given league-season, $\lambda + \alpha$ represents the mean runs allowed in a half-inning at a given ballpark during that league-season.
So, to adjust for ballpark, we use $\lambda + \alpha$ in place of $\lambda$ in our Poisson model~\eqref{eqn:Apoisson_model_2post} and positive Normal prior~\eqref{eqn:Apoisson_model_2prior_tuned}.
In Section~\ref{sec:parkFxFinal} we estimate the park effects $\alpha$.

\begin{figure}[htb!]
    \centering
    \includegraphics[width=10cm]{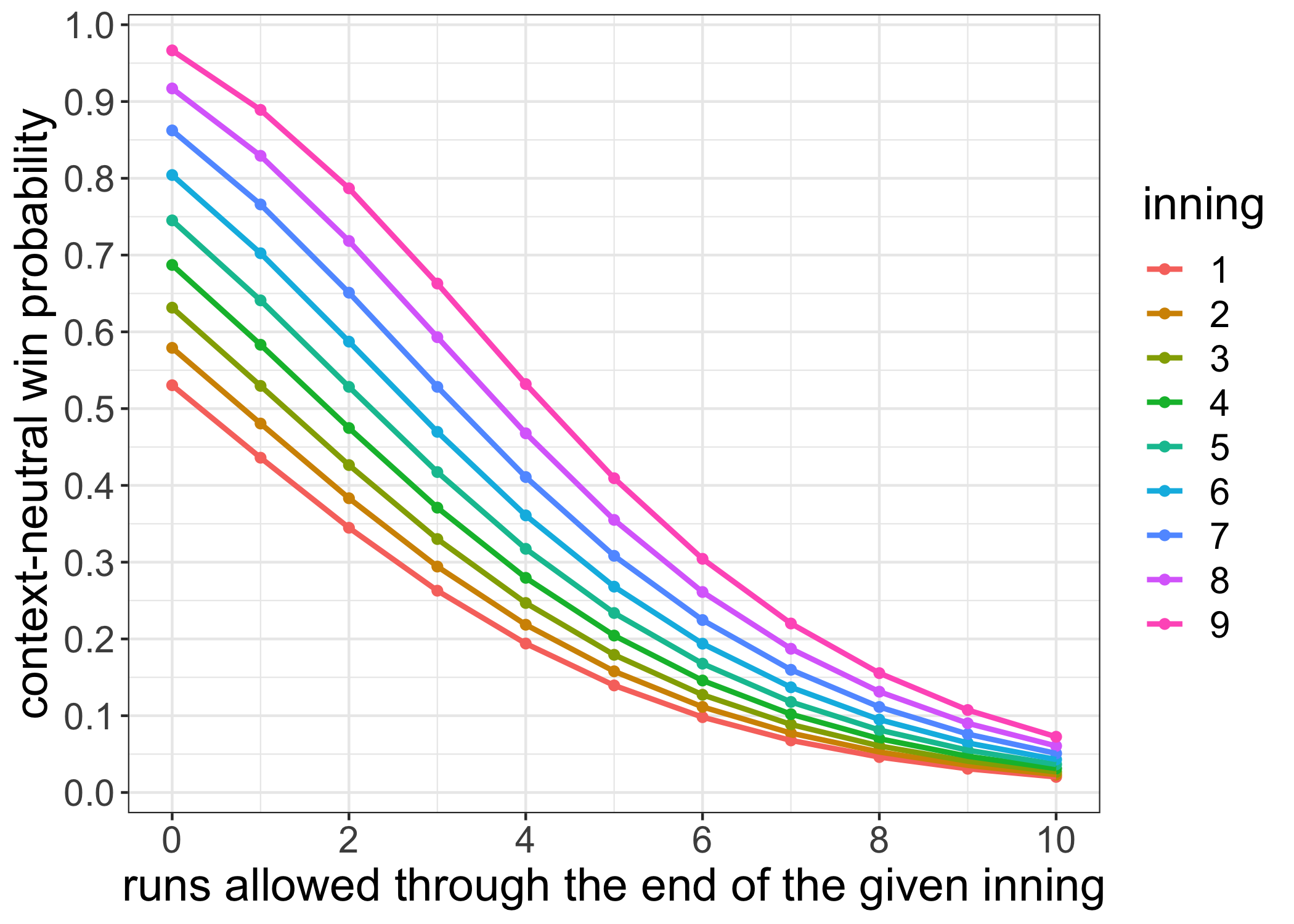}
    \caption{
    Context-neutral win probability ($y$-axis) if a starter allowed $R$ runs ($x$-axis) through $I$ complete innings (color) according to the 2019 National League grid function $f$, fit from our Poisson model~\eqref{eqn:Apoisson_model_2post} with positive Normal prior~\eqref{eqn:Apoisson_model_2prior_tuned}.
    % Context-neutral win probability ($y$-axis) as a function of runs allowed $R$ ($x$-axis) through the end of inning $I$ (color) according to the 2019 National League grid function $f$, fit from our Poisson model~\eqref{eqn:Apoisson_model_2post} with positive Normal prior~\eqref{eqn:Apoisson_model_2prior_tuned}.
    % Estimates of the 2019 National League function $R \mapsto f(I,R)$ using our Poisson model~\eqref{eqn:Apoisson_model_2post} with positive Normal prior~\eqref{eqn:Apoisson_model_2prior_tuned}.
    } 
    \label{fig:fir}
\end{figure}

In Figure~\ref{fig:fir} we visualize the estimated grid $f$ according to our Poisson model~\eqref{eqn:Apoisson_model_2post}, with prior~\eqref{eqn:Apoisson_model_2prior_tuned}, using the 2019 $\NL$ $\lambda$ and $\sigma^2_\lambda$, without a park adjustment.
Note that the $f$ grid for other league-seasons are similar, but differ slightly according to the differing run environments $\lambda$ and $\sigma^2_\lambda$.
We see that $f$ is monotonic decreasing in $R$ because as a pitcher allows more runs through a fixed number of innings, his team is less likely to win the game. 
Also, $f$ is monotonic increasing in $I$ because giving up $R$ runs through $I$ innings is worse than giving up $R$ runs through $I+i$ innings for $i > 0$, since giving up $R$ runs through $I+i$ innings implies a pitcher gave up no more than $R$ runs through $I$ innings.
Further, $f$ is convex in $R$ for large values of $R$ because the marginal impact of allowing an additional run diminishes to zero as $R$ increases because, after giving up a certain number of runs, the game has essentially already been lost.
Succinctly, ``you can only lose a game once''.
Finally, the grid $f$ is smooth.

%%%%%%%%%%%%%%%%%%%%%
%%%%%%%%%%%%%%%%%%%%%

%%%%%%%%%%%%%%%%%%%%%%%%%%%%%%%%%%%%%%%%%%%%%%%%%%%%%%%%%%%%%%%%%%%%%%%%%%%
\subsection{Estimating the grid function $g$}\label{sec:estimate_g}

\begin{figure}[htb!]
    \centering{}
    \includegraphics[width=10cm]{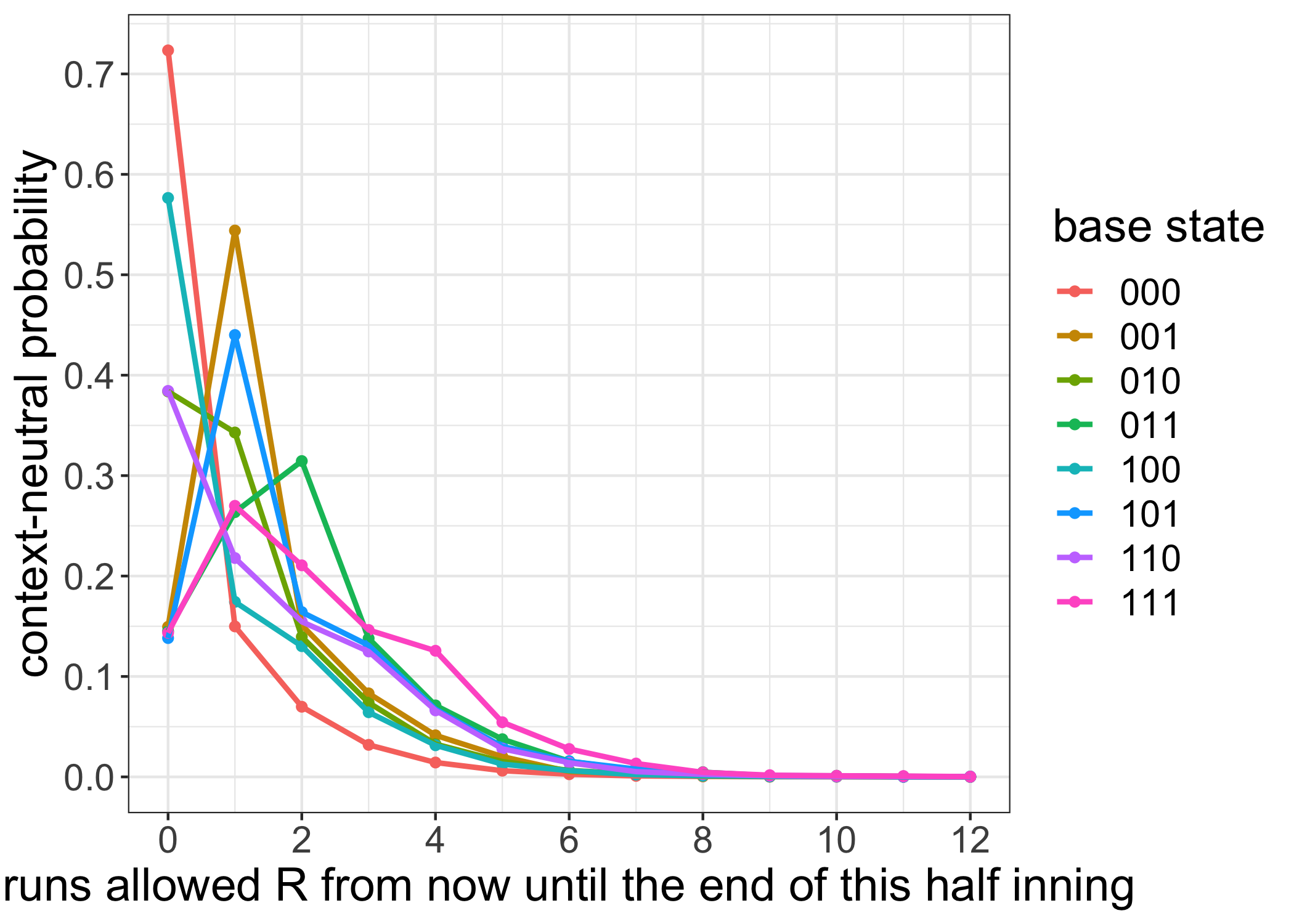}
    \caption{
    From base-state $S$ (color) and $O=0$ outs through the end of an inning, the context-neutral probability ($y$-axis) that the pitcher allows $R$ runs ($x$-axis) according to the grid function $g$.
    % The discrete probability distribution $R \mapsto g(r|S,O=0)$ for each base-state $S$. 
    } 
    \label{fig:g0}
\end{figure}

Now, we estimate the function $g=g(r|S,O)$ which, assuming both teams have league-average offenses, computes the probability that, starting midway through an inning with $O \in \{0,1,2\}$ outs and base-state 
$$S \in \{000,100,010,001,110,101,011,111\},$$
a team scores exactly $r$ runs through the end of the inning. We estimate $g(r|S,O)$ using the empirical distribution.
%, for $R \in \{0,...,13\}$. 
Specifically, we bin and average over the variables $(r,S,O)$, using data from every game from 2010 to 2019. Because $g$ isn't significantly different across innings, we use data from each of the first eight innings.
In Figure \ref{fig:g0} we visualize $g(r|S,O=0)$, with $O=0$ outs, for each base-state $S$.

%%%%%%%%%%%%%%%%%%%%%
%%%%%%%%%%%%%%%%%%%%%

\subsection{Estimating the constant $\wrep$}\label{sec:estimate_wrep}

% To estimate wins \textit{above replacement}, we need to compare a starting pitcher's context-neutral win contribution to that of a potential replacement-level pitcher. 
% Thus we estimate a constant $\wrep$ which represents the context-neutral probability a team wins a game with a replacement-level starting pitcher, assuming both teams have a league-average offense and league-average fielding. 
% \citet{ReplacementLevel} defines \textit{replacement-level} as the ``level of production you could get from a player that would cost you nothing but the league minimum salary to acquire.''
% We estimate $\wrep$ so as to match FanGraphs' definition of replacement-level.
% In particular, we choose $\wrep = 0.428$ so that the sum of $\GWAR$ across all starting pitchers from 2010 to 2019 equals the sum of $\FWARr$.  

To estimate wins \textit{above replacement}, we need to compare a starting pitcher's context-neutral win contribution to that of a potential replacement-level pitcher. 
Thus we estimate a constant $\wrep$ which represents the context-neutral probability a team wins a game with a replacement-level starting pitcher, assuming both teams have a league-average offense and league-average fielding. 
\citet{ReplacementLevel} defines \textit{replacement-level} as the ``level of production you could get from a player that would cost you nothing but the league minimum salary to acquire.''
We could estimate $\wrep$ in a given season by averaging the context neutral win probability at the point when the starter exits the game across all games featuring a replacement-level starting pitcher.
To facilitate fair comparison of Grid $\WAR$ to FanGraphs $\WAR$, we instead estimate $\wrep$ so as to match FanGraphs' definition of replacement-level.
In particular, we choose $\wrep = 0.428$ so that the sum of $\GWAR$ across all starting pitchers from 2010 to 2019 equals the sum of $\FWARr$.  

%%%%%%%%%%%%%%%%%%%%%
\begin{figure}[htb!]
    \centering{}
    \includegraphics[width=15cm]{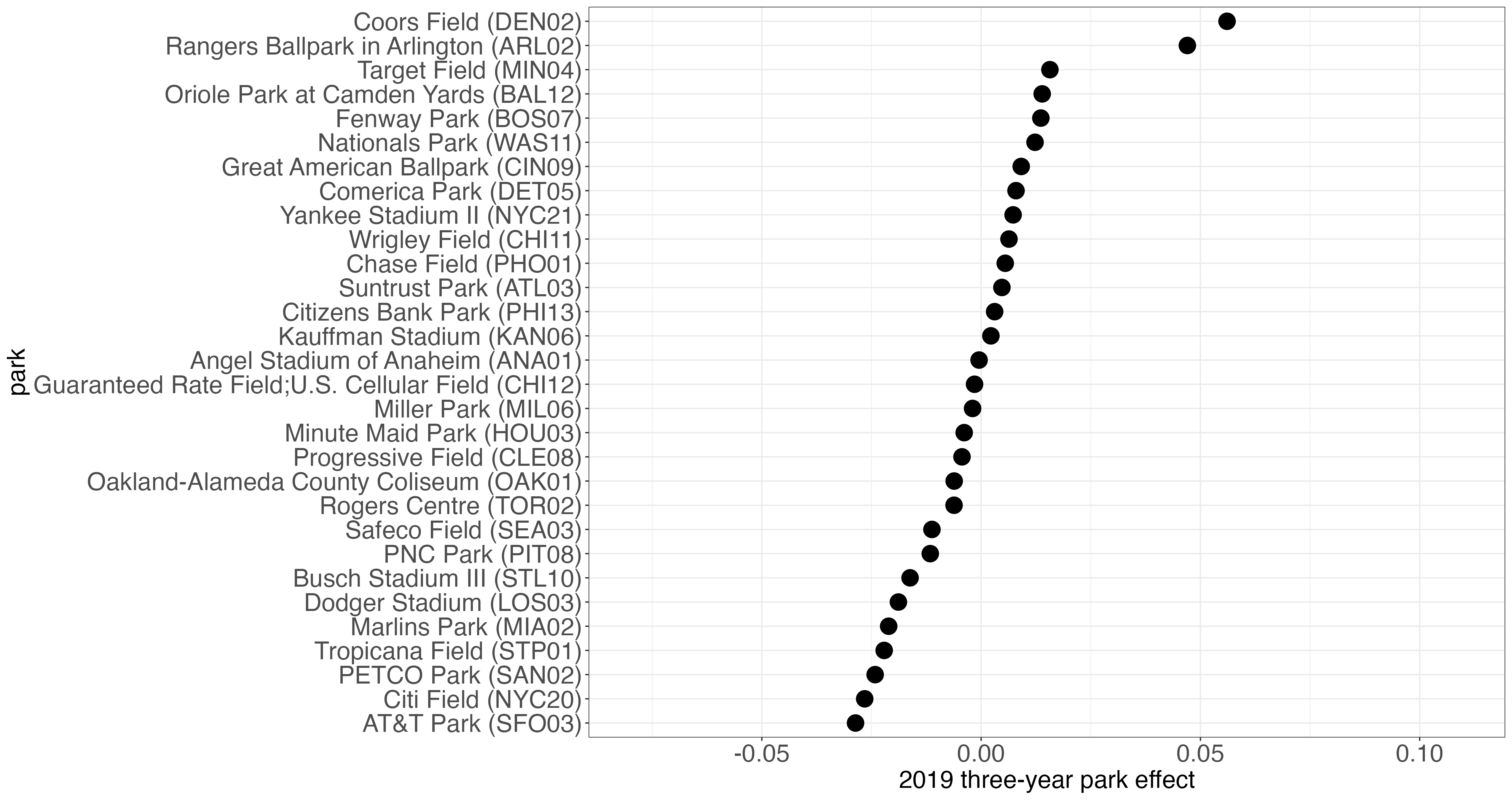}
    \caption{
    Our 2019 three-year park effects ($x$-axis), fit from half-inning data from 2017 to 2019, for each ballpark ($y$-axis).  The abbreviations are Retrosheet ballpark codes.
    } 
    \label{fig:final_parkFx}
\end{figure}
%%%%%%%%%%%%%%%%%%%%%

\subsection{Estimating the park effects $\alpha$}\label{sec:parkFxFinal}

Finally, we estimate the park effect $\alpha$ of each ballpark, which measures the expected runs scored in one half-inning at that park above that of an average park, if an average offense faces an average defense. 
To compute the park effects for 2019, we take all half-innings from 2017 to 2019 and fit a ridge regression, using cross validation to tune the ridge hyperparameter, where the outcome is runs scored during the half-inning and the covariates are fixed effects for each park, team-offensive-season, and team-defensive-season. 
We compute similar three-year park effects for other seasons.
We visualize the 2019 park effects in Figure \ref{fig:final_parkFx}. 
We use ridge regression, as opposed to ordinary least squares or existing park effects from ESPN, FanGraphs, or Baseball Reference, because, as detailed in Appendix~\ref{sec:park_effects}, it performs the best in two simulation studies and has the best out-of-sample predictive performance on observed data.

%%%%%%%%%%%%%%%%%%%%%%%%%%%%%%%%%%%%%%%%%%%%%%%%%%%%%%%%%%%%%%%%%%%%%%%%%%%
%%%%%%%%%%%%%%%%%%%%%%%%%%%%%%%%%%%%%%%%%%%%%%%%%%%%%%%%%%%%%%%%%%%%%%%%%%%
\section{Results}\label{sec:Results}

\begin{figure}[p]
\centering
\includegraphics[width=15cm]{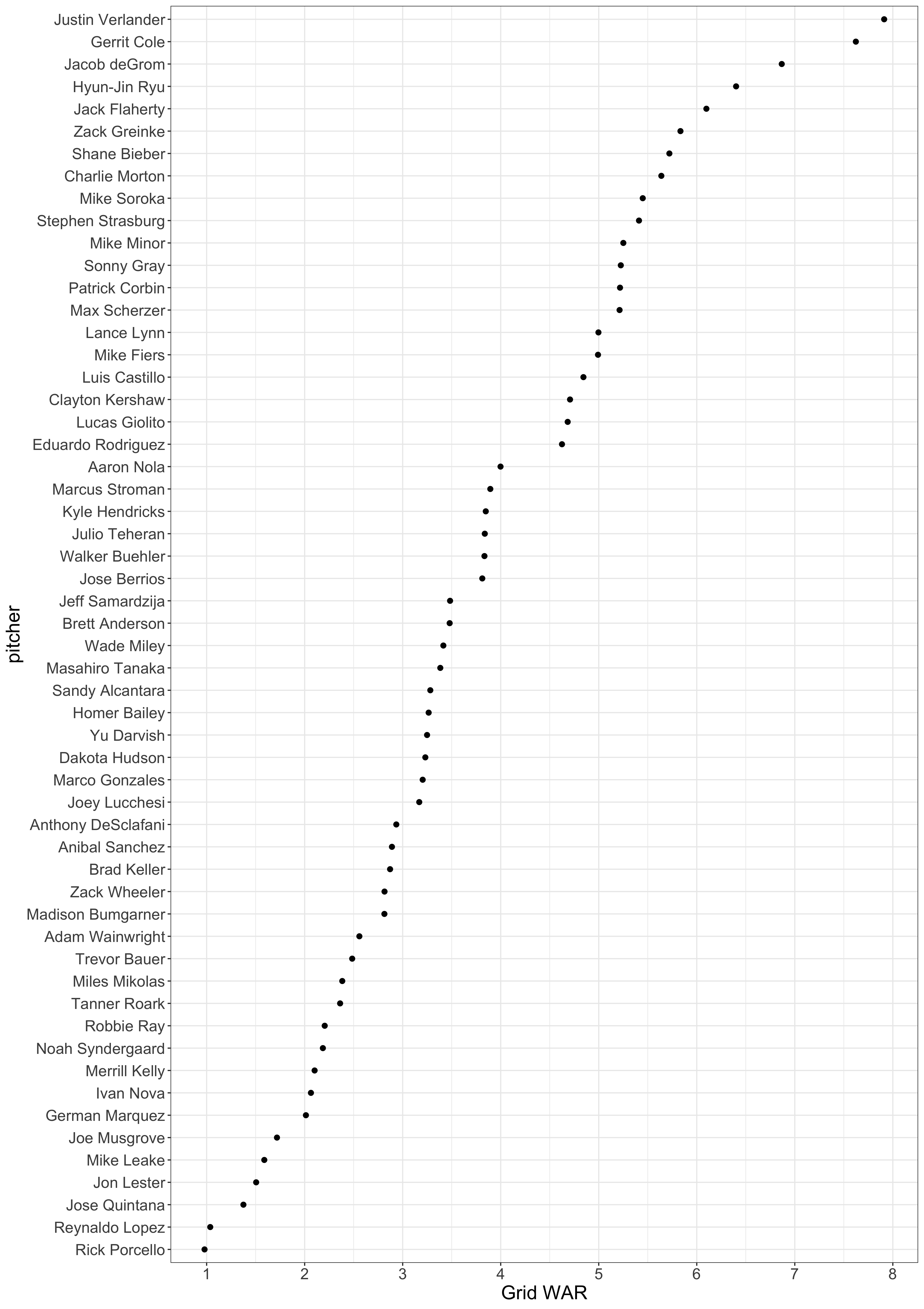}
\caption{Ranking starting pitchers in 2019 by Grid $\WAR$.} 
\label{fig:gwar2019rankings}
\end{figure}

\begin{figure}[p]
\centering
\includegraphics[width=15cm]{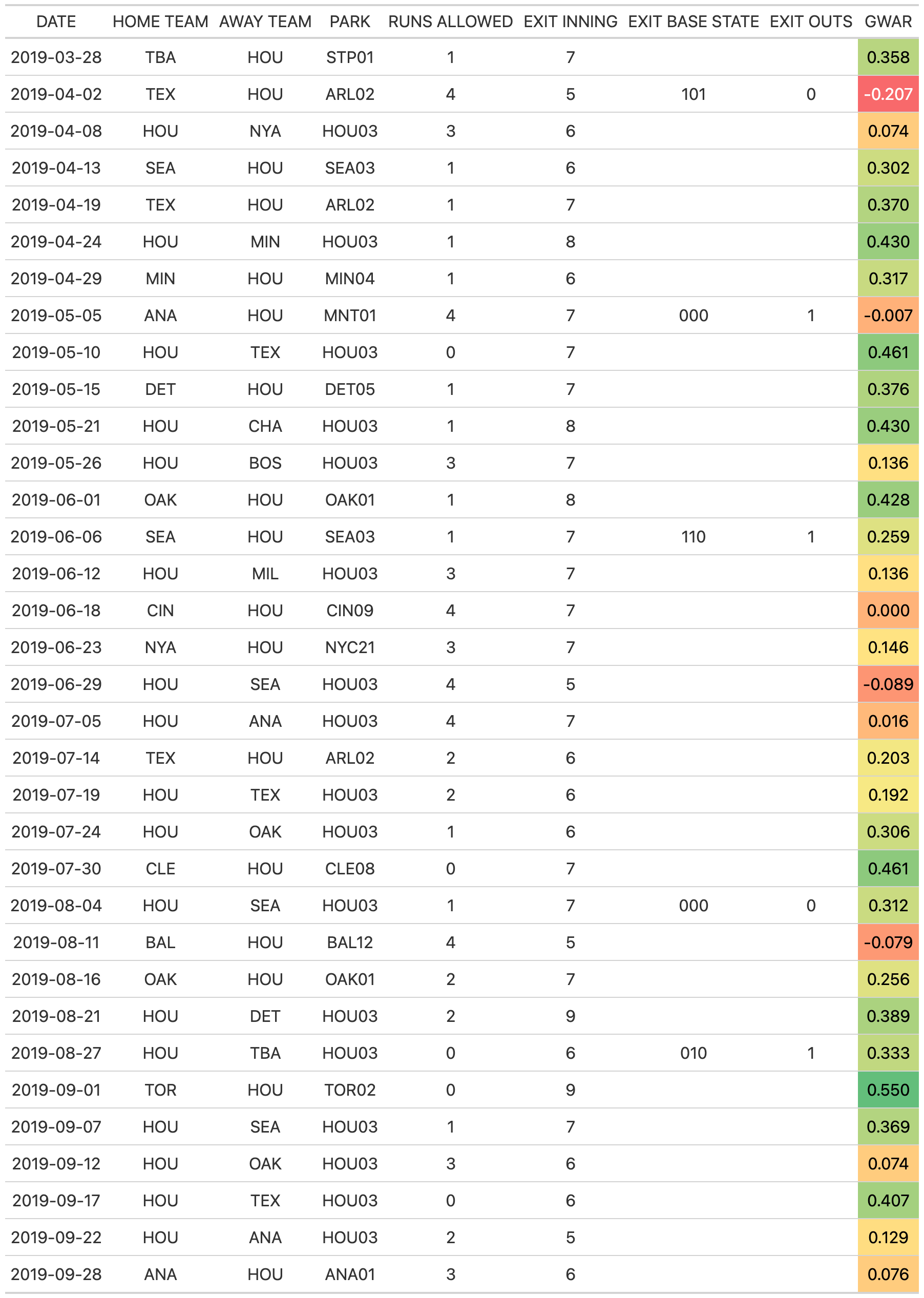}
\caption{A game-by-game breakdown of Justin Verlander's 2019 season.} 
\label{fig:jverlander19gbg}
\end{figure}

After estimating the grid functions $f$ and $g$, the constant $\wrep$, and the park effects $\alpha$, we compute Grid $\WAR$ for each starting pitcher in each game from 2010 to 2019. 
As a quick exposition, in the Appendix we show the full 2019 Grid $\WAR$ rankings (Figure~\ref{fig:gwar2019rankings}) and a full game-by-game breakdown of Justin Verlander's 2019 season (Figure~\ref{fig:jverlander19gbg}).
The remainder of this section is organized as follows.
In Section~\ref{sec:compare_fwar_gwar_2019} we compare $\GWAR$ to FanGraphs $\WAR$ ($\FWAR$) in order to understand the effect of averaging pitcher performance over the entire season on valuing pitchers.
We find that averaging over the season allows a pitcher's terrible performances to dilute the contributions of his great ones.
% This is because the convexity of $\GWAR$ diminishes the impact of games in which a pitcher allows many runs, and weaker pitchers tend to have more of these games.
This is because the convexity of $\GWAR$ diminishes the impact of games in which a pitcher allows many runs, whereas averaging across games doesn't.
Thus traditional measures of $\WAR$ have generally undervalued mediocre pitchers and higher variance pitchers who tend to have more of these terrible games.
Then in Section~\ref{sec:predictive_value_gwar} we quantify the value lost by using $\FWAR$ to estimate pitcher quality rather than using $\GWAR$.
We find that pitcher rankings built from past $\GWAR$ are better than pitcher rankings built from past $\FWAR$ at predicting future pitcher rankings according to $\GWAR$.
This provides evidence that game-by-game variability in pitcher performance is a fundamental and measurable trait.

\subsection{Averaging pitcher performance across games dilutes the contributions of his great games}\label{sec:compare_fwar_gwar_2019}

In this section we compare $\GWAR$ to FanGraphs $\WAR$ ($\FWAR$) in order to understand the effect of averaging pitcher performance over the entire season on valuing pitchers.
We find that averaging over the season allows a pitcher's terrible performances to dilute the contributions of his great ones.
Note that to compare the \textit{relative} value of starting pitchers according to $\GWAR$ relative to $\FWARr$, in this section we rescale $\GWAR$ in each year so that the sum of $\GWAR$ across all starters in each season equals the corresponding sum in $\FWARr$.

%%%%%%%%%%%%%%%%%%
\begin{figure}[hbt!]%[H]
\centering
\includegraphics[width=10cm]{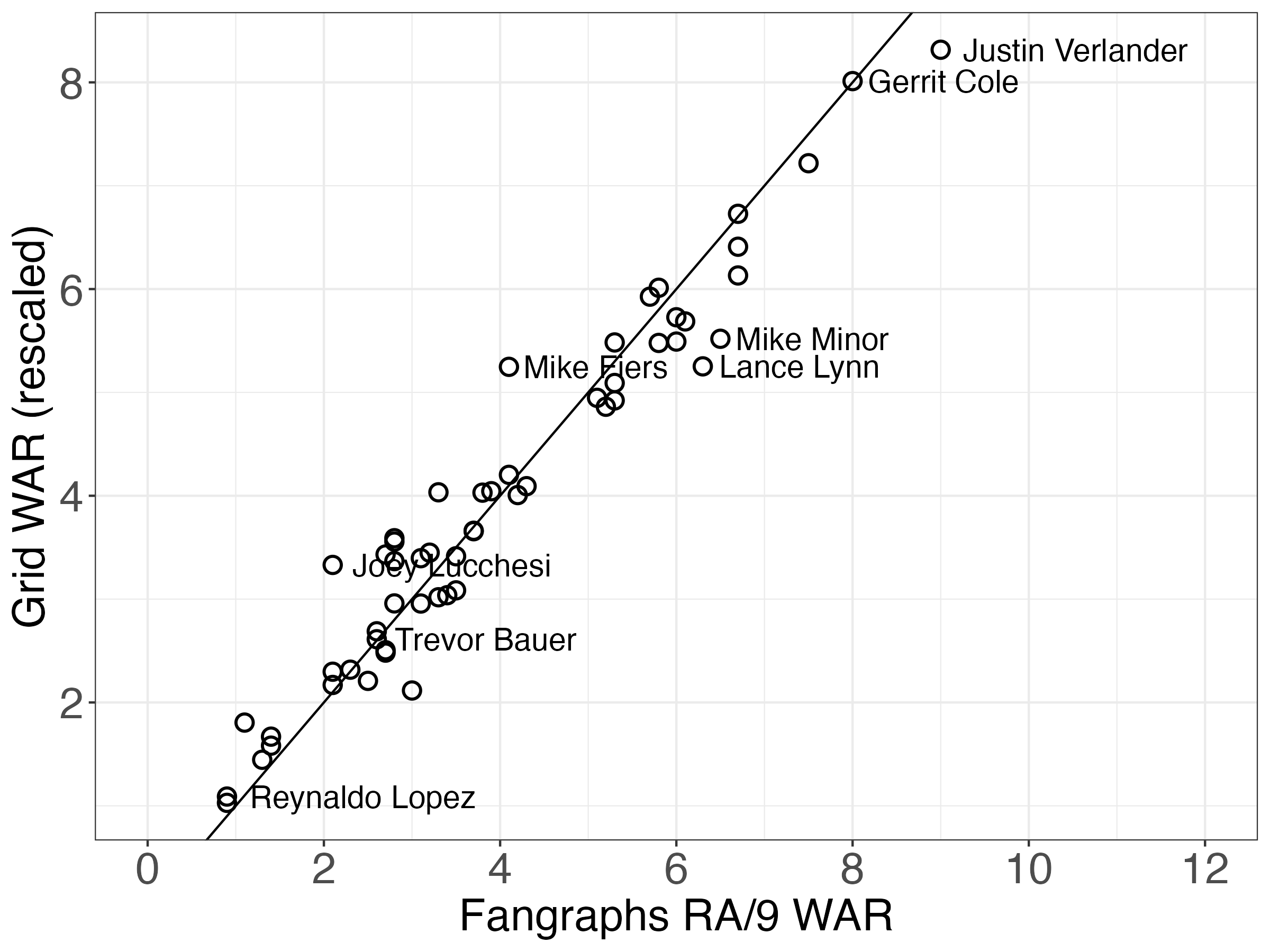}
\caption{
Grid $\WAR$ ($y$-axis) versus FanGraphs $\WAR$ ($\RA/9$) ($x$-axis) for each pitcher-season in 2019.  The pitcher name refers to the dot on its immediate left.
} 
\label{fig:gwarVfwar19}
\end{figure}
%%%%%%%%%%%%%%%%%%

\begin{figure}[hbt!]
\centering
\includegraphics[width=15cm]{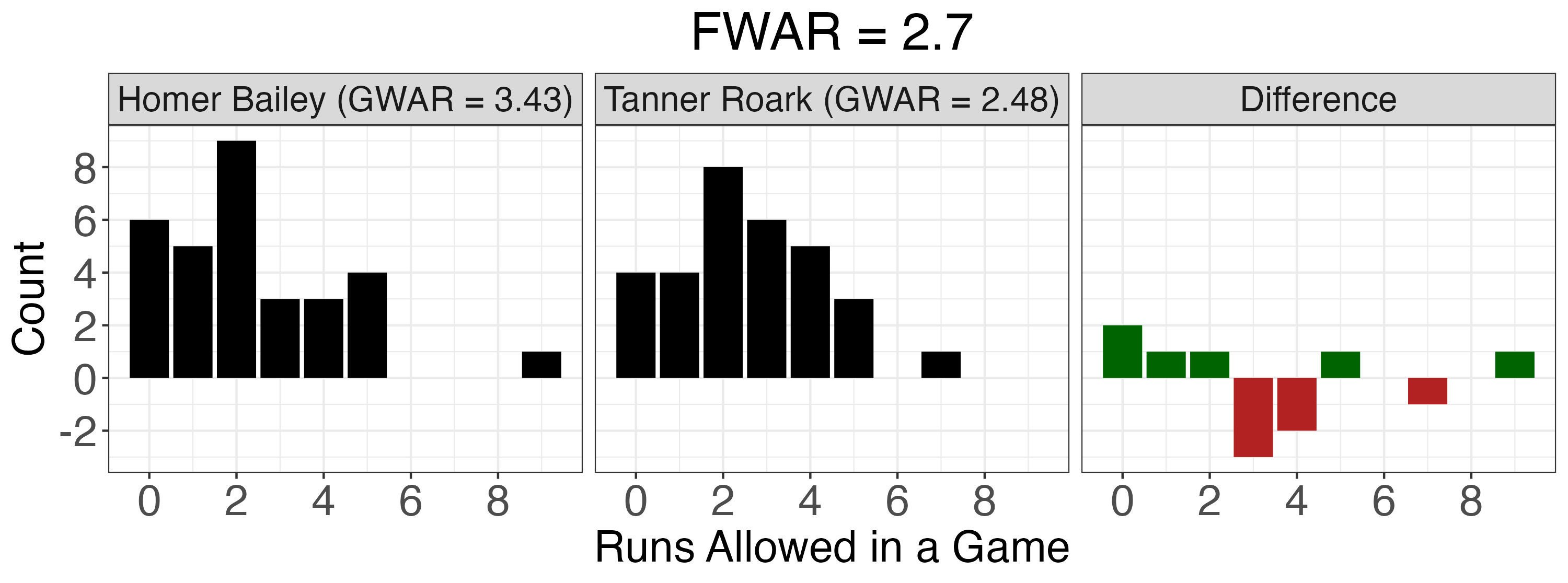}
\caption{
Histogram of runs allowed in a game in 2019 for Homer Bailey (left), Tanner Roark (middle), and the difference between these two histograms (right). 
Even though they have the same $\FWAR$, Bailey has a higher $\GWAR$ than Roark because he has more games in which he allows fewer runs.
} 
\label{fig:pf2}
\end{figure}

\begin{figure}[hbt!]
\centering
\includegraphics[width=15cm]{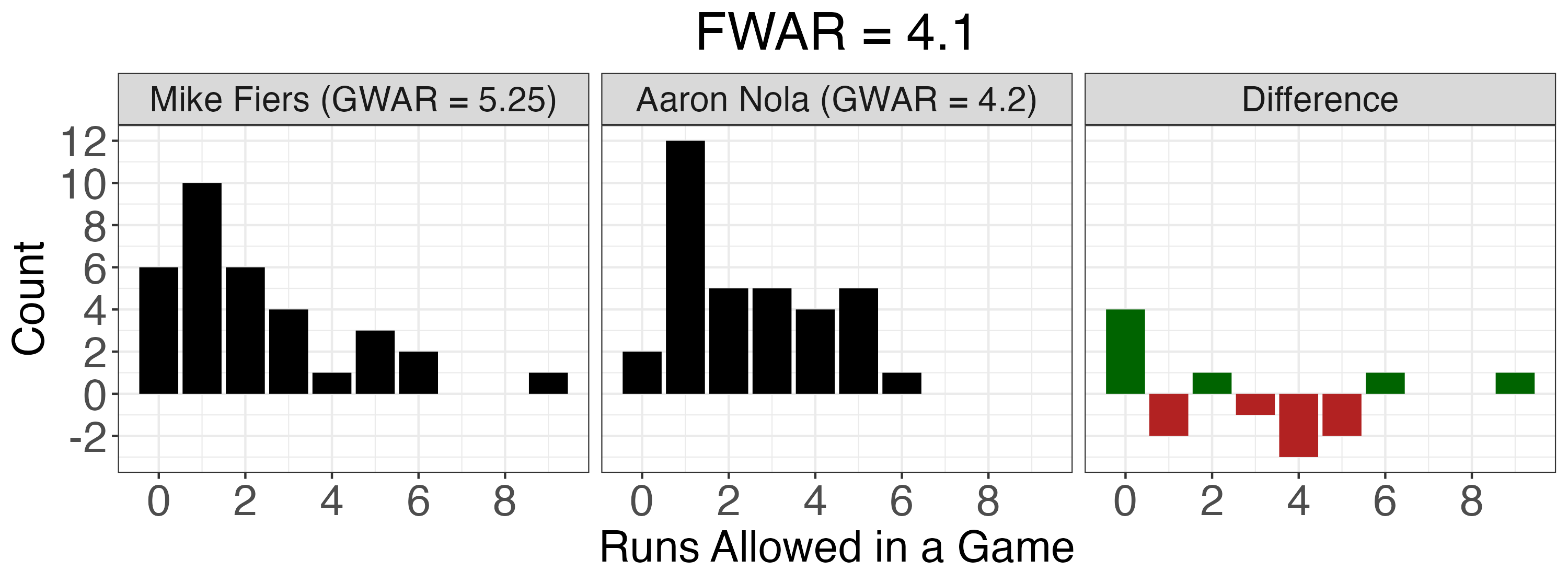}
\caption{
Histogram of runs allowed in a game in 2019 for Mike Fiers (left), Aaron Nola (middle), and the difference between these two histograms (right). 
Even though they have the same $\FWAR$, Fiers has a higher $\GWAR$ than Nola because he has more games in which he allows fewer runs.
} 
\label{fig:pf1}
\end{figure}

We begin with Figure~\ref{fig:gwarVfwar19}, which visualizes $\GWAR$ vs. $\FWARr$ for starting pitchers in 2019.
Pitchers who lie above the line $y=x$ are undervalued according to $\GWAR$ relative to $\FWAR$ and pitchers who lie below the line are overvalued.
To understand why some players are undervalued and others are overvalued, we compare players who have similar $\FWAR$ but different $\GWAR$ values in 2019. 
In Figure~\ref{fig:pf2} we compare Homer Bailey's 2019 season to Tanner Roark's.
They have the same $\FWARr$, $2.7$, but Bailey has a much higher $\GWAR$ (Bailey $3.43$, Roark $2.48$).
Similarly, in Figure~\ref{fig:pf1} we compare Mike Fiers's 2019 season to Aaron Nola's.
They have the same $\FWARr$, $4.1$, but Fiers has a higher $\GWAR$ (Fiers $5.25$, Nola $4.2$).

%%%%%%%%%%%%%%%%%%
%%%%%%%%%%%%%%%%%%

%%%%%%%%%%%%%%%%%%
%%%%%%%%%%%%%%%%%%

In each of these comparisons, we see a similar trend explaining the differences in $\GWAR$. 
The pitcher with higher $\GWAR$ allows fewer runs in more games and allows more runs in fewer games. 
This is depicted graphically in the ``Difference'' histograms, which show the difference between the histogram on the left and the histogram on the right. 
The green bars denote positive differences (i.e., the pitcher on the left has more games with a given number of runs allowed than the pitcher on the right) and the red bars denote negative differences (i.e., the pitcher on the left has fewer games with a given number of runs allowed than the pitcher on the right). 
In each of these examples, the green bars are shifted towards the left (pitchers with higher $\GWAR$ allow few runs in more games) and the red bars are shifted towards the right (pitchers with lower $\GWAR$ allow many runs in more games). 
In Figure~\ref{fig:pf2}, Bailey pitches four more games than Roark in which he allows two runs or fewer and Roark pitches four more games than Bailey in which he allows three runs or more. 
Similarly, in Figure~\ref{fig:pf1}, Fiers pitches four more games than Nola in which he allows zero runs and Nola pitches five more games than Fiers in which he allows one run or more.
The convexity of $\GWAR$ diminishes the impact of games in which a pitcher allows many runs, explaining why Bailey has more $\GWAR$ than Roark and why Fiers has more $\GWAR$ than Nola.

%%%%%%%%%%%%%%%%%%
\begin{figure}[hbt!]
\centering
\includegraphics[width=10cm]{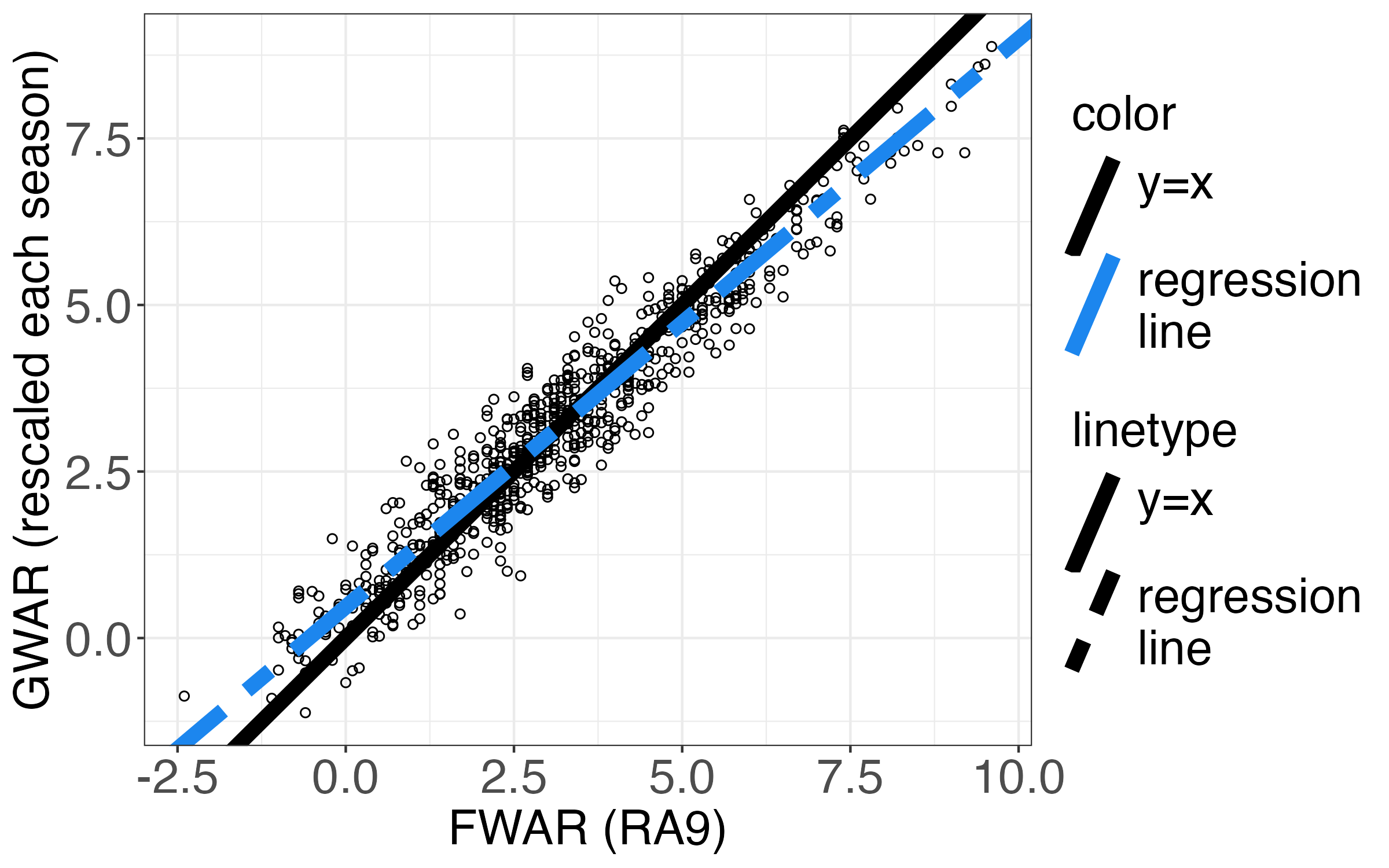}
\caption{
Grid $\WAR$ ($y$-axis) versus FanGraphs $\WAR$ ($\RA/9$) ($x$-axis) for each pitcher-season from 2010 to 2019. 
The solid black line is $y=x$ and the dashed blue line is the regression line $y = 0.47 + 0.85x$.
The slope of the regression line is less than 1, indicating that worse pitchers are generally undervalued and better pitchers are generally overvalued by $\FWARr$ relative to $\GWAR$.
} 
\label{fig:plot_career_fwar_v_gwar}
\end{figure}
%%%%%%%%%%%%%%%%%%

Next, in Figure~\ref{fig:plot_career_fwar_v_gwar} we visualize $\GWAR$ versus $\FWARr$ for each starting pitcher-season across all years from 2010 to 2019.
Pitchers who lie above the line $y=x$ (the solid black line) are undervalued according to $\GWAR$ relative to $\FWAR$ and pitchers who lie below the line are overvalued.
%%%%%%%%%%%%%
Worse pitchers generally lie above the line $y=x$ (the solid black line), and so are undervalued, whereas better pitchers generally lie below the line.
The regression line $y = 0.47 + 0.85x$ (the blue dashed line), which has slope less than one, summarizes this phenomenon.
This occurs because $\FWAR$ averages pitcher performance across an entire season, which dilutes the contributions of his good games.
Since Grid $\WAR$ is (mostly) convex in runs allowed, $\GWAR$ downweights the contribution of the $R^{th}$ run of a game for large $R$, whereas $\FWAR$ weighs all runs allowed in a season equally.
Since worse pitchers have many more occurrences than better pitchers of allowing a large number of runs allowed in a game, $\FWARr$ undervalues worse pitchers in general.
As we have constrained $\GWAR$ and $\FWAR$ to have the same sum, $\FWARr$ must therefore overvalue better pitchers in general.

\begin{figure}[htb!]
    \centering{}
    \subfloat[]{{\includegraphics[height=3.5cm]{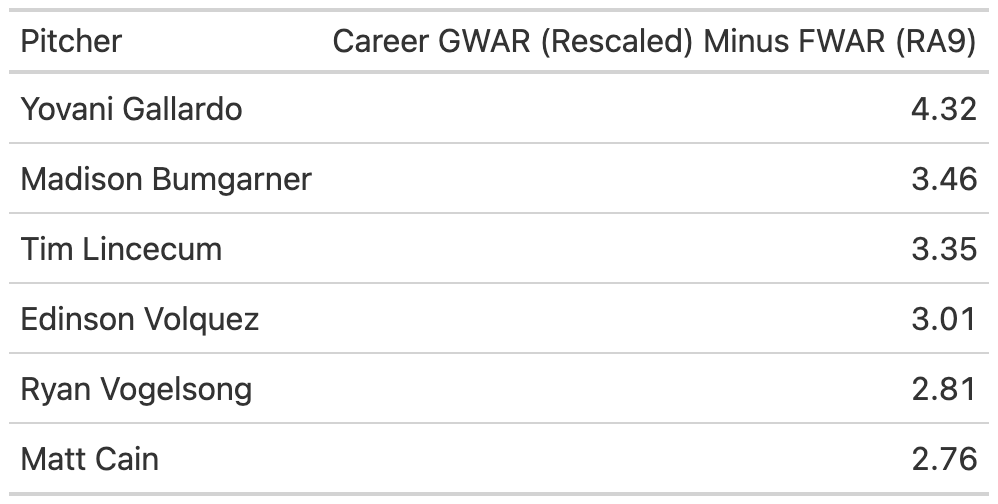}}\label{fig:undervalued_pits}}%
    \qquad
    \subfloat[]{{\includegraphics[height=3.5cm]{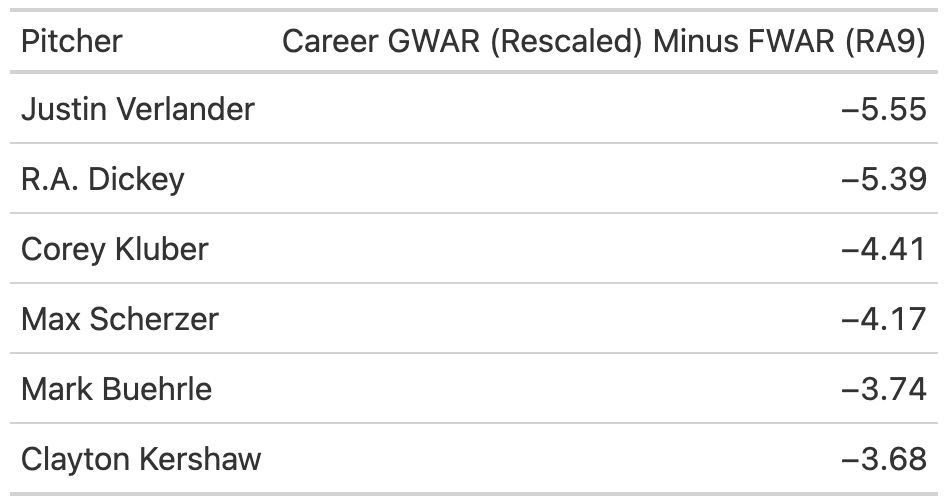}}\label{fig:overvalued_pits}}%
    \caption{
    The six most undervalued (Figure (a)) and overvalued (Figure (b)) pitchers according to $\GWAR$ relative to $\FWARr$, aggregated across all seasons from 2010 to 2019.  The undervalued pitchers are generally considered worse pitchers and the overvalued pitchers are generally considered better pitchers.
    }
    \label{fig:underOverValuedPits}
\end{figure}

\begin{figure}[hbt!]
\centering
\includegraphics[width=15cm]{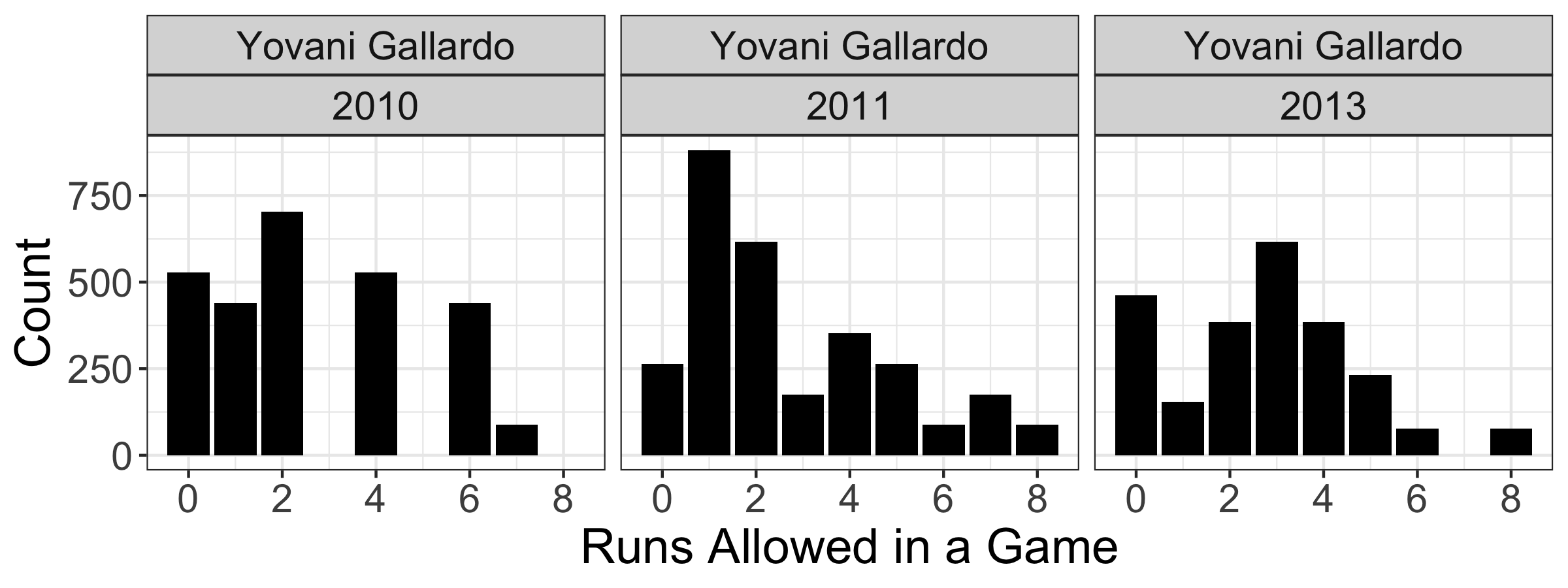}
\caption{
Histogram of Yovani Gallardo's runs allowed in each game across three seasons.
Gallardo has many great and terrible games. Traditional $\WAR$ metrics allow his terrible performances to dilute his great ones.
} 
\label{fig:plot_Yovani_Gallardo_undervalued_3_pit_hists}
\end{figure}

In Figure~\ref{fig:underOverValuedPits} we show the six most undervalued and overvalued pitchers according to $\GWAR$ relative to $\FWARr$, aggregated across all seasons from 2010 to 2019.
As expected, the undervalued pitchers are generally considered worse pitchers and the overvalued pitchers are generally considered better pitchers.
In Figure~\ref{fig:plot_Yovani_Gallardo_undervalued_3_pit_hists} we visualize the runs allowed distribution of the most undervalued pitcher, Yovani Gallardo, in his three most undervalued seasons.
Gallardo has many great games (say, zero or one run allowed) and many terrible games (say, six or more runs allowed).
$\GWAR$ diminishes the impact of these terrible games and magnifies the impact of these great games, increasing his estimated contribution to winning.

%%%%%%%%%%%%%%%%%%%%%
%%%%%%%%%%%%%%%%%%%%%

%%%%%%%%%%%%%%%%%%
%%%%%%%%%%%%%%%%%%

\begin{figure}[htb!]
    \centering
    \includegraphics[width=10cm]{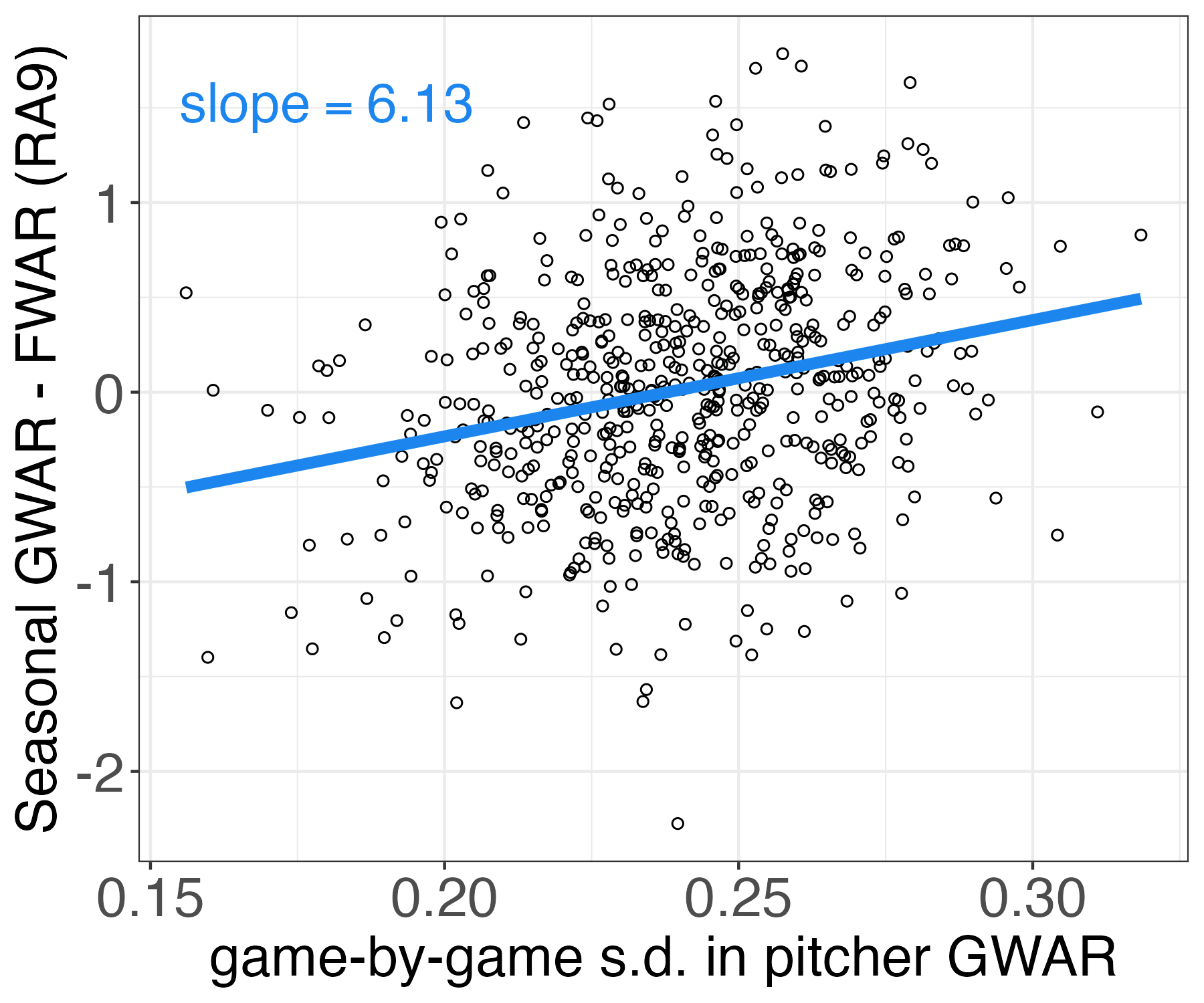}
    \caption{
    For each starting pitcher-season from 2010 to 2019 consisting of at least 30 games, the difference between his seasonal $\GWAR$ and $\FWARr$ ($y$-axis) versus the game-by-game standard deviation of his $\GWAR$ ($x$-axis).  Higher variance pitchers are undervalued by $\FWARr$ because they allow a pitcher's worse performances to dilute his better performances.
    } 
    \label{fig:plot_var_vs_diff}
\end{figure}

As averaging pitcher performance across games allows a pitcher's worse performances to dilute his better performances, we also expect FanGraphs $\WAR$ to undervalue higher variance pitchers. We see this in Figure~\ref{fig:plot_var_vs_diff} in which we visualize the relationship between a pitcher's variability in performance across games and the extent to which he is undervalued by FanGraphs $\WAR$.  For each starting pitcher-season from 2010 to 2019 consisting of at least 30 games, we plot the difference between his seasonal $\GWAR$ and $\FWARr$ against the game-by-game standard deviation of his $\GWAR$.  
Higher variance pitchers indeed generally have higher $\GWAR$ than $\FWAR$.

%%%%%%%%%%%%%%%%%%%%%
%%%%%%%%%%%%%%%%%%%%%

%%%%%%%%%%%%%%%%%%%%%%%%%%%%%%%%%%%%%%%%%%%%%%%%%%%%%%%%%%%%%%%%%%%%%%%%%%%
%%%%%%%%%%%%%%%%%%%%%%%%%%%%%%%%%%%%%%%%%%%%%%%%%%%%%%%%%%%%%%%%%%%%%%%%%%%
\subsection{Grid $\WAR$ has predictive value}\label{sec:predictive_value_gwar}

Recall that it is wrong to average pitcher performance across games in estimating a starting pitcher's $\WAR$.
This is why we devised Grid $\WAR$, which estimates $\WAR$ on a per-game basis and is the right way to estimate historical $\WAR$ for starting pitchers.
Nonetheless, it is possible \textit{a priori} that averaging provides a stabler estimate of pitcher quality which is more predictive of future $\WAR$.
In particular, if game-by-game variability in pitcher performance is due mostly to chance, accounting for these variations introduces noise into estimates of pitcher quality. 

Because a pitcher’s historical $\WAR$ at the end of a season defines how valuable he was during the season, we want a predictive pitcher quality measure to predict his next season’s historical $\WAR$ as best as possible. 
In other words, pitcher quality is simply predicted future historical $\WAR$.
Hence our goal is to predict a starting pitcher's future Grid $\WAR$.
\textit{A priori}, it is not immediately obvious whether a pitcher's past Grid $\WAR$ is predictive of his future Grid $\WAR$.
In particular, if a pitcher's game-by-game variance in runs allowed is due mostly to randomness rather than a fundamental identifiable trait, a $\WAR$ which averages pitcher performance over the season may be more predictive than Grid $\WAR$ of future Grid $\WAR$.
Thus, in this section, we compare the predictive capabilities of Grid $\WAR$ and FanGraphs $\WAR$.
We find that, in predicting future pitcher rankings according to Grid $\WAR$, our predicted pitcher ranking built from Grid $\WAR$ is more predictive than that built from FanGraphs $\WAR$.
This suggests that some pitchers' game-by-game variance in performance is a fundamental trait.

To value a starting pitcher using his previous seasons' $\WAR$ and number of games played, we could simply use his mean game $\WAR$.
The fewer games a pitcher has played, however, the less reliable his mean game $\WAR$ is in predicting his latent pitcher quality. 
Therefore, we use shrinkage estimation to construct a pitcher quality metric.
In calculating pitcher $p$'s quality estimate $\widehat\mu_p$, the fewer games he has played, the more his mean game $\WAR$ is shrunk towards the overall mean pitcher quality.
Specifically, we construct three shrinkage estimators of pitcher $p$'s quality, denoted $\widehat{\mu}_p^{\GWAR}$, $\widehat{\mu}_p^{\FWARf}$, and $\widehat{\mu}_p^{\FWARr}$, built from the three respective $\WAR$ metrics.
We use a parametric Empirical Bayes approach in the spirit of \citet{brown2008} to formulate these shrinkage estimators, detailed in Appendix~\ref{sec:empiricalBayes}.

Recall that our goal is to predict each starting pitcher's next season's cumulative Grid $\WAR$, which at the end of next season will represent his historical value added.
So, using the 2019 season as a hold-out validation set, our goal is to predict each starting pitcher's 2019 Grid $\WAR$.
We use our remaining data from 2010 to 2018 to estimate pitcher quality, built separately from $\GWAR$ and $\FWAR$, in order to predict 2019 Grid $\WAR$.
Thus we restrict our analysis to the set of starting pitchers who have a FanGraphs' $\WAR$ in at least one season from 2010 to 2018 (so, they must have at least 25 starts in that season).
Our pitcher quality estimators, however, are on different scales since each $\WAR$ metric is on its own scale.
To ensure fair comparison of Grid $\WAR$ and FanGraphs $\WAR$, we map each estimator to a starting pitcher ranking, ranking each pitcher from one (best) to the number of pitchers (worst).
% Hence we use each of the three estimators to construct three starting pitcher rankings.
We denote the three ranks of pitcher $p$ according to $\widehat{\mu}_p^{\GWAR}$, $\widehat{\mu}_p^{\FWARf}$, and $\widehat{\mu}_p^{\FWARr}$ by $\widehat{R}_p^{\GWAR}$, $\widehat{R}_p^{\FWARf}$, and $\widehat{R}_p^{\FWARr}$, respectively.
In Figure~\ref{fig:plot_EB_pitRankings} of Appendix~\ref{sec:empiricalBayes} we visualize these starting pitcher rankings prior to the 2019 season according to these estimators $\widehat\mu_p$ (left) and their associated ranks $\widehat{R}_p$ (right).
% Finally, to ensure a fair comparison of Grid $\WAR$ and FanGraphs $\WAR$, 
Additionally, we rank pitchers in 2019 by their observed cumulative 2019 Grid $\WAR$, denoted $R_p^{\GWAR}$.
Finally, we use root mean squared error ($\rmse$) to measure how well the predicted pitcher rankings $\widehat R$ predict the observed rankings $R$, shown in Table~\ref{table:pred_war_rmse}.
We see that pitcher rankings built from Grid $\WAR$ are more predictive than those built from FanGraphs $\WAR$.
Formally, $A < B$ and $A < C$, where $A$, $B$, and $C$ are defined in Table~\ref{table:pred_war_rmse}.
In other words, baseball analysts lose value by not using Grid $\WAR$ to value pitchers.

%%%%%%%%%%%%%%%%%%%%%%%%%%%%%%%%%%%%
\begin{table}[hbt!]
\centering
\begin{tabular}{ llll } \hline 
symbol & observed ranking $R$ & predicted ranking $\widehat{R}$ & $\rmse(R,\widehat{R})$ \\ \hline
A & ${R}^{\GWAR}$ & $\widehat{R}^{\GWAR}$ & 10.2 \\ %[0.3cm]
B & ${R}^{\GWAR}$ & $\widehat{R}^{\FWARr}$ & 12.4 \\ %[0.3cm]
C & ${R}^{\GWAR}$ & $\widehat{R}^{\FWARf}$ & 13.1 \\ %[0.3cm]
% B & ${R}^{\FWARr}$ & $\widehat{R}^{\FWARr}$ & 14.47 \\ [0.3cm]
% C & ${R}^{\FWARf}$ & $\widehat{R}^{\FWARf}$ & 13.83 \\ [0.3cm]
\hline
\end{tabular}
\caption{The $\rmse$ of the observed pitcher ranking ${R}^{\GWAR}$ in 2019 and three pitcher ranking estimates $\widehat{R}$ computed from three different $\WAR$ metrics.
$\GWAR$ is more predictive than $\FWAR$ of future $\GWAR$.
}
\label{table:pred_war_rmse}
\end{table}
%%%%%%%%%%%%%%%%%%%%%%%%%%%%%%%%%%%%

\begin{figure}[htb!]
    \centering{}
    \includegraphics[width=15cm]{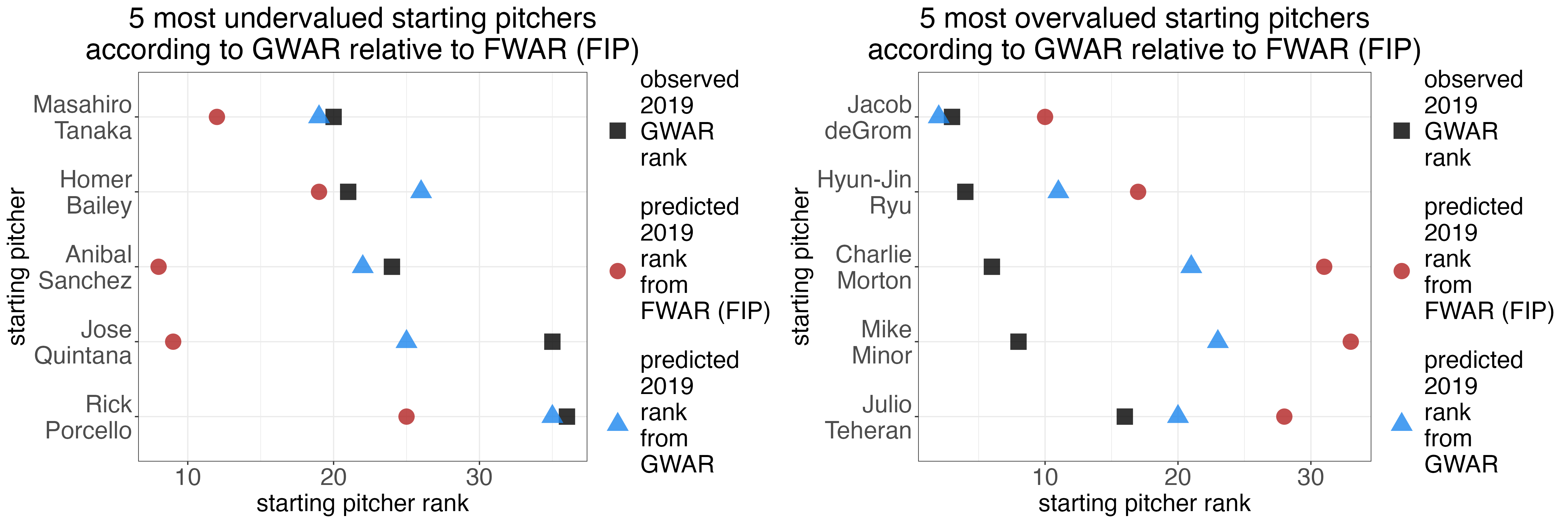}
    \caption{
    Visualizing the observed pitcher rankings ${R}^{\GWAR}$ in 2019 and pitcher ranking estimates $\widehat{R}^{\FWARf}$ of the five most undervalued (left) and overvalued (right) starting pitchers in 2019 (according to $\widehat{R}^{\GWAR}$ relative to $\widehat{R}^{\FWARf}$). 
    As the blue triangles (our 2019 $\GWAR$-based predictions) are closer to the black squares (the observed 2019 pitcher rankings according to $\GWAR$) than the red dots (our 2019 $\FWARf$-based predictions), $\GWAR$ is more predictive than $\FWAR$ of future $\GWAR$.
    }
    \label{fig:pred_war_rmse_underOvervalued}
\end{figure}

In Tables \ref{table:pred_war_rmse_undervalued_r} and \ref{table:pred_war_rmse_undervalued_f} we conduct a similar analysis, but restricting the test set to just the five most \textit{undervalued} starting pitchers in 2019 according to $\widehat{R}^{\GWAR}$ relative to $\widehat{R}^{\FWARr}$ (in Table~\ref{table:pred_war_rmse_undervalued_r}) and relative to $\widehat{R}^{\FWARf}$ (in Table~\ref{table:pred_war_rmse_undervalued_f}).
Conversely, in Tables \ref{table:pred_war_rmse_overvalued_r} and \ref{table:pred_war_rmse_overvalued_f} we conduct a similar analysis, but restricting the test set to just the five most \textit{overvalued} starting pitchers in 2019 according to $\widehat{R}^{\GWAR}$ relative to $\widehat{R}^{\FWARr}$ (in Table~\ref{table:pred_war_rmse_overvalued_r}) and relative to $\widehat{R}^{\FWARf}$ (in Table~\ref{table:pred_war_rmse_overvalued_f}).
We again find that baseball analysts lose value by not using Grid $\WAR$ to estimate pitcher quality.
In particular, for these ``extreme'' pitchers who are highly undervalued or highly overvalued, analysts do worse predicting their quality when they use $\FWAR$ rather than $\GWAR$. 
%%%%%%%%%%%%%
In Figure~\ref{fig:pred_war_rmse_underOvervalued} we visualize how our $\GWAR$- and $\FWAR$-based pitcher ranking predictions fare against the observed 2019 $\GWAR$ pitcher rankings.
Specifically, the blue triangles (our 2019 $\GWAR$-based predictions) are closer to the black squares (the observed 2019 pitcher rankings according to $\GWAR$) than the red dots (our 2019 $\FWARf$-based predictions).

%%%%%%%%%%%%%%%%%%%%%%%%%%%%%%%%%%%%
\begin{table}[hbt!]
\centering
\begin{tabular}{ lll } \hline 
observed ranking $R$ & predicted ranking $\widehat{R}$ & $\rmse(R,\widehat{R})$ \\ \hline
${R}^{\GWAR}$ & $\widehat{R}^{\GWAR}$ & 7.2 \\ %[0.3cm]
${R}^{\GWAR}$ & $\widehat{R}^{\FWARr}$ & 15.7  \\ %[0.3cm]
\hline
\end{tabular}
\caption{The $\rmse$ of the observed pitcher ranking ${R}^{\GWAR}$ in 2019 and pitcher ranking estimates $\widehat{R}$ computed from three different $\WAR$ metrics, using just the five most \textit{undervalued} starting pitchers in 2019 according to $\widehat{R}^{\GWAR}$ relative to each $\widehat{R}^{\FWARr}$.
$\GWAR$ is more predictive than $\FWARr$ of future $\GWAR$, particularly for undervalued pitchers. }
\label{table:pred_war_rmse_undervalued_r}
\end{table}
%%%%%%%%%%%%%%%%%%%%%%%%%%%%%%%%%%%%

%%%%%%%%%%%%%%%%%%%%%%%%%%%%%%%%%%%%
\begin{table}[hbt!]
\centering
\begin{tabular}{ lll } \hline 
observed ranking $R$ & predicted ranking $\widehat{R}$ & $\rmse(R,\widehat{R})$ \\ \hline 
${R}^{\GWAR}$ & $\widehat{R}^{\GWAR}$ & 5.1 \\ %[0.3cm]
${R}^{\GWAR}$ & $\widehat{R}^{\FWARf}$ & 15.0 \\ %[0.3cm]
\hline
\end{tabular}
\caption{The $\rmse$ of the observed pitcher ranking ${R}^{\GWAR}$ in 2019 and pitcher ranking estimates $\widehat{R}$ computed from three different $\WAR$ metrics, using just the five most \textit{undervalued} starting pitchers in 2019 according to $\widehat{R}^{\GWAR}$ relative to each $\widehat{R}^{\FWARf}$. 
$\GWAR$ is more predictive than $\FWARf$ of future $\GWAR$, particularly for undervalued pitchers.}
\label{table:pred_war_rmse_undervalued_f}
\end{table}
%%%%%%%%%%%%%%%%%%%%%%%%%%%%%%%%%%%%

%%%%%%%%%%%%%%%%%%%%%%%%%%%%%%%%%%%%
\begin{table}[hbt!]
\centering
\begin{tabular}{ lll } \hline 
observed ranking $R$ & predicted ranking $\widehat{R}$ & $\rmse(R,\widehat{R})$ \\ \hline
${R}^{\GWAR}$ & $\widehat{R}^{\GWAR}$ & 10.7 \\ %[0.3cm]
${R}^{\GWAR}$ & $\widehat{R}^{\FWARr}$ & 14.0 \\ %[0.3cm]
\hline
\end{tabular}
\caption{The $\rmse$ of the observed pitcher ranking ${R}^{\GWAR}$ in 2019 and pitcher ranking estimates $\widehat{R}$ computed from three different $\WAR$ metrics, using just the five most \textit{overvalued} starting pitchers in 2019 according to $\widehat{R}^{\GWAR}$ relative to each $\widehat{R}^{\FWARr}$. 
$\GWAR$ is more predictive than $\FWARr$ of future $\GWAR$, particularly for overvalued pitchers.}
\label{table:pred_war_rmse_overvalued_r}
\end{table}
%%%%%%%%%%%%%%%%%%%%%%%%%%%%%%%%%%%%

%%%%%%%%%%%%%%%%%%%%%%%%%%%%%%%%%%%%
\begin{table}[hbt!]
\centering
\begin{tabular}{ lll } \hline 
observed ranking $R$ & predicted ranking $\widehat{R}$ & $\rmse(R,\widehat{R})$ \\ \hline
${R}^{\GWAR}$ & $\widehat{R}^{\GWAR}$ & 10.2 \\ %[0.1cm]
${R}^{\GWAR}$ & $\widehat{R}^{\FWARf}$ & 18.0 \\ %[0.3cm]
\hline
\end{tabular}
\caption{The $\rmse$ of the observed pitcher ranking ${R}^{\GWAR}$ in 2019 and pitcher ranking estimates $\widehat{R}$ computed from three different $\WAR$ metrics, using just the five most \textit{overvalued} starting pitchers in 2019 according to $\widehat{R}^{\GWAR}$ relative to each $\widehat{R}^{\FWARf}$. 
$\GWAR$ is more predictive than $\FWARf$ of future $\GWAR$, particularly for overvalued pitchers.}
\label{table:pred_war_rmse_overvalued_f}
\end{table}
%%%%%%%%%%%%%%%%%%%%%%%%%%%%%%%%%%%%

%%%%%%%%%%%%%%%%%%%%%
%%%%%%%%%%%%%%%%%%%%%

\begin{figure}[htb!]
    \centering
    \includegraphics[width=15cm]{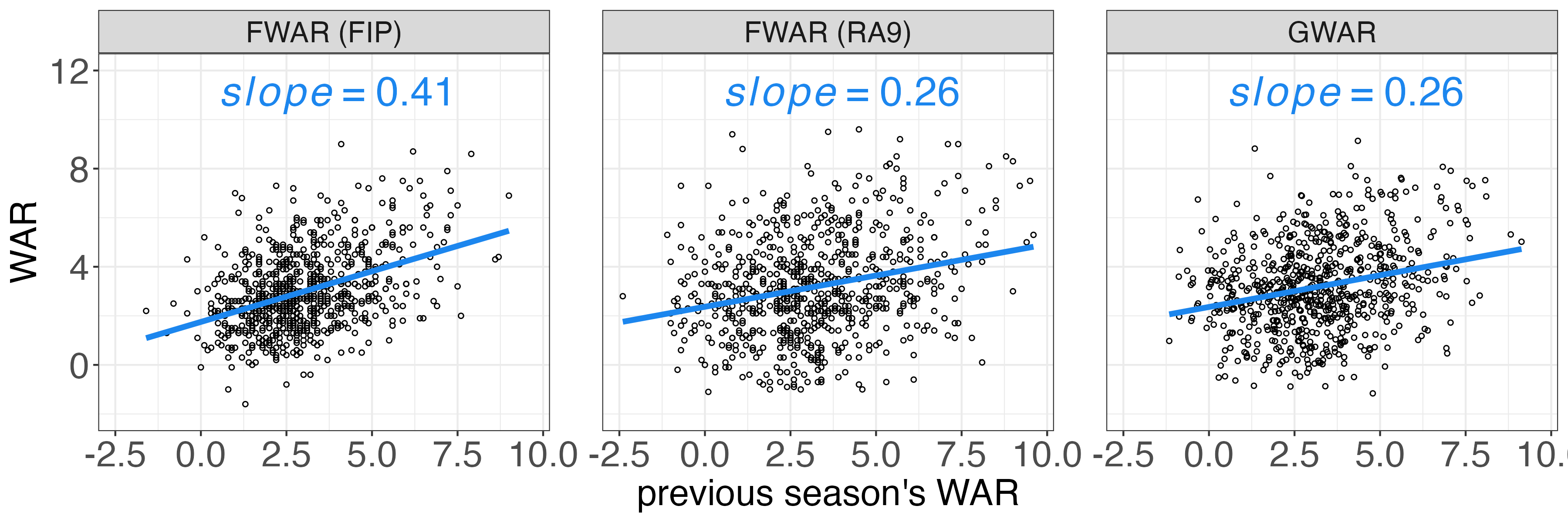}
    \caption{
    For each pitcher-season from 2011 to 2019, a pitcher's seasonal $\WAR$ ($y$-axis) versus his previous season's $\WAR$ ($x$-axis). $\FWARf$ is more stable from year to year than $\GWAR$ or $\FWARr$, which have similar levels of season-to-season variability.
    } 
    \label{fig:szn_by_szn_WAR_variability}
\end{figure}

Further, in Figure~\ref{fig:szn_by_szn_WAR_variability} we visualize the variability of Grid $\WAR$ and FanGraphs $\WAR$ from season to season.  For each pitcher-season from 2011 to 2019, we plot a pitcher's seasonal $\WAR$ against his previous season's $\WAR$.  We see that $\FWARf$ is more stable from season to season than $\GWAR$ and $\FWARr$, which have similar levels of season-to-season variability. This makes sense: runs allowed, from which $\GWAR$ and $\FWARr$ are constructed, is inherently more noisy than $\FIP$, which doesn't account for balls in play. Having greater stability across seasons, however, does not mean that $\FWARf$ is a better pitcher valuation metric than $\GWAR$. As shown previously in this section, an estimate of latent pitcher quality built from $\GWAR$ is more predictive of future $\GWAR$ than estimates built from $\FWARf$ and $\FWARr$.  Predictiveness of future Grid $\WAR$ is a more important quality than year-to-year stability, as a general manager should want to acquire a starting pitcher who will have more Grid $\WAR$ in future seasons.

%%%%%%%%%%%%%%%%%%%%%
%%%%%%%%%%%%%%%%%%%%%

% %%%%%%%%%%%%%%%%%%%%%%%%%%%%%%%%%%%%%%%%%%%%%%%%%%%%%%%%%%%%%%%%%%%%%%%%%%%
% %%%%%%%%%%%%%%%%%%%%%%%%%%%%%%%%%%%%%%%%%%%%%%%%%%%%%%%%%%%%%%%%%%%%%%%%%%%
% \subsection{Traditional $\WAR$ metrics undervalue mediocrity}\label{sec:pitcherQuality_gameByGameGridWar}

So, a game-by-game measure of $\WAR$ like Grid $\WAR$ is not only the right way to measure historical $\WAR$ for starting pitchers, but is also predictive of future Grid $\WAR$.
In particular, an estimator of latent pitcher talent should be built using Grid $\WAR$ or some other game-by-game metric. %, not just using FanGraphs $\WAR$.
Now that we have such an estimator of pitcher talent at our disposal, we explore the relationship between a pitcher's talent according to $\widehat\mu_p^{\GWAR}$ and his game-by-game performance. 
We find that all pitchers have great games, but great pitchers have few terrible games.
Therefore, averaging pitcher performance over the entire season dilutes the value of mediocre pitchers' good games, causing existing $\WAR$ metrics to undervalue mediocrity.
This agrees with our assessment from the previous section.

\begin{figure}[htb!]
    \centering{}
    \subfloat[]{{\includegraphics[width=7cm]{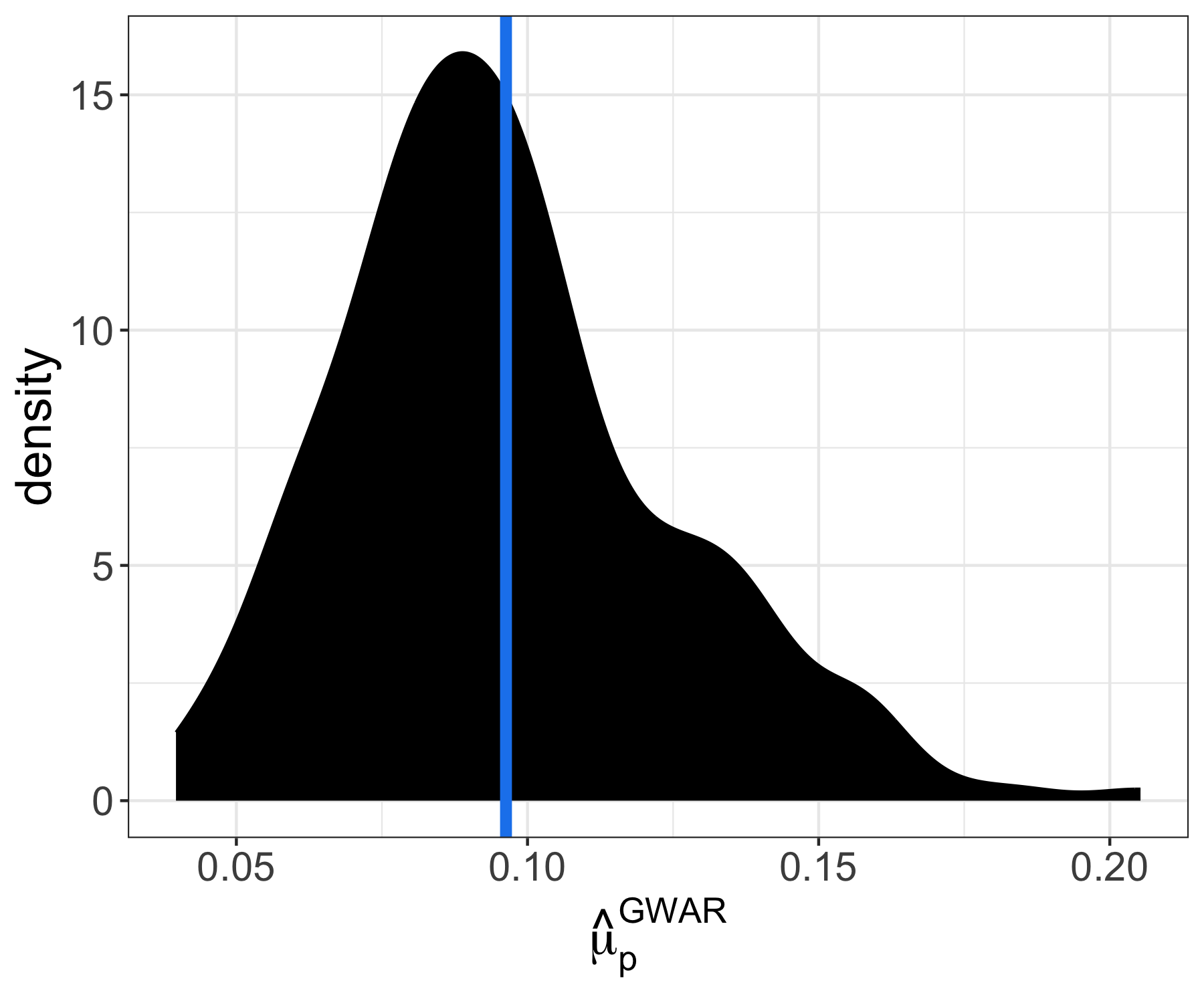}}\label{fig:plot_EB_gwarTalentDist}}%
    \qquad
    \subfloat[]{{\includegraphics[width=7cm]{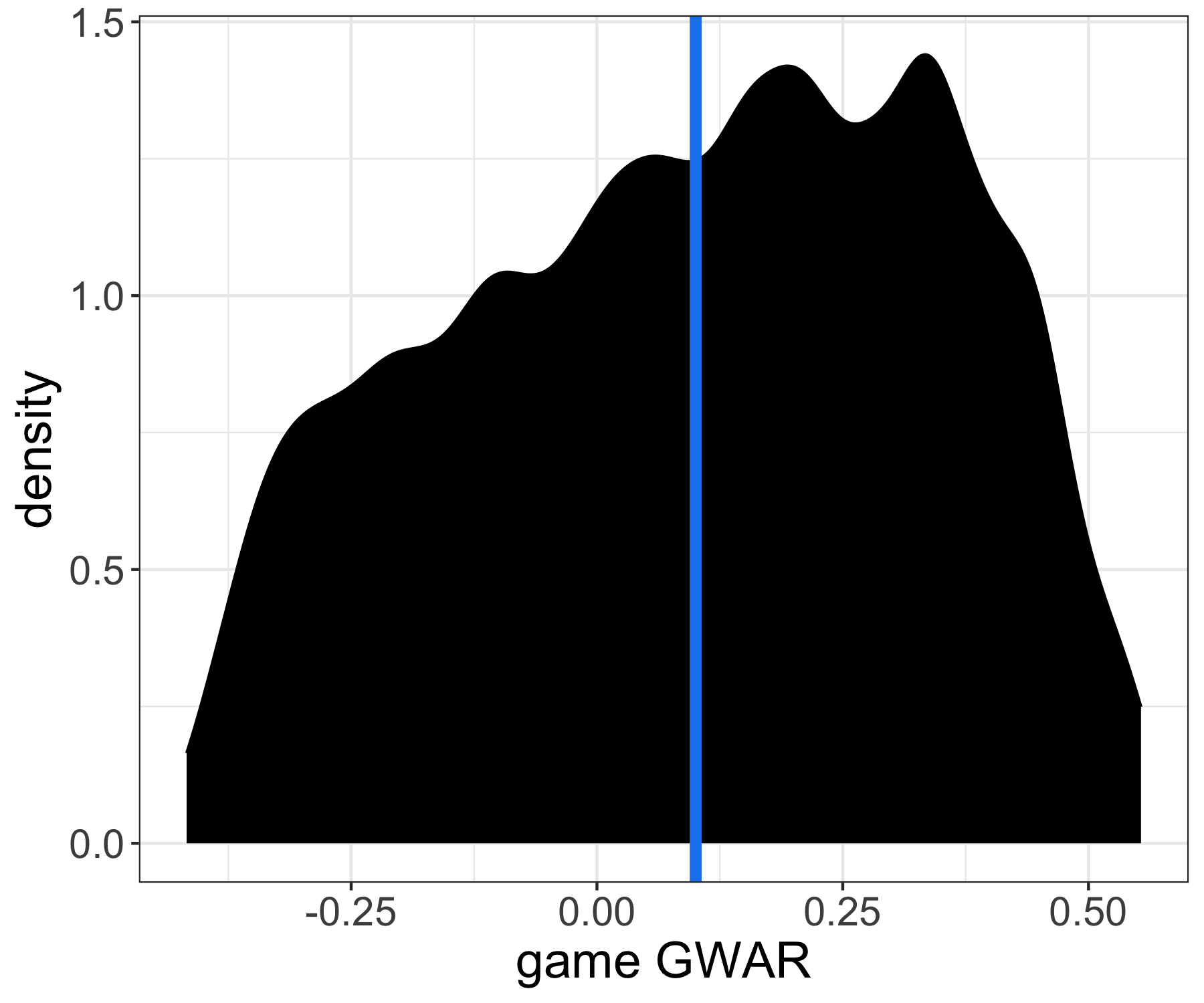}}\label{fig:plot_EB_gwarGameDist}}%
    \caption{
    Distribution of estimated pitcher talent $\widehat\mu_p^{\GWAR}$ (Figure (a)) and game-by-game Grid $\WAR$ (Figure (b)). The blue lines denote the respective means. 
    }
    \label{fig:plot_EB_gwarTalentAndGameDists}
\end{figure}

\begin{figure}[htb!]
    \centering{}
    \subfloat[]{{\includegraphics[width=7cm]{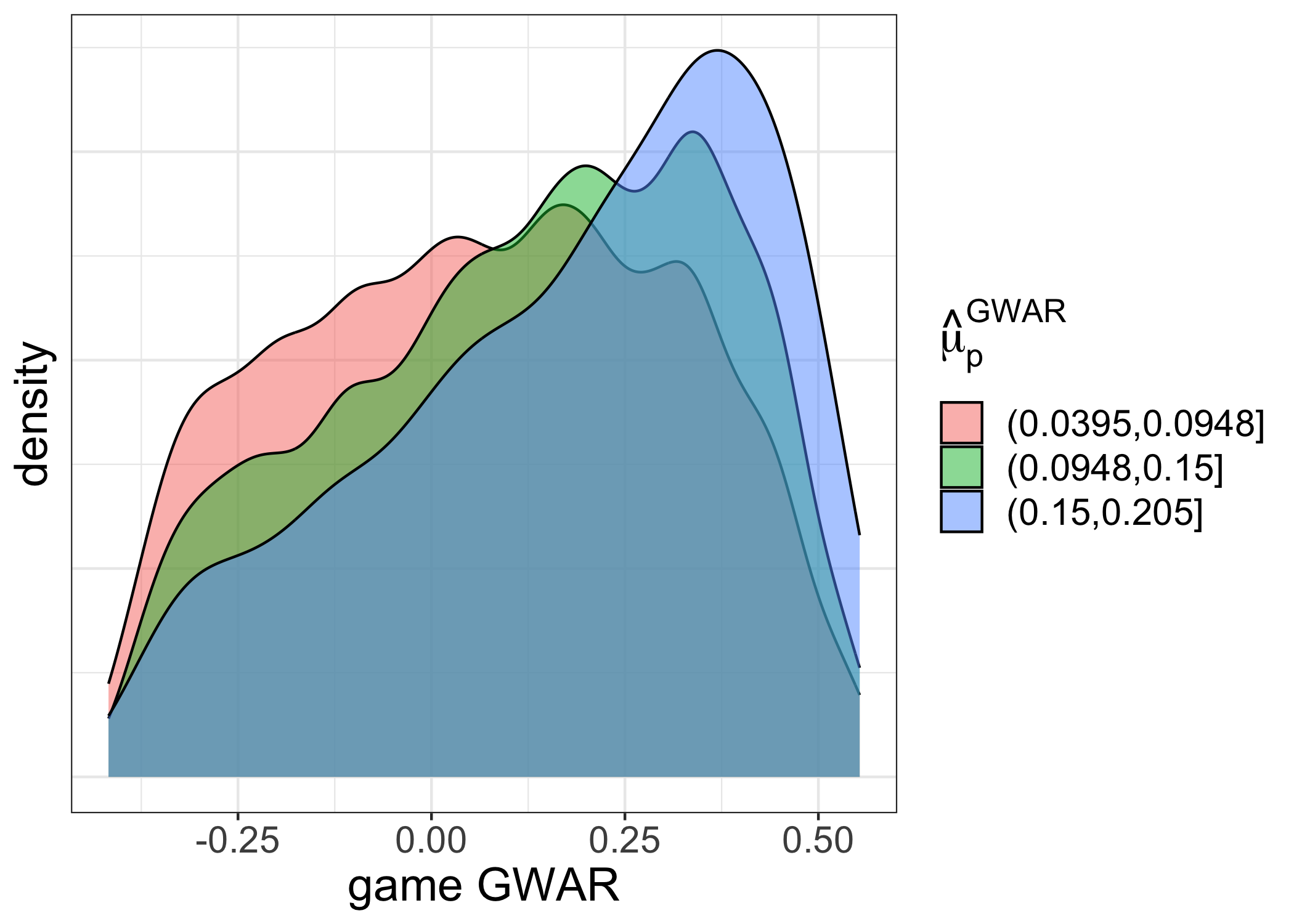}}\label{fig:plot_EB_gameGwarDist_givenTalent}}%
    % \qquad
    \subfloat[]{{\includegraphics[width=7cm]{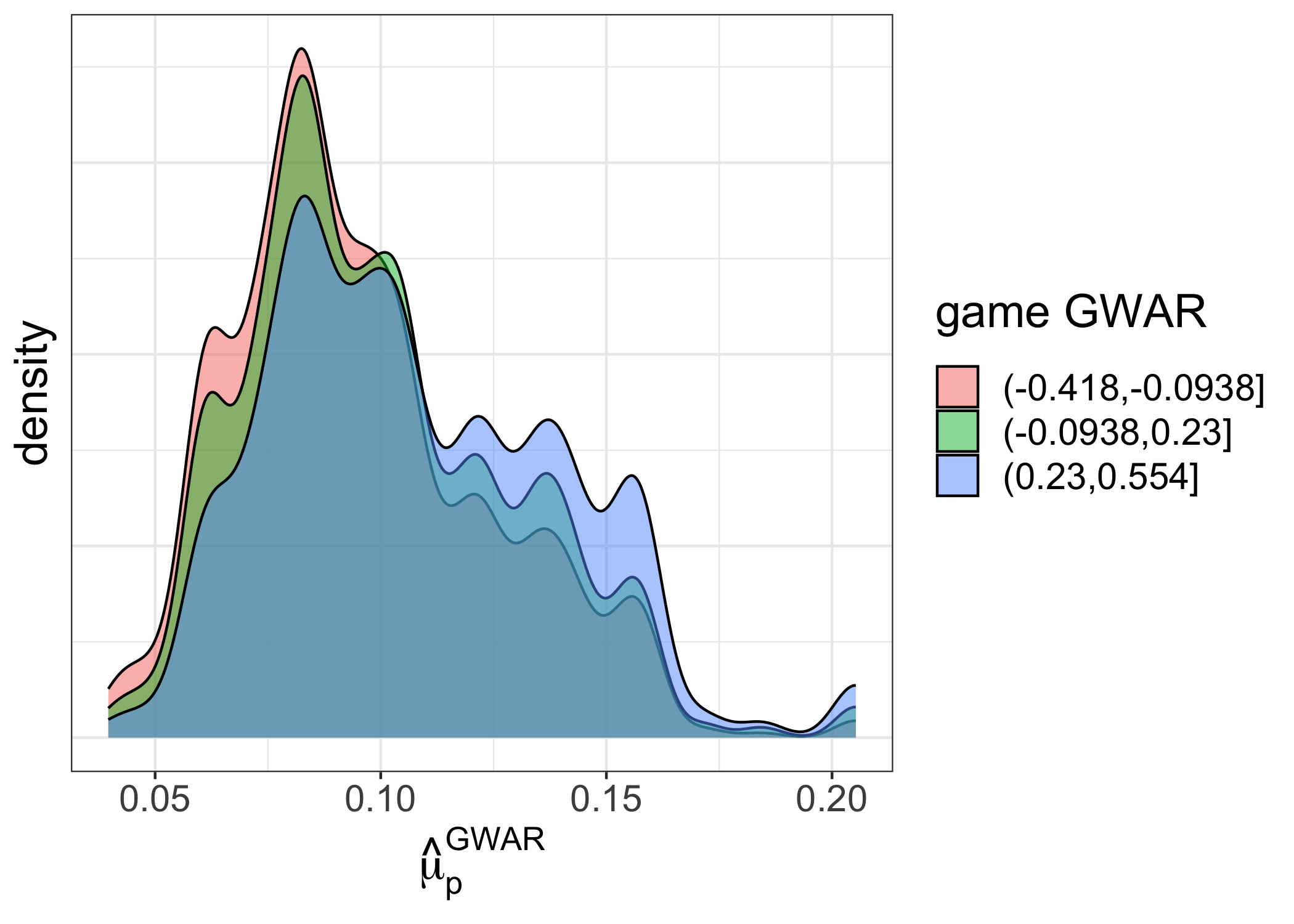}}\label{fig:plot_EB_plot_gwarTalentDist_givenGameGwar}}%
    \caption{
        Figure (a) shows the distribution of game-by-game Grid $\WAR$ conditional on being a bad pitcher (red), a typical pitcher (green), and a great pitcher (blue), according to $\widehat\mu_p^{\GWAR}$.
        Figure (b) shows the distribution of estimated pitcher talent $\widehat\mu_p^{\GWAR}$ conditional on having a bad game (red), a typical game (green), and a great game (blue).
        All pitchers have terrible games, but great pitchers have fewer of them.
    }
    \label{fig:plot_EB_Conditional_gwarTalentAndGameDists}
\end{figure}

In Figure~\ref{fig:plot_EB_gwarTalentAndGameDists} we provide a sense of the distribution of pitcher talent $\widehat\mu_p^{\GWAR}$ (left) and of the distribution of game-by-game Grid $\WAR$ (right).
Then, in Figure~\ref{fig:plot_EB_gameGwarDist_givenTalent}, we visualize the distribution of game-by-game Grid $\WAR$ conditional on being a bad pitcher (red), a typical pitcher (green), and a great pitcher (blue), according to $\widehat\mu_p^{\GWAR}$.
Bad pitchers, typical pitchers, and great pitchers all have great games.
Great pitchers, on the other hand, pitch many fewer bad games than bad and mediocre pitchers do.
In Figure \ref{fig:plot_EB_plot_gwarTalentDist_givenGameGwar} we view this phenomenon through another lens.
We visualize the distribution of pitcher quality $\widehat\mu_p^{\GWAR}$ conditional on having a bad game (red), a typical game (green), and a great game (blue).
Bad pitchers, typical pitchers, and great pitchers all have great games, but bad games feature a higher proportion of bad pitchers.

Averaging pitcher performance over the season allows a pitcher's bad performances to dilute the value of his good ones.
Consequently, such $\WAR$ metrics like FanGraphs $\WAR$ devalue the contributions of mediocre and bad pitchers, who have many more bad games than great pitchers.
In short, the baseball community has been undervaluing the contributions of the mediocre.

%%%%%%%%%%%%%%%%%%%%%
%%%%%%%%%%%%%%%%%%%%%

%%%%%%%%%%%%%%%%%%%%%
%%%%%%%%%%%%%%%%%%%%%

%%%%%%%%%%%%%%%%%%%%%%%%%%%%%%%%%%%%%%%%%%%%%%%%%%%%%%%%%%%%%%%%%%%%%%%%%%%
%%%%%%%%%%%%%%%%%%%%%%%%%%%%%%%%%%%%%%%%%%%%%%%%%%%%%%%%%%%%%%%%%%%%%%%%%%%
%%%%%%%%%%%%%%%%%%%%%%%%%%%%%%%%%%%%%%%%%%%%%%%%%%%%%%%%%%%%%%%%%%%%%%%%%%%
\section{Discussion}\label{sec:Discussion}

Traditional implementations of $\WAR$ for starting pitchers estimate $\WAR$ as a function of pitcher performance averaged over the entire season.
Averaging pitcher performance, however, allows a pitcher's bad games to dilute the performances of his good games.
One bad ``blow-up'' game after averaging can reduce a pitcher's $\WAR$ by more than minimum possible $\WAR$ in a game.
Therefore, a starters' seasonal $\WAR$ should be the sum of the $\WAR$ of each of his individual games. 
Hence we devise Grid $\WAR$, which estimates a starting pitcher's $\WAR$ in each of his games.
Grid $\WAR$ estimates the context-neutral win probability added above replacement at the point when a pitcher exits the game.
Grid $\WAR$ is convex in runs allowed, capturing the fundamental baseball principle that you can only lose a game once.

Comparing starting pitchers' Grid $\WAR$ to his FanGraphs $\WAR$ from 2010 to 2019, 
we find that standard $\WAR$ calculations undervalue mediocrity and variability relative to Grid $\WAR$.
Because all starters pitch great games, but great starters don't pitch many terrible games, averaging pitcher performance over a season discounts the contributions of great games by mediocre and bad pitchers.
We also show that past performance according to Grid $\WAR$ is predictive of future Grid $\WAR$, providing evidence that a pitcher's runs allowed profile is not entirely  random, but is the result of an identifiable game-by-game variation.

To compare starting pitchers across baseball history through the lens of Grid $\WAR$, we created an interactive Shiny app,\footnote{
    The website is built using pre-2008 play-by-play data from \cite{retroRaw} and play-by-play data since 2008 from \cite{statcast}.
    We use the \texttt{baseballr} package in \textsf{R} to scrape from each of these data sources \citep{petti_gilani_2021}.
    We automatically scrape Statcast data each morning, so the website is up-to-date. 
} hosted at \url{https://gridwar.xyz}, which displays the Grid $\WAR$ results of every starting pitcher game, season, and career since 1952.
For many starters, Grid $\WAR$ is similar to FanGraphs $\WAR$.\footnote{
    \url{https://www.fangraphs.com/leaders/major-league?pos=all&lg=all&qual=y&type=8&month=0&ind=0&team=0&rost=0&players=0&startdate=&enddate=&season1=1952&season=2023&stats=sta&sortcol=20&sortdir=default&pagenum=1&pageitems=2000000000}
}
Grid $\WAR$, however, looks much more favorably upon the careers of some starters with intrinsic game-by-game variance (that is, the occasional tendency to pitch an awful game).\footnote{
    There is an asymmetry due to selection bias: pitchers that are usually awful and occasionally brilliant don't pitch for long.
} There are many starters that have substantial differences including 
Whitey Ford and Catfish Hunter. Whitey Ford (resp., Catfish Hunter) has a whopping $25$ (resp., $15$) more career $\GWAR$ than $\FWAR$!
Ford is the $49^{th}$ best starter since 1952 according to FanGraphs (53 $\FWAR$) but is the $19^{th}$ best according to Grid $\WAR$ (78 $\GWAR$).
Similarly, Hunter is the $107^{th}$ best starter since 1952 according to FanGraphs (37 $\FWAR$) but is the $32^{nd}$ best according to Grid $\WAR$ (52 $\GWAR$).
What drives this difference? Ford and Hunter are extreme boom-bust pitchers. Standard $\WAR$, from either FanGraphs and Baseball Reference, average pitcher performance across games, allowing these pitchers' many blow-up games to dilute their great performances
which accumulate into huge discrepancies across careers, significantly devaluing them relative to other starters.
To understand the fundamental game-by-game variance of these pitchers, consider Ford's 1961 season and Hunter's 1967 season.
In 1961 Ford (7.2 $\GWAR$) started 39 games with six complete game shutouts accompanied by seven blow-up games (lower than $-0.1$ $\GWAR$).
Similarly, in 1967 Hunter (4.5 $\GWAR$) started 35 games with five complete game shutouts and four one-run complete games accompanied by eight blow-up games (lower than $-0.1$ $\GWAR$).
Grid $\WAR$, which correctly values pitcher performance in each individual game, sees the value of pitchers having such strong variability across games.
The public agrees with us: Catfish Hunter made the Hall-of-Fame (despite having just the $107^{th}$ best career according to FanGraphs) and Whitey Ford won a Cy Young award in 1961. We continue our discussion of Grid $\WAR$ across baseball history in Appendix~\ref{sec:GWARhistory}.

While we focus in this paper on developing $\WAR$ for starting pitchers, some of the principles we discussed should be used to estimate $\WAR$ for relief pitchers, opening pitchers, and batters and some should not.
The primary difference between valuing starting and relief pitchers is that opposing batter context matters for valuing the latter but not the former.
A relief pitcher who pitches the ninth inning up or down, say, three runs only marginally impacts the game, as its outcome has essentially already been decided.
Conversely, a relief pitcher who pitches the ninth inning up or down, say, one run hugely impacts the game. 
Grid $\WAR$ for starting pitchers, as developed in this paper, doesn't account for opposing batter context and so is not appropriate to value relief pitchers.
Nonetheless, due to the massive variability in leverage for relief pitchers across games, relief pitcher $\WAR$ should be estimated separately for each game.
Perhaps we should construct relief pitcher game $\WAR$ using a context-neutral version of win probability added that adjusts for score differential at the time the relief pitcher enters the game.

Opening pitchers are starting pitchers, so we can use Grid $\WAR$ to estimate their value.
Grid $\WAR$ provides a strong justification for the use of an opener.
For concreteness, suppose you are the manager of the 2023 Padres, featuring closer Josh Hader who had a 1.28 ERA through 56 innings in 61 games.
For simplicity, suppose Hader pitched for exactly one inning, the ninth inning, in each of 56 games, resulting in $1.7$ FanGraphs $\WAR$.\footnote{
    \url{https://www.fangraphs.com/players/josh-hader/14212/stats?position=P}
}
Now suppose he had instead pitched just the first inning in 56 games with that ERA, allowing an average of about $0.14$ runs per inning, or $0$ runs in $86\%$ of innings and $1$ run in $14\%$ of innings.
His average context neutral win probability per game would be about 
$$(0.86)\cdot f(I=1,R=0) + (0.14)\cdot f(I=1,R=1) \approx (0.86)\cdot(0.53) + (0.14)\cdot(0.44) = 0.52,$$
taking these values of $f$ from Figure~\ref{fig:fir}, resulting in
$$(0.52 - \wrep)\cdot(56 \text{ games}) \approx 5.15$$
seasonal Grid $\WAR$.
Our preliminary analysis suggests that Hader would be much more valuable as an opener than as a closer.
What drives the difference?
Closers aren't that valuable because many of the games they pitch in are low leverage situations in which the outcome of the game has essentially already been decided.
Conversely, the performance of an opening pitcher \textit{always matters}.
Allowing few runs through the first inning is very valuable since those innings always impact the outcome of the game.
Of course, our initial analysis is overly simplistic and is intended to be just an impetus for further research.
A closing pitcher's dominant ERA doesn't necessarily extrapolate to the first inning because, for instance, in each first inning but not necessarily in each ninth inning, a pitcher faces the opposing team's best batters.
A more thorough analysis would also need to account for the following telescoping phenomenon: if we moved the closing pitcher to the opening pitcher position, the setup man would become the closer and the middle relief pitcher would become the setup man; but, we expect the resulting potential loss of $\WAR$ to be much smaller than the positive gain in $\WAR$ from switching the closer to the opener position.
Finally, if many teams used an opening pitcher, replacement-level $\wrep$ for openers may change.

Game-by-game $\WAR$, on the other hand, doesn't make sense for batters.
This largely stems from the fact that each individual batter has just three to five plate appearances in a typical nine-inning game.
A batter doesn't influence any individual game enough for the non-convexity of game-level $\WAR$ to meaningfully change his valuation over the course of a season.
Also, these aren't enough plate appearances to allow strong evidence of fundamental variance in batter performance across games to appear in the data.
Finally, in each game the starting pitcher is responsible for the vast majority of his team's (defensive) context (until he is pulled), which is not true for any individual batter.  
Having said that, in estimating seasonal batter $\WAR$, batter performance should still be adjusted for opposing pitcher quality and park.

Although Grid $\WAR$ improves substantially upon existing estimates of $\WAR$ for starting pitchers, our analysis is not without limitations.
In particular, the version of Grid $\WAR$ described in this paper, as well $\WAR$ estimates from FanGraphs and Baseball Reference, doesn't adjust for opposing batter quality.
Thus, for a pitcher who faces good offensive teams more often than other pitchers do, Grid $\WAR$ underestimates his $\WAR$.
In our updated version of Grid $\WAR$ at \url{gridwar.xyz}, we adjust for opposing batter quality via $\GWAR+$ (details are included on the website).
Additionally, the current version of Grid $\WAR$ doesn't adjust for the pitcher's team's fielding.
Thus, for a pitcher who plays with great fielders who reduce his runs allowed, Grid $\WAR$ overestimates his $\WAR$.
Nonetheless, we expect these adjustments to have a very small impact.
In particular, we expect the effect of fielding to have a smaller total  impact than ballpark, which itself has a small impact, except at extreme parks like Coors Field.
This can be seen in Figure~\ref{fig:gwar_parkFx_comp}: Grid $\WAR$ computed with our ridge-adjusted park effects is extremely similar to Grid $\WAR$ without park effects. 
We leave the addition of batting and fielding adjustments to future work.

Moreover, the distribution of runs scored in a half-inning is not Poisson; more likely it is a zero-inflated Poisson, a more general Conway-Maxwell-Poisson, or a similar distribution on the non-negative integers.
Computationally, it is straightforward to modify the $f$ grid formula (Equation~\ref{eqn:Apoisson_model}) to accommodate  different distributions.
One interesting modification would adjust to allow different parameters for each inning depending on when a starting pitcher is pulled.
In particular, middle relievers tend to be worse than starting pitchers, suggesting a higher value of $\lambda$ for those innings, and closers are often very good pitchers, suggesting a lower value of $\lambda$. 
%%%
But there are  several substantial benefits to sticking with a simpler Poisson model.
First, it produces a closed-form formula which is quick to evaluate.
Second,  a simple parametric model makes it easier to adjust for ballpark (and other confounders like batting quality and fielding quality).
Finally, the resulting $f$ grid is reasonable and  quite accurate for our purposes. For example, while the Poisson model systematically underestimates the probability of big-deficit late inning comeback, these differences  have an insignificant impact on Grid WAR.
We leave any adjustment to the half-inning runs distribution as future work.

Finally, there is a flaw in our Empirical Bayes shrinkage estimator of latent pitcher quality $\mu_p$.  Formula~\eqref{eqn:post_mean_pq} assumes that $\mu_p$ remains constant over the entire decade from 2010 to 2019, but player quality is non-stationary over time, and a more elaborate estimator should account for this. In future work we suggest using a similar Empirical Bayes approach to estimate pitcher quality, with weights that decay further back in time (e.g., using exponential decay weighting as in \citet{darko}) in the posterior mean Formulas \eqref{eqn:post_mean_pq} and \eqref{eqn:post_mean_pq_FWAR}.

% %%%% REFERENCES
% \bibliography{refs}

% \end{document}

\section*{Acknowledgments}
% Acknowledgements

The authors thank Eric Babitz and Sam Bauman, who first calculated Grid $\WAR$ for starting pitchers and created the name Grid $\WAR$, and Justin Lipitz, Emma Segerman, and Ezra Troy, who contributed to an early version of this paper.

\bibliography{refs}

\begin{thebibliography}{}

\bibitem[Acharya et~al., 2008]{AcharyaParkFactors}
Acharya, R.~A., Ahmed, A.~J., D'Amour, A.~N., Lu, H., Morris, C.~N., Oglevee,
  B.~D., Peterson, A.~W., and Swift, R.~N. (2008).
\newblock Improving major league baseball park factor estimates.
\newblock {\em Journal of Quantitative Analysis in Sports}, 4(2).

\bibitem[Albert and Bennett, 2003]{curveBall}
Albert, J. and Bennett, J. (2003).
\newblock {\em {Curve Ball: Baseball, Statistics, and the Role of Chance in the
  Game}}.
\newblock Copernicus Books.

\bibitem[Appelman, 2016]{FanGraphs_pf}
Appelman, D. (2016).
\newblock {Park Factors – 5 Year Regressed}.
\newblock \\ \url{https://library.fangraphs.com/park-factors-5-year-regressed/
  }.

\bibitem[{Baseball Reference}, 2011]{war_BR}
{Baseball Reference} (2011).
\newblock {Pitcher WAR Calculations and Details}.
\newblock \\
  \url{https://www.baseball-reference.com/about/war_explained_pitch.shtml}.

\bibitem[{Baseball Reference}, 2022]{BR_pf}
{Baseball Reference} (2022).
\newblock {Park Adjustments}.
\newblock \\ \url{https://www.baseball-reference.com/about/parkadjust.shtml }.

\bibitem[Baumer and Zimbalist, 2014]{saberRev}
Baumer, B. and Zimbalist, A. (2014).
\newblock {\em {The Sabermetric Revolution: Assessing the Growth of Analytics
  in Baseball}}.
\newblock University of Pennsylvania Press.

\bibitem[Baumer et~al., 2015]{BaumerJensenMatthews+2015+69+84}
Baumer, B.~S., Jensen, S.~T., and Matthews, G.~J. (2015).
\newblock {openWAR: An open source system for evaluating overall player
  performance in major league baseball}.
\newblock {\em Journal of Quantitative Analysis in Sports}, 11(2):69--84.

\bibitem[Brill et~al., 2023]{BrillDeshpandeWyner+2023}
Brill, R.~S., Deshpande, S.~K., and Wyner, A.~J. (2023).
\newblock A bayesian analysis of the time through the order penalty in
  baseball.
\newblock {\em Journal of Quantitative Analysis in Sports}.

\bibitem[Brown, 2008]{brown2008}
Brown, L.~D. (2008).
\newblock {In-season prediction of batting averages: A field test of empirical
  Bayes and Bayes methodologies}.
\newblock {\em The Annals of Applied Statistics}, 2(1):113 -- 152.

\bibitem[Chen and Guestrin, 2016]{xgboost}
Chen, T. and Guestrin, C. (2016).
\newblock {XGBoost}: A scalable tree boosting system.
\newblock In {\em Proceedings of the 22nd ACM SIGKDD International Conference
  on Knowledge Discovery and Data Mining}, KDD '16, pages 785--794, New York,
  NY, USA. ACM.

\bibitem[ESPN, 2014]{Scherzer}
ESPN (2014).
\newblock {Max Scherzer 2014 Game Log}.
\newblock \\
  \url{https://www.espn.com/mlb/player/gamelog/_/id/28976/year/2014/category/pitching}.

\bibitem[ESPN, 2022]{espn_pf}
ESPN (2022).
\newblock {MLB Park Factors - 2019}.
\newblock \\ \url{https://www.espn.com/mlb/stats/parkfactor/_/year/2019 }.

\bibitem[Fangraphs, 2010]{ReplacementLevel}
Fangraphs (2010).
\newblock {Replacement Level}.
\newblock \\ \url{https://library.fangraphs.com/misc/war/replacement-level/}.

\bibitem[Hoerl and Kennard, 1970]{ridge}
Hoerl, A.~E. and Kennard, R.~W. (1970).
\newblock {Ridge Regression: Biased Estimation for Nonorthogonal Problems}.
\newblock {\em Technometrics}, 12(1):55--67.

\bibitem[Lewis, 2003]{moneyball}
Lewis, M. (2003).
\newblock {\em {Moneyball: The Art of Winning an Unfair Game}}.
\newblock WW Norton \& Company.

\bibitem[Medvedovsky and Patton, 2022]{darko}
Medvedovsky, K. and Patton, A. (2022).
\newblock {\em {Daily Adjusted and Regressed Kalman Optimized projections |
  DARKO}}.
\newblock https://apanalytics.shinyapps.io/DARKO/.

\bibitem[Petti and Gilani, 2021]{petti_gilani_2021}
Petti, B. and Gilani, S. (2021).
\newblock baseballr: The sportsdataverse's r package for baseball data.

\bibitem[Retrosheet, 2021]{retroRaw}
Retrosheet (2021).
\newblock {Retrosheet Play-by-Play Data Files (Event Files)}.
\newblock \\ \url{https://www.retrosheet.org/game.htm}.

\bibitem[Schwarz and Gammons, 2005]{numbersGame}
Schwarz, A. and Gammons, P. (2005).
\newblock {\em {The Numbers Game: Baseball's Lifelong Fascination with
  Statistics}}.
\newblock Thomas Dunne Books.

\bibitem[Slowinski, 2012]{war_FG}
Slowinski, P. (2012).
\newblock {WAR for Pitchers}.
\newblock \\ \url{https://library.fangraphs.com/war/calculating-war-pitchers/}.

\bibitem[Statcast, 2023]{statcast}
Statcast (2023).
\newblock {Statcast Search}.
\newblock \\ \url{https://baseballsavant.mlb.com/statcast_search}.

\bibitem[Tango et~al., 2007]{theBook}
Tango, T., Lichtman, M., and Dolphin, A. (2007).
\newblock {\em {The Book: Playing the Percentages in Baseball}}.
\newblock Potomac Books.

\bibitem[Thorn and Palmer, 1984]{hiddenBaseball}
Thorn, J. and Palmer, P. (1984).
\newblock {\em {The Hidden Game of Baseball: A Revolutionary Approach to
  Baseball and Its Statistics}}.
\newblock Doubleday, Garden City, NY.

\bibitem[{Wikipedia}, 2023]{rules_of_baseball}
{Wikipedia} (2023).
\newblock {Baseball}.
\newblock \\ \url{https://en.wikipedia.org/wiki/Baseball}.

\bibitem[Wolverton, 1993]{wolvertonSNWL}
Wolverton, M. (1993).
\newblock {``Support-Neutral'' Statistics -- A Method of Evaluating the True
  Quality of a Pitcher's Start}.
\newblock {\em By The Numbers}, 5(4):4 -- 14.

\bibitem[Wolverton, 1999]{wolvertonSNWLbp2}
Wolverton, M. (1999).
\newblock {The Top Pitchers of the 1990s: A Support-Neutral Approach}.
\newblock \\
  \url{https://www.baseballprospectus.com/news/article/416/the-top-pitchers-of-the-1990s-a-support-neutral-approach/}.

\bibitem[Wolverton, 2004]{wolvertonSNWLbp1}
Wolverton, M. (2004).
\newblock {Baseball Prospectus Basics: The Support-Neutral Stats}.
\newblock \\
  \url{https://www.baseballprospectus.com/news/article/2590/baseball-prospectus-basics-the-support-neutral-stats/}.

\end{thebibliography}
% \bibliography{refs.bib}
% %%%%%%%%%%%%%%%%%%%%%%%%%%%%%
% \documentclass[12pt]{article}
% \input{../header}
% \usepackage{fullpage, parskip}
% \onehalfspacing
% %%%%%%%%%%%%%%%%%%%%%%%%%%%%%

% \begin{document}

%%%%%%%%%%%%%%%%%%%%%%%%%%%%%%%%%%%%%%%%%%%%%%%%%%%%%%%%%
%%%%%%%%%%%%%%%%%%%%%%%%%%%%%%%%%%%%%%%%%%%%%%%%%%%%%%%%%
%%%%%%%%%%%%%%%%%%%%%%%%%%%%%%%%%%%%%%%%%%%%%%%%%%%%%%%%%

%%%%%%%%%%%%%%%%%
{}
%%%%%%%%%%%%%%%%%

%%%%%%%%%%%%%%%%%
%%%%%%%%%%%%%%%%%

\clearpage
\appendix

% %%%%%%%%%%%%%%%%%%%%%%%%%%%%%%%%%%%%%%%%%%%%%%%%%%%%%%%%%%%%%%%%%%%%%%%%%%%
% \section{Our Code \& Data}\label{sec:data_and_code}

% Our code is available on github at \url{https://github.com/snoopryan123/grid_war}. Our data is available on Dropbox at \url{https://upenn.app.box.com/v/retrosheet-pa-1990-2000}.

%%%%%%%%%%%%%%%%%%%%%%%%%%%%%%%%%%%%%%%%%%%%%%%%%%%%%%%%%%%%%%%%%%%%%
\section{A review of the rules of baseball}\label{app:rules_of_baseball}

In a baseball game, two teams of nine players each take turns on offense (batting and baserunning) and defense (fielding and pitching).
The game occurs as a sequence of plays, with each play beginning when the pitcher throws a ball that a batter tries to hit with a bat.
The objective of the offensive team (batting team) is to hit the ball into the field of play, away from the other team's players, in order to get on base.
The goal of a baserunner is to eventually advance counter-clockwise around all four bases to score a ``run'', which occurs when he touches home plate (the position where the batter initially began batting).
The defensive team tries to prevent batters from reaching base and scoring runs by getting batters or baserunners ``out'', which forces them out of the field of play. 
The pitcher can get the batter out by throwing three pitches which result in ``strikes''. 
Fielders can get the batter out by catching a batted ball before it touches the ground, and can get a runner out by tagging them with the ball while the runner is not touching a base.
The batting team's turn to bat is over once the defensive team records three outs.
The two teams switch back and forth between batting and fielding;  one turn batting for each team constitutes an inning. 
A game is usually composed of nine innings, and the team with the greater number of runs at the end of the game wins. 
Most games end after the ninth inning, but if scores are tied at that point, extra innings are played \citep{rules_of_baseball}.

%%%%%%%%%%%%%%%%%%%%%%%%%%%%%%%%%%%%%%%%%%%%%%%%%%%%%%%%%%%%%%%%%%%%%%%%%%%
\section{Estimating $f$ using a mathematical, not a statistical, model}\label{app:estimate_f}

In this section, we detail our modeling process for estimating the grid function $f=f(I,R)$ which, assuming both teams have randomly drawn offenses, computes the probability a team wins a game after giving up $R$ runs through $I$ complete innings. 
In particular, we compare statistical models fit from observational data to mathematical probability models, which are superior.

To account for different run environments across different seasons and leagues ($\NL$ vs. $\AL$), we estimate a different grid for each league-season.
% In other words, the grid function $f$ is implicitly also a function of league and season.
We begin by estimating $f$ from our observational dataset of half-innings from 2010 to 2019.
The response variable is a binary indicator denoting whether the pitcher's team won the game, and the features are the inning number $I$, the runs allowed through that half-inning $R$, the league, and the season.
Note that if a home team leads after the top of the $9^{th}$ inning, then the bottom of the $9^{th}$ is not played.
Therefore, to avoid selection bias, we exclude all $9^{th}$ inning instances in which a pitcher pitches at home.

\begin{figure}[htb!]
    \centering{}
    \subfloat[]{{\includegraphics[width=7cm]{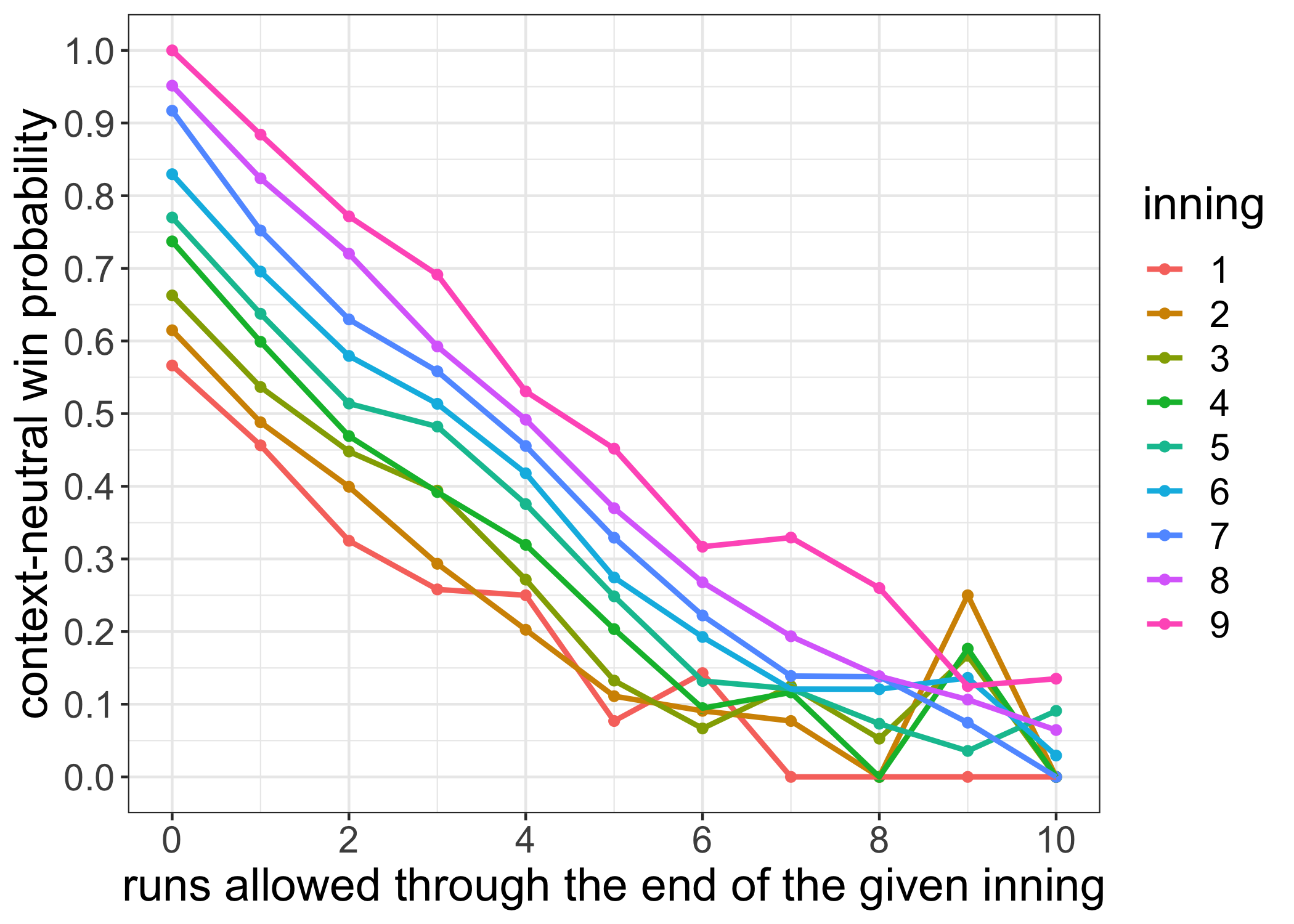}}\label{fig:fRge}}%
    \qquad
    \subfloat[]{{\includegraphics[width=7cm]{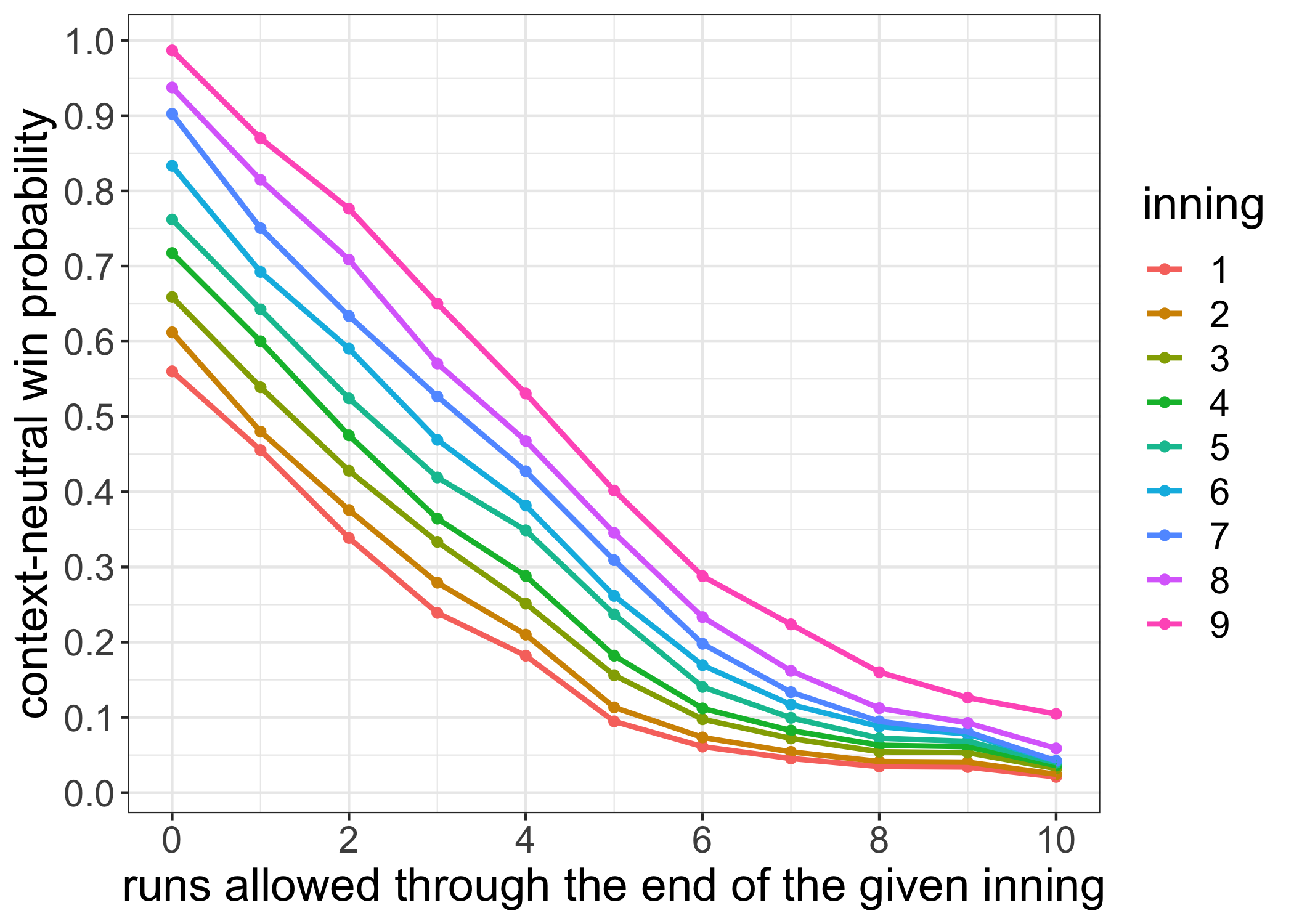}}\label{fig:fRxgb}}%
    \caption{
    Estimates of the 2019 National League function $R \mapsto f(I,R)$ using the empirical grid (Figure (a)) and XGBoost with monotonic constraints (Figure (b)).
    } 
    \label{fig:fir1}
\end{figure}

With enough data, the empirical grid (e.g., binning and averaging over all combinations of $I$ and $R$ within a league-season) is a great estimator of $f$.
In Figure~\ref{fig:fRge} we visualize the empirical grid fit on a dataset of 
all half-innings from 2019 in which the home team is in the National League.
The function $f$ should be monotonic decreasing in $R$.
In particular, as a pitcher allows more runs through a fixed number of innings, his team is less likely to win the game. 
It should also be monotonic increasing in $I$ because giving up $R$ runs through $I$ innings is worse than giving up $R$ runs through $I+i$ innings for $i > 0$, since giving up $R$ runs through $I+i$ innings implies a pitcher gave up no more than $R$ runs through $I$ innings.
The empirical grid, however, is not monotonic in either $R$ or $I$ because each league-season dataset is not large enough.
Moreover, even when we use our entire dataset of all half-innings from 2010 to 2019, the empirical grid is still not monotonic in $R$ or $I$.

%%%%%%%%%%%%%%%%%%%%%
%%%%%%%%%%%%%%%%%%%%%

To force our fitted $f$ to be monotonic, we use $\xgb$ with monotonic constraints, tuned using cross validation \citep{xgboost}.
We visualize our 2019 $\NL$ $\xgb$ fit in Figure~\ref{fig:fRxgb}. 
% We visualize our $\xgb$ model fit on the full dataset in Figure~\ref{fig:fRxgb}.
We indeed see that the fitted $f$ is decreasing in $R$ and increasing in $I$.
Additionally, $R \mapsto f(I,R)$ is mostly convex: if a pitcher has already allowed a high number of runs, there is a lesser relative impact of throwing an additional run on winning the game. 
% Conversely, if a pitcher has allowed few runs thus far, there is a high relative impact of throwing an additional run. 
Nonetheless, $\xgb$ overfits, especially towards the tails (e.g., for $R$ large).
For instance, the 2019 $\NL$ $\xgb$ model indicates that the probability of winning a game after allowing 10 runs through 9 innings is about $0.11$, which is too large.

As there is not enough data to use machine learning to fit a separate grid for each league-season without overfitting, we turn to a parametric mathematical model.
Indeed, the power of parameterization is that it distills the information of a dataset into a concise form (e.g., into a few parameters), allowing us create a strong model from limited data. 
Because the runs allowed in a half-inning is a natural number, we begin our parametric quest by supposing that the runs allowed in a half-inning is a $\Poisson(\lambda)$ random variable.
In particular, denoting the runs allowed by the pitcher's team's batters in inning $i$ by $X_i$ and the runs allowed by the opposing team in inning $i$ (for innings $i$ after the pitcher exits the game), we assume
\begin{equation}
    X_i, Y_i \overset{i.i.d.}{\sim} \Poisson(\lambda).
\end{equation}
Then the probability that a pitcher wins the game after allowing $R$ runs through $I$ innings, assuming win probability in overtime is $1/2$, is 
\begin{align}
\label{eqn:poisson_model_constant_lambda}
    & f(I,R|\lambda) := \Pr\bigg(\sum_{i=1}^{9} X_i > R + \sum_{i=I+1}^{9} Y_i\bigg) + \frac{1}{2}\cdot\Pr\bigg(\sum_{i=1}^{9} X_i = R + \sum_{i=I+1}^{9} Y_i\bigg).
\end{align}
If $I = 9$, this is equal to
\begin{align}
    & \Pr\big(\Poisson(9\lambda) > R\big) + \frac{1}{2}\cdot\Pr\big(\Poisson(\lambda) = R\big).
\end{align}
If $I < 9$, it is equal to
\begin{align}
    & \Pr\big(\Skellam(9\lambda, (9-I-1)\lambda) > R\big) + \frac{1}{2}\cdot\Pr\big(\Skellam(9\lambda, (9-I-1)\lambda) = R\big),
\end{align}
noting that the $\Skellam$ distribution arises as a difference of two independent $\Poisson$ distributed random variables.
Then, we estimate $\lambda$ separately for each league-season by computing each team's mean runs allowed in each half inning, and then averaging over all teams.

\begin{figure}[htb!]
    \centering{}
    \subfloat[]{{\includegraphics[width=7cm]{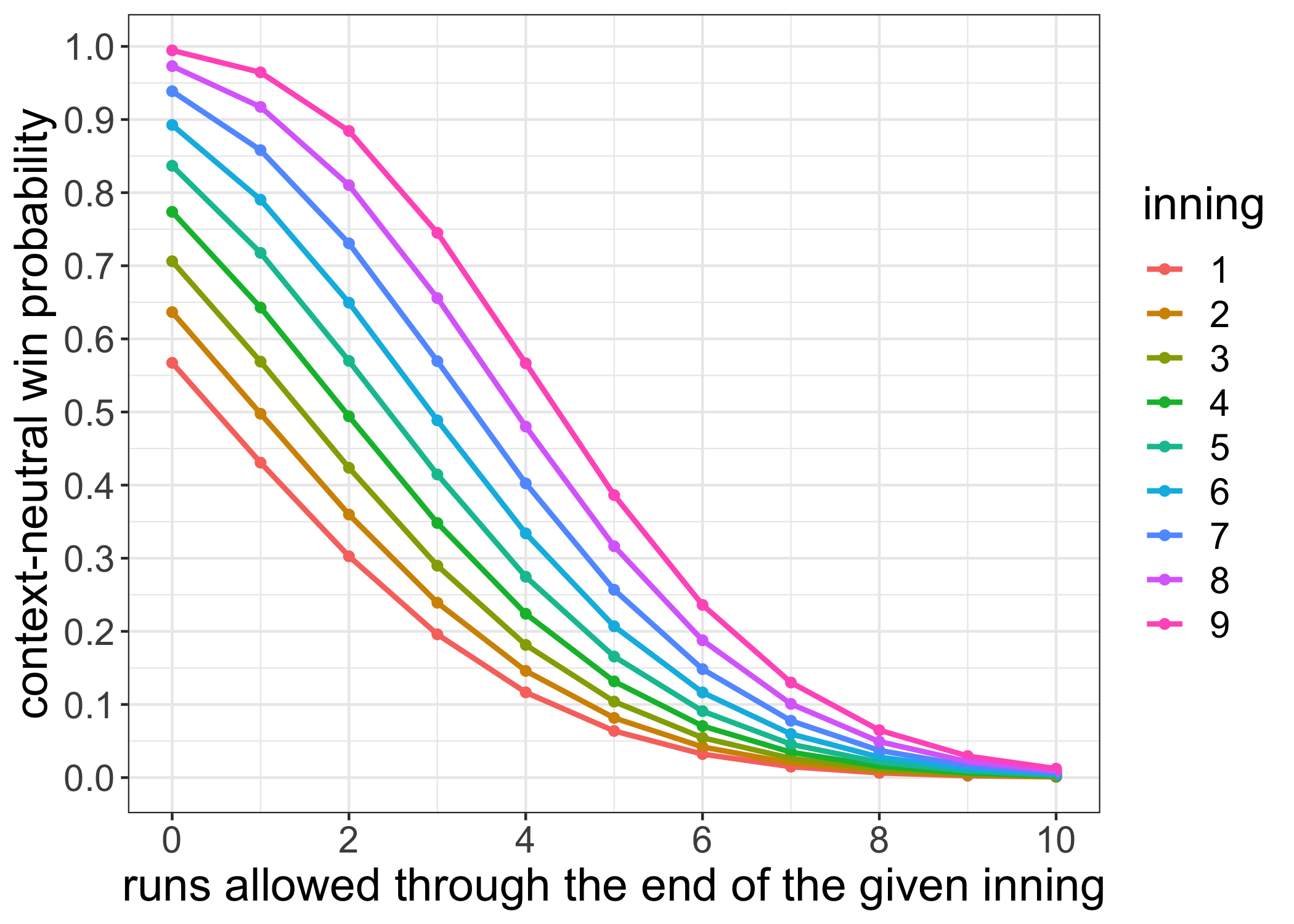}}\label{fig:fRskellam}}%
    \qquad
    \subfloat[]{{\includegraphics[width=7cm]{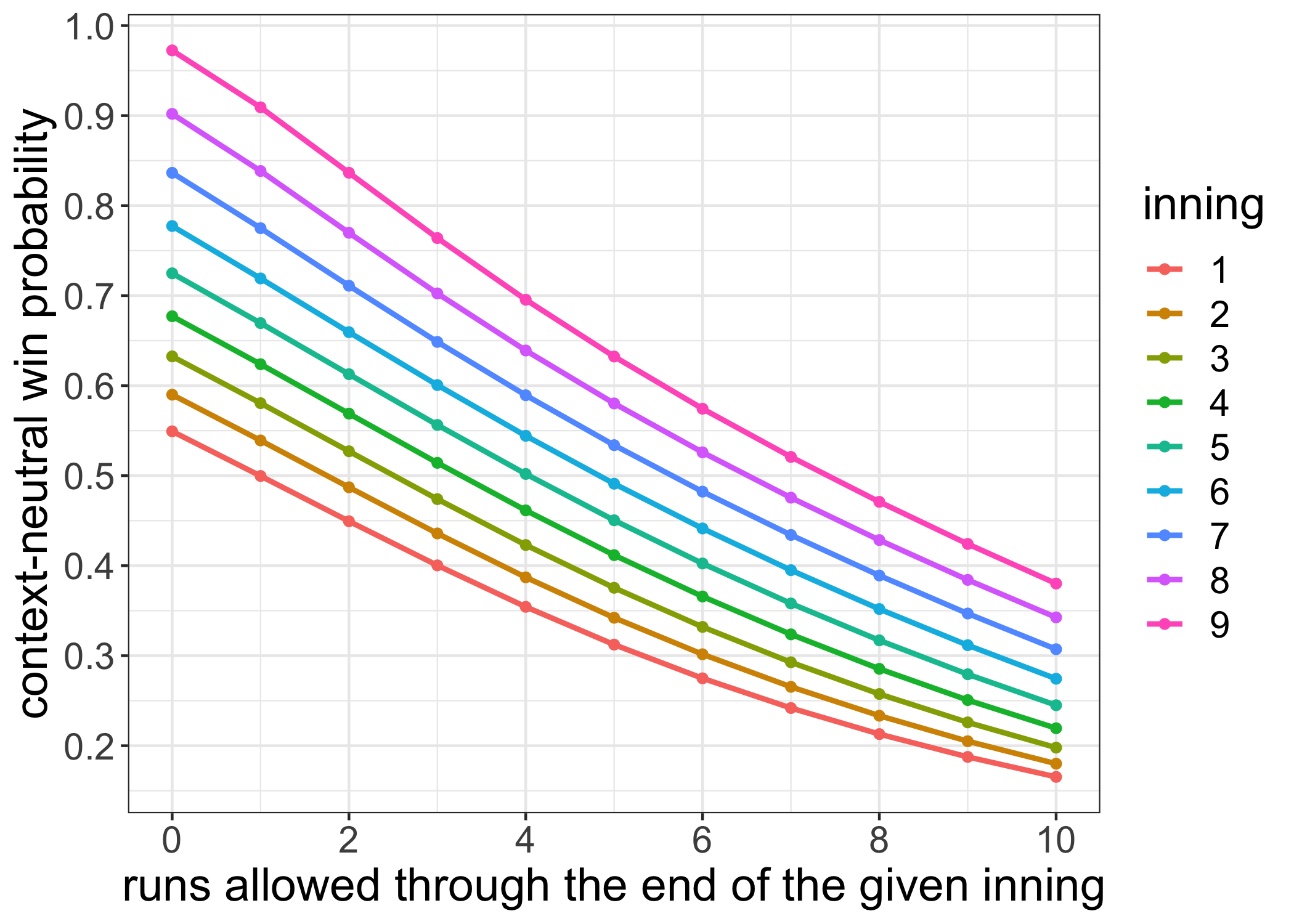}}\label{fig:fRdskellam0}}%
    \caption{
    Estimates of the 2019 National League function $R \mapsto f(I,R)$ using our Poisson model~\eqref{eqn:poisson_model_constant_lambda} with constant $\lambda$ (Figure (a)) and our Poisson model~\eqref{eqn:poisson_model_2post} with a truncated normal prior~\eqref{eqn:poisson_model_2prior} on two team strength parameters $\lambda_X$ and $\lambda_Y$ (Figure (b)). 
    } 
    \label{fig:fir_skellam}
\end{figure}

In Figure~\ref{fig:fRskellam} we visualize the estimated $f$ according to our Poisson model~\eqref{eqn:poisson_model_constant_lambda} using the 2019 $\NL$ $\lambda$.
We see that $f$ is decreasing in $R$, increasing in $I$, convex in the tails of $R$, and is smooth.
Nonetheless, some of the win probability values from this model are unrealistic.
For instance, it implies the probability of winning the game after shutting out the opposing team through 9 innings is about $99\%$, which is too high, and the probability of winning the game after allowing $10$ runs through $9$ innings is about $1\%$, which is too low.
% For instance, the latter value is too low
% Succintly, the win probability values at both tails of $R$ are too extreme.

%%%%%%%%%%%%%%%%%%%%%
%%%%%%%%%%%%%%%%%%%%%

The win probability values at both tails of $R$ are too extreme in our original Poisson model~\eqref{eqn:poisson_model_constant_lambda} because we assume both teams have the same mean runs per inning $\lambda$. 
This is an unrealistic assumption:  in real life, a baseball season involves teams of varying strength playing against each other.
When teams of differing batting strength play each other, win probabilities differ.
For instance, a great hitting team down seven runs has a larger probability of coming back to win the game than a worse hitting team would.
Thus, accounting for random differences in team strength across games should flatten the $f(I,R)$ grid.

On this view, it is more realistic to assume the pitcher's team and the opposing team have their own runs scored per inning parameters,
\begin{equation}
    X_i \overset{i.i.d.}{\sim} \Poisson(\lambda_X) \quad \text{and} \quad Y_i \overset{i.i.d.}{\sim} \Poisson(\lambda_Y),
\end{equation}
and
\begin{align}
\label{eqn:poisson_model}
    & f(I,R|\lambda_X,\lambda_Y) := \Pr\bigg(\sum_{i=1}^{9} X_i > R + \sum_{i=I+1}^{9} Y_i\bigg) + \frac{1}{2}\cdot\Pr\bigg(\sum_{i=1}^{9} X_i = R + \sum_{i=I+1}^{9} Y_i\bigg).
\end{align}
Moreover, to capture the variability in team strength across each of the 30 MLB teams, we impose a positive normal prior,
\begin{equation}
\label{eqn:poisson_model_2prior}
    \lambda_X, \lambda_Y \sim \N_{+}(\lambda, \sigma^2_\lambda).
\end{equation}
We estimate the prior hyperparameters $\lambda$ and $\sigma_\lambda$ separately for each league-season by computing each team's mean and s.d. of the runs allowed in each half inning, respectively, and then averaging over all teams.

Given $\lambda_X$ and $\lambda_Y$, we compute Formula~\eqref{eqn:poisson_model} similarly as before using the $\Poisson$ and $\Skellam$ distributions.
We use Monte Carlo integration with $B=100$ samples to estimate the posterior mean grid,
\begin{align}
\label{eqn:poisson_model_2post}
    f(I,R|\lambda,\sigma^2_\lambda) \approx \frac{1}{B} \sum_{b=1}^{B} f(I,R|\lambda_X^{(b)}, \lambda_Y^{(b)}),
\end{align}
where $\lambda_X^{(b)}$ and $\lambda_Y^{(b)}$ are i.i.d. samples from the prior distribution~\eqref{eqn:poisson_model_2prior}.

In Figure~\ref{fig:fRdskellam0} we visualize the estimated $f$ according to this Poisson model~\eqref{eqn:poisson_model_2post}, with prior~\eqref{eqn:poisson_model_2prior}, using the 2019 $\NL$ $\lambda$ and $\sigma^2_\lambda$.
We see that $f$ is mostly linear in $R$, rather than convex, and the values of $f$ when $R$ is large are highly unrealistic.
For instance, this model indicates that the probability of winning the game after allowing 10 runs through 9 innings is about $38\%$, which is way too high.
This is because our model is overdispersed, i.e. the estimated prior variance $\sigma^2_\lambda$ is too large.
For example, too large of a $\sigma^2_\lambda$ allows $\lambda_X$ and $\lambda_Y$ to be very far apart, so if a pitcher allows 10 runs through 9 innings and $\lambda_X$ is much larger than $\lambda_Y$, then his team will have a significant chance of coming back to win.
% ~\eqref{eqn:poisson_model_2prior} has too much 

\begin{figure}[htb!]
    \centering
    \includegraphics[width=10cm]{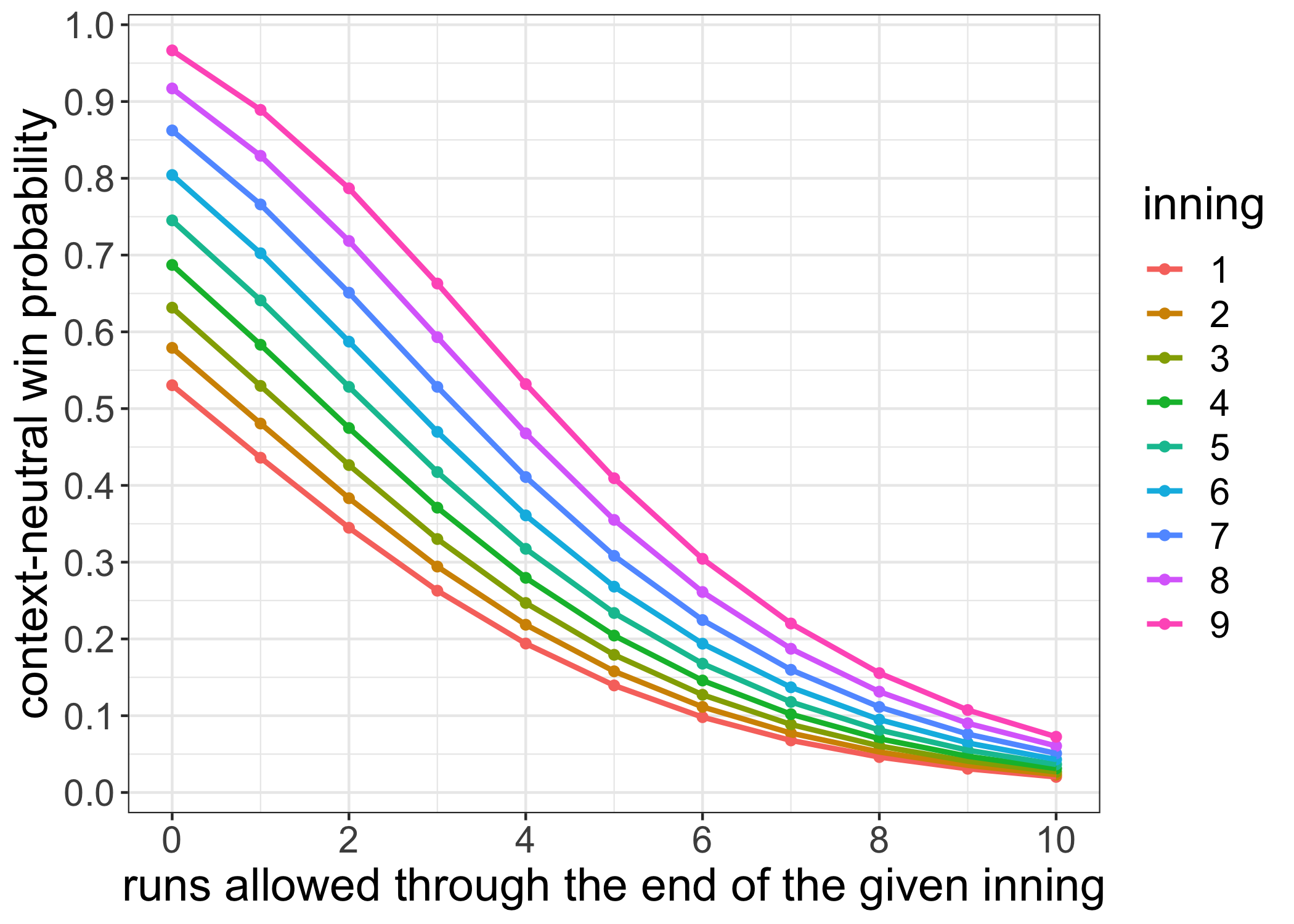}
    \caption{
    Estimates of the 2019 National League function $R \mapsto f(I,R)$ using our Poisson model~\eqref{eqn:poisson_model_2post} with tuned dispersion prior~\eqref{eqn:poisson_model_2prior_tuned}.
    } 
    \label{fig:fir_APP}
\end{figure}

To resolve the overdispersion issue, we introduce a tuning parameter $k$ designed to tune the dispersion across team strengths to match observed data,
\begin{equation}
\label{eqn:poisson_model_2prior_tuned}
    \lambda_X, \lambda_Y \sim \N_{+}(\lambda, k\cdot\sigma^2_\lambda).
\end{equation}
In particular, we use $k = 0.28$, which minimizes the log-loss between the observed win/loss column and predictions from the induced grid $f(I,R|\lambda, \sigma^2_\lambda, k)$.
In Figure~\ref{fig:fir_APP} we visualize the estimated $f$ according to our Poisson model~\eqref{eqn:poisson_model_2post}, with tuned dispersion prior~\eqref{eqn:poisson_model_2prior_tuned}, using the 2019 $\NL$ $\lambda$ and $\sigma^2_\lambda$.
We see that $f$ is decreasing in $R$, increasing in $I$, and convex when $R$ is large.
In particular, it looks like a smoothed version of the $\xgb$ grid from Figure~\ref{fig:fRxgb}.
Additionally, the values of the grid at both tails of $R$ seem reasonable.
For instance, the model indicates that allowing 0 runs through 9 innings has about a $97\%$ win probability, which is more reasonable than before.
For all of these reasons, we use this model for the grid $f$ to compute Grid $\WAR$ for starting pitchers. 

%%%%%%%%%%%%%%%%%%%%%
%%%%%%%%%%%%%%%%%%%%%

%%%%%%%%%%%%%%%%%%%%%%%%%%%%%%%%%%%%%%%%%%%%%%%%%%%%%%%%%%%%%%%%%%%%%%%%%%%%%%%%%%%%
\section{Estimating pitcher quality using Empirical Bayes}\label{sec:empiricalBayes}

In this section, we describe how we estimate pitcher quality.
Given enough data, a pitcher's mean game $\WAR$ would suffice to capture his quality.
In the MLB, however, a pitcher starts just a finite number of games per season, so for many pitchers there is not enough data to use just his mean game $\WAR$ to represent his quality.
% On this view, the more games a pitcher has started, the more we should regress his mean game $\WAR$ to the overall mean.
Therefore, in this section we use a parametric Empirical Bayes approach in the spirit of \citet{brown2008} to devise shrinkage estimators $\widehat{\mu}^{\GWAR}_p$ and $\widehat{\mu}^{\FWAR}_p$, built from Grid $\WAR$ and FanGraphs $\WAR$ respectively, to represent pitcher $p$'s quality.
In particular, $\widehat\mu_p$ shrinks his mean game $\WAR$ to the overall mean in proportion to his number of games played.

% \fixme{} We begin by crafting a pitcher quality metric $\widehat{\mu}_p^{\GWAR}$ built from starting pitcher $p$'s previous games' Grid $\WAR$ and number of games played.

%%%%%%%%%%%%%%%%%%%%%%%%%%%%%%%%%%%%%%%%%%%%%%%%%%%%%%%%%%%%%%%%%%%%%%%%%%%%%%%%%%%%
\subsection{Empirical Bayes estimator of pitcher quality built from Grid $\WAR$}

\begin{figure}[hbt!]
\centering
\includegraphics[width=10cm]{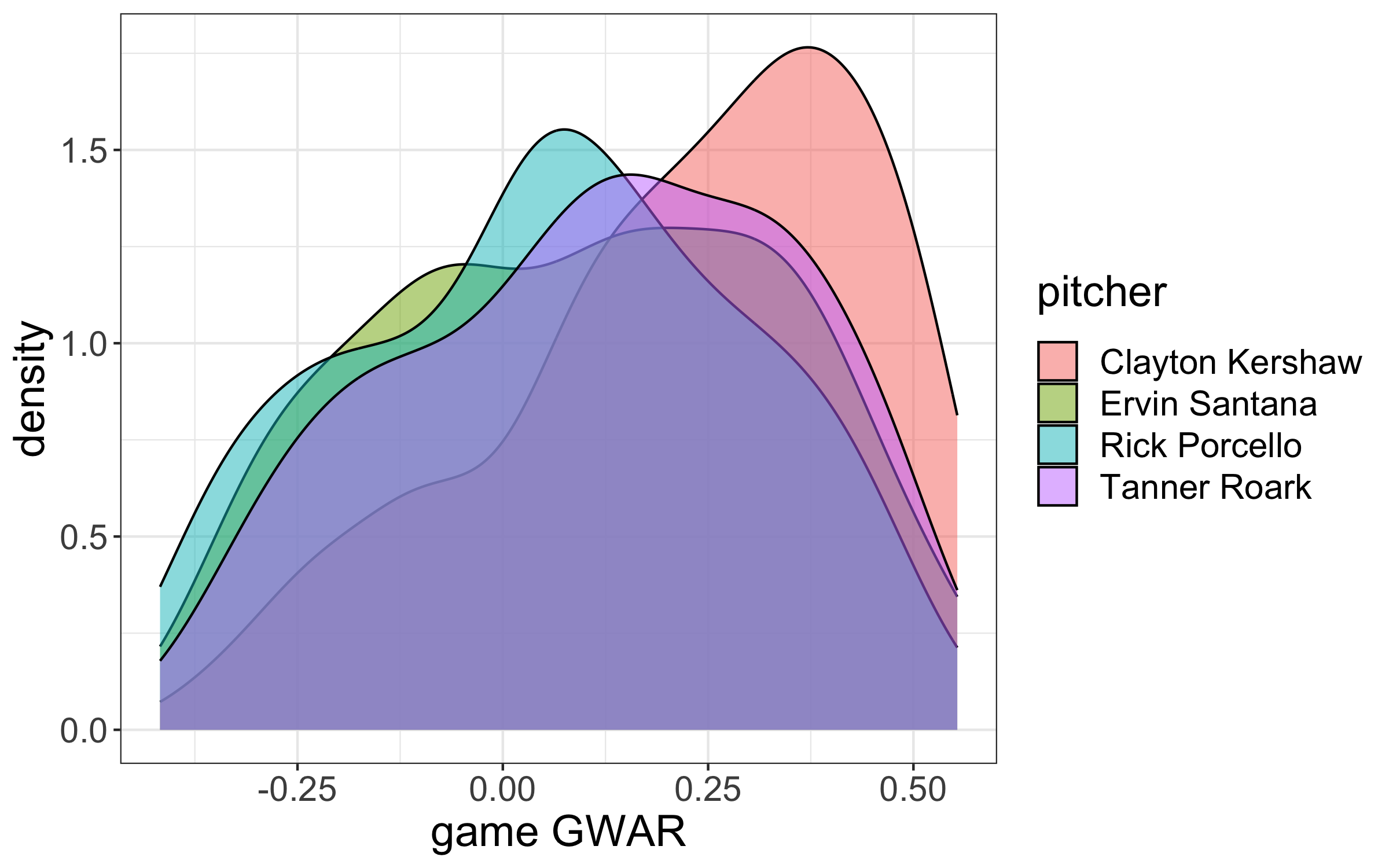}
\caption{
For four starting pitchers $p$, the distribution of his game-level $\GWAR$ $\{X_{pg}\}$ from 2010 to 2018.
} 
\label{fig:plot_Xpg_GWAR}
\end{figure}

To begin, index each starting pitcher by $p \in \{1,...,\bP\}$ and index pitcher $p$'s games by $g \in \{1,...,N_p\}$.
Let $X_{pg}$ denote pitcher $p$'s observed Grid $\WAR$ in game $g$.
After observing his $N_p$ games, we model
\begin{align}
\begin{split}
\label{eqn:emp_bayes_pq}
    X_{pg} &\sim \N(\mu_p, \sigma^2_p), \\
    \mu_p  &\sim \N(\mu, \tau^2).
\end{split}
\end{align}
In this model, $\mu_p$ represents pitcher $p$'s unobservable ``true'' pitcher quality, or his latent underlying mean game Grid $\WAR$.
Similarly, $\sigma^2_p$ represents pitcher $p$'s latent game-by-game variance in pitcher quality, or his game-by-game variance in mean game Grid $\WAR$.
% Although in real life the Grid $\WAR$ of a game $X_{pg}$ is not normally distributed, we model it as a Gaussian because it produces a good and interpretable estimator of pitcher $p$'s latent pitcher quality.
The prior parameters $\mu$ and $\tau^2$ represent the mean and variance, respectively, of pitcher quality across all pitchers.
In Figure~\ref{fig:plot_Xpg_GWAR} we visualize the game-level Grid $\WAR$ of four starting pitchers.
While Grid $\WAR$ is not actually normally distributed, it isn't too unreasonable an approximation (particularly for typical pitchers).
In particular, we use a Gaussian model because it produces a good and interpretable estimator of pitcher $p$'s latent pitcher quality, not because of accuracy.

%%%%%%%%%%%%%%%%%
%%%%%%%%%%%%%%%%%

% Then, we 
We estimate pitcher $p$'s pitcher quality $\mu_p$ using the posterior mean, which as a result of our normal-normal conjugate model~\eqref{eqn:emp_bayes_pq} is
\begin{align}
\begin{split}
\label{eqn:post_mean_pq}
    \widehat{\mu}_p := \E[\mu_p | X_{p1},...,X_{p{N_p}}] = \frac{\frac{1}{\sigma^2_p}\sum_{g=1}^{N_p} X_{pg} + \frac{\mu}{\tau^2}}{\frac{N_p}{\sigma^2_p} + \frac{1}{\tau^2}}.
\end{split}
\end{align}
The posterior mean is a weighted sum between the observed total Grid $\WAR$ and the overall mean pitcher quality, weighted by the variances $\sigma^2_p$ and $\tau^2$ and the number of games played $N_p$.
In particular, the more games a pitcher has played, the closer his estimated quality is to his observed mean game Grid $\WAR$.
Conversely, the fewer games he has played, the closer his estimated quality is to the overall mean quality.

Estimator~\eqref{eqn:post_mean_pq}, however, is defined in terms of unknown parameters $\mu$, $\tau^2$, and $\sigma^2_p$.
Thus, to effectively use this estimator, we employ an Empirical Bayes approach in the spirit of \citet{brown2008}.
Specifically, in place of these estimators in Equation~\eqref{eqn:post_mean_pq}, we plug in their maximum likelihood estimates (MLE) $\widehat\mu$, $\widehat\tau^2$, and $\widehat\sigma^2_p$, estimated from the data $\{X_{pg}\}$.

We being finding the MLE by noting the marginal distribution of $X_{pg}$ according to model~\eqref{eqn:emp_bayes_pq},
\begin{equation}
\label{eqn:marginal_EB}
    X_{pg} \sim \N(\mu, \tau^2 + \sigma^2_p).
\end{equation}
Thus the likelihood of pitcher $p$'s data $\{X_{pg}:1\leq g\leq N_p\}$ is
\begin{equation}
    \Pr(\{X_{pg}:1 \leq g \leq N_p\} | \mu,\tau^2,\sigma^2_p) = \bigg( \frac{1}{\sqrt{2\pi(\\\tau^2+\sigma^2)}} \bigg)^{N_p} \cdot \exp\bigg( \frac{-1}{2(\tau^2+\sigma^2_p)} \sum_{g=1}^{N_p} (X_{pg}-\mu)^2 \bigg).
\end{equation}
Therefore the log-likelihood of the full dataset $\{X_{pg}:1\leq g\leq N_p, 1 \leq p \leq \bP\}$ is %proportional to
\begin{align}
\begin{split}
    \ell(\{X_{pg}\}|\mu,\tau^2,\sigma^2_p) &= \sum_{p=1}^{\bP} \log \Pr(\{X_{pg}:1 \leq g \leq N_p\} | \mu,\tau^2,\sigma^2_p)    \\
    &\propto -\sum_{p=1}^{\bP} \frac{N_p}{2} \cdot \log(\tau^2 + \sigma^2_p) -\sum_{p=1}^{\bP} \frac{1}{2(\tau^2 + \sigma^2_p)} \sum_{g=1}^{N_p} (X_{pg} - \mu)^2.
\end{split}
\end{align}
To find the MLE of $\mu$, we set the derivative of the log-likelihood with respect to $\mu$ equal to 0 and solve for $\mu$,
\begin{equation}
    \frac{\partial \ell}{\partial \mu} = -\sum_{p=1}^{\bP} \frac{1}{2(\tau^2 + \sigma^2_p)} \sum_{g=1}^{N_p} 2(X_{pg} - \mu) = 0,
\end{equation}
which yields
\begin{equation}
    \label{eqn:mu_mle_EB_gwar}
    \mu = \frac{\sum_{p,g} X_{pg} / (\tau^2 + \sigma^2_p)}{\sum_{p,g} 1 / (\tau^2 + \sigma^2_p)}.
\end{equation}
We use a similar approach to find the MLE of $\tau^2$ and $\sigma^2_p$.
In particular,
\begin{equation}
    \frac{\partial \ell}{\partial \tau^2} = -\frac{1}{2} \sum_{p,g} \frac{1}{\tau^2+\sigma^2_p} + \frac{1}{2}  \sum_{p,g} \frac{(X_{pg}-\mu)^2}{(\tau^2+\sigma^2_p)^2},
\end{equation}
or equivalently
\begin{equation}
    \label{eqn:tau_sq_mle_EB_gwar}
    \sum_{p,g} \frac{1}{\tau^2+\sigma^2_p} = \sum_{p,g} \frac{(X_{pg}-\mu)^2}{(\tau^2+\sigma^2_p)^2}.
\end{equation}
Additionally, for each pitcher $p$,
\begin{equation}
    \frac{\partial \ell}{\partial \sigma_p^2} = -\frac{1}{2} \sum_{g=1}^{N_p} \frac{1}{\tau^2+\sigma^2_p} + \frac{1}{2} \sum_{g=1}^{N_p} \frac{(X_{pg}-\mu)^2}{(\tau^2+\sigma^2_p)^2},
\end{equation}
or equivalently
\begin{equation}
    \label{eqn:sig_sq_mle_EB_gwar}
    \sum_{g=1}^{N_p} \frac{1}{\tau^2+\sigma^2_p} = \sum_{g=1}^{N_p} \frac{(X_{pg}-\mu)^2}{(\tau^2+\sigma^2_p)^2}.
\end{equation}
This process yields $\bP+2$ equations \eqref{eqn:mu_mle_EB_gwar}, \eqref{eqn:tau_sq_mle_EB_gwar}, and \eqref{eqn:sig_sq_mle_EB_gwar} in $\bP+2$ unknown variables $\mu$, $\tau^2$, and $\{\sigma^2_p\}$.
As suggested in \citet{brown2008}, we solve these equations by iterating until convergence, detailed in iterative Algorithm~\ref{algo:mle_EB_gwar}.

%%%%%%%%%%%%%%%%
\begin{algorithm}[hbt!]
    \caption{Compute the MLE of $\mu$, $\tau^2$, and $\{\sigma^2_p\}$ from model~\eqref{eqn:marginal_EB} }  %from Model~\eqref{eqn:emp_bayes_pq}}
    \label{algo:mle_EB_gwar}
    \begin{algorithmic}[1]
        % \Procedure{ExampleProcedure}{}
        \State \textbf{Input:} Grid $\WAR$ $\{X_{pg}: 1 \leq p \leq \bP, 1 \leq g \leq N_p\}$, \ $\epsilon$

        \State \textbf{Initialization:}
            \State $\mu(t=0) = \frac{1}{\bP} \sum_{p} \frac{1}{N_p} \sum_{g} X_{pg}$
            \State $\sigma^2_p(t=0) = \var(\{X_{pg}: 1\leq g \leq N_p\}) $
            \State $\tau^2(t=0) = \var(\{X_{pg}: 1\leq p \leq \bP, 1\leq g \leq N_p\}) - \frac{1}{\bP} \sum_p \sigma^2_p(t=0)$    %\hspace{\algorithmicindent} 
            \State $t = 1$

        \While{TRUE}
            \State \textbf{Step 1.} Solve for $\mu$ and save the result as $\mu(t)$:
            $$\mu(t) = \frac{\sum_{p,g} X_{pg} / (\tau^2(t-1) + \sigma^2_p(t-1))}{\sum_{p,g} 1 / (\tau^2(t-1) + \sigma^2_p(t-1))}.$$
            \State \textbf{Step 2.} Solve for $\tau^2$ (e.g., using a root finder) and save the result as $\tau^2(t)$:
            $$\sum_{p,g} \frac{1}{\tau^2+\sigma^2_p(t-1)} = \sum_{p,g} \frac{(X_{pg}-\mu(t))^2}{(\tau^2+\sigma^2_p(t-1))^2}.$$
            \For{$p = 1$ to $\bP$}
                \State \textbf{Step 3.} Solve for $\sigma^2_p$ (e.g., using a root finder) and save the result as $\sigma^2_p(t)$: 
                $$\sum_{g=1}^{N_p} \frac{1}{\tau^2(t)+\sigma^2_p} = \sum_{g=1}^{N_p} \frac{(X_{pg}-\mu(t))^2}{(\tau^2(t)+\sigma^2_p)^2}.$$
            \EndFor
            \If{$|\mu(t)-\mu(t-1)| < \epsilon$, $|\tau^2(t)-\tau^2(t-1)| < \epsilon$, and $|\sigma^2_p(t)-\sigma^2_p(t-1)| < \epsilon$ $\forall p$}
                \State \textbf{break} the while loop
            \Else
                \State $t = t + 1$
            \EndIf
        \EndWhile
        \State \textbf{Output:} $\widehat\mu = \mu(t)$, $\widehat\tau^2 = \tau^{2}(t)$, $\widehat\sigma^2_p = \sigma_p^{2}(t)$
        % \EndProcedure
    \end{algorithmic}
\end{algorithm}
%%%%%%%%%%%%%%%%

\begin{figure}[hbt!]
\centering
\includegraphics[width=10cm]{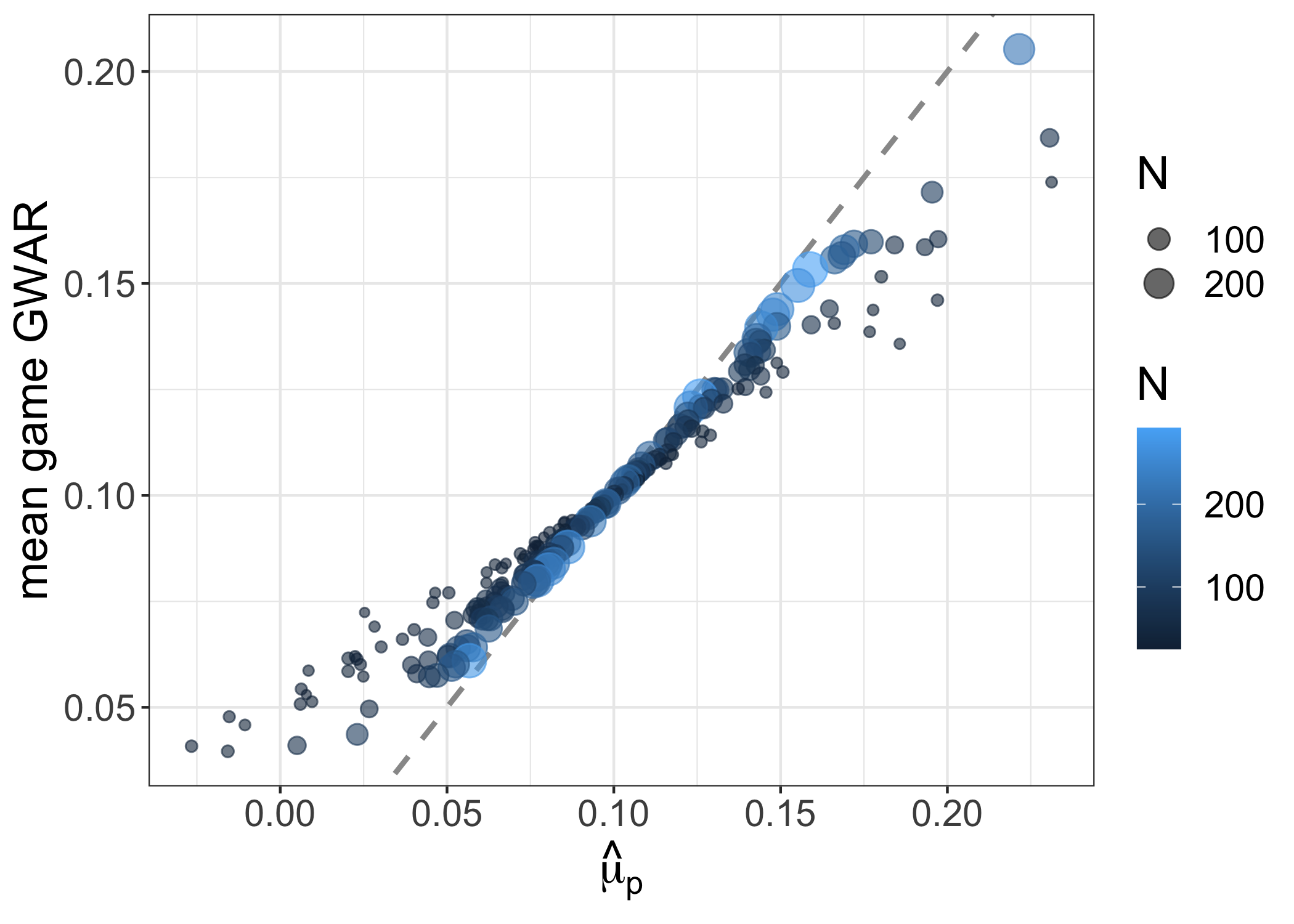}
\caption{
For starting pitchers $p$ from 2010 to 2018, his mean game $\GWAR$ versus his Empirical Bayes estimator $\widehat\mu_p^{\GWAR}$.
The dashed gray line is the line $y=x$.
} 
\label{fig:plot_EB_GWAR}
\end{figure}

Using our dataset of all starting pitchers from 2010 to 2018,\footnote{
    Here we just use starting pitcher-seasons whose FanGraphs $\WAR$ is available online, because the purpose of crafting our pitcher quality estimates $\widehat\mu_p$ is to compare the predictive capability of Grid $\WAR$ to FanGraphs $\WAR$.  Specifically, these are the starting pitcher-seasons with at least 25 starts in a season.
}
we run Algorithm~\ref{algo:mle_EB_gwar}, yielding maximum likelihood estimators of $\mu$, $\tau^2$, and $\{\sigma^2_p\}$.
With $\epsilon = 10^{-5}$, the algorithm converges after just four iterations.
Then, we plug these estimators into Formula~\eqref{eqn:post_mean_pq}, yielding parametric Empirical Bayes estimators of $\{\mu_p\}$.
In Figure~\ref{fig:plot_EB_GWAR} we compare these estimates $\{\widehat\mu^{GWAR}_p\}$ to each pitcher's observed mean game Grid $\WAR$.
For players with fewer games played (small gray dots), $\widehat\mu_p$ is shrunk towards the overall mean $\widehat\mu$. 
For players with enough games played (large blue dots), $\widehat\mu_p$ is essentially pitcher $p$'s mean game $\GWAR$, lying on the line $y=x$.

%%%%%%%%%%%%%%%%%
%%%%%%%%%%%%%%%%%

\begin{figure}[htb!]
    \centering{}
    \subfloat[]{{\includegraphics[width=7.5cm]{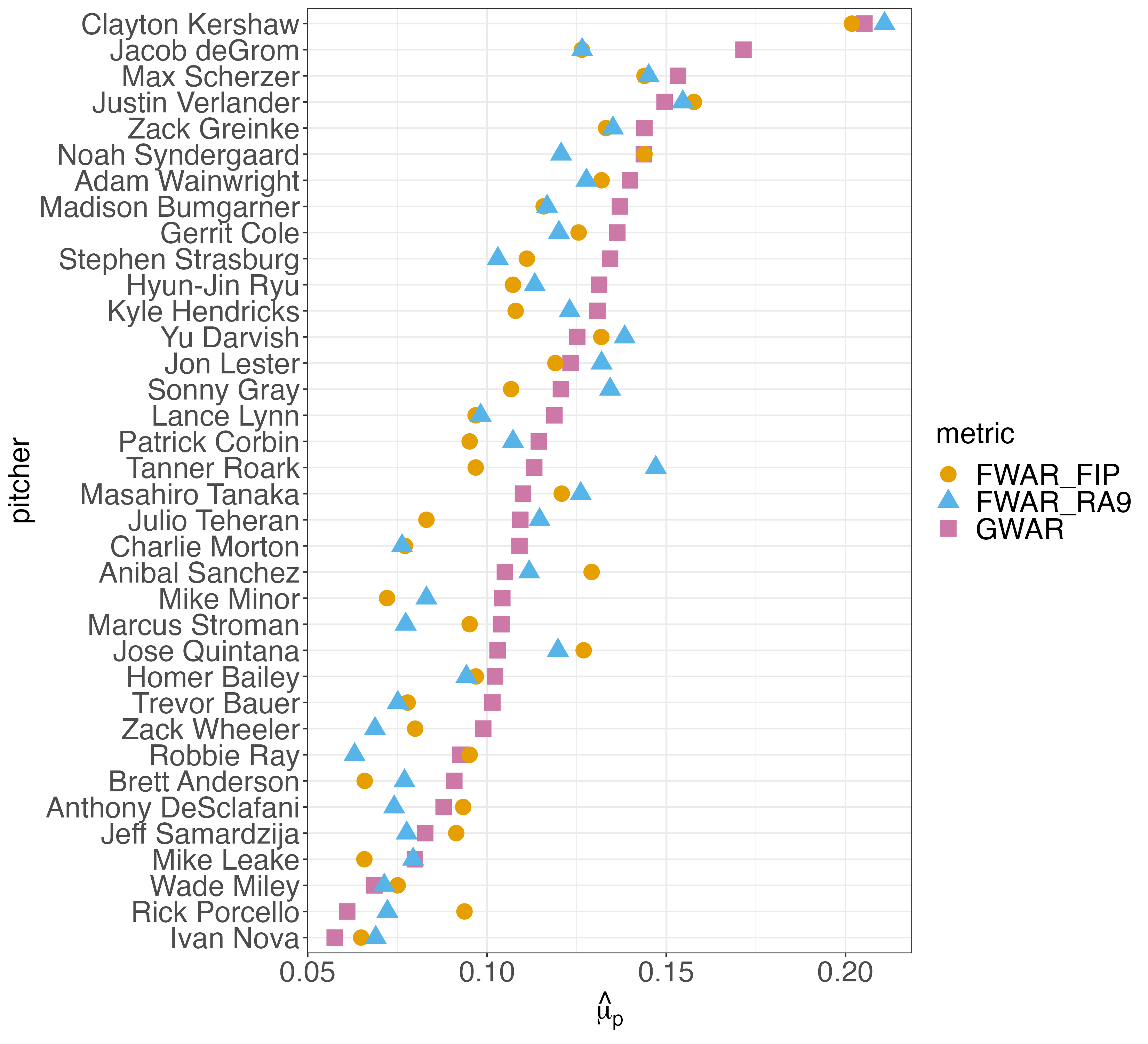}}\label{fig:plot_EB_pitRankingsMu}}%
    % \qquad
    \subfloat[]{{\includegraphics[width=7.5cm]{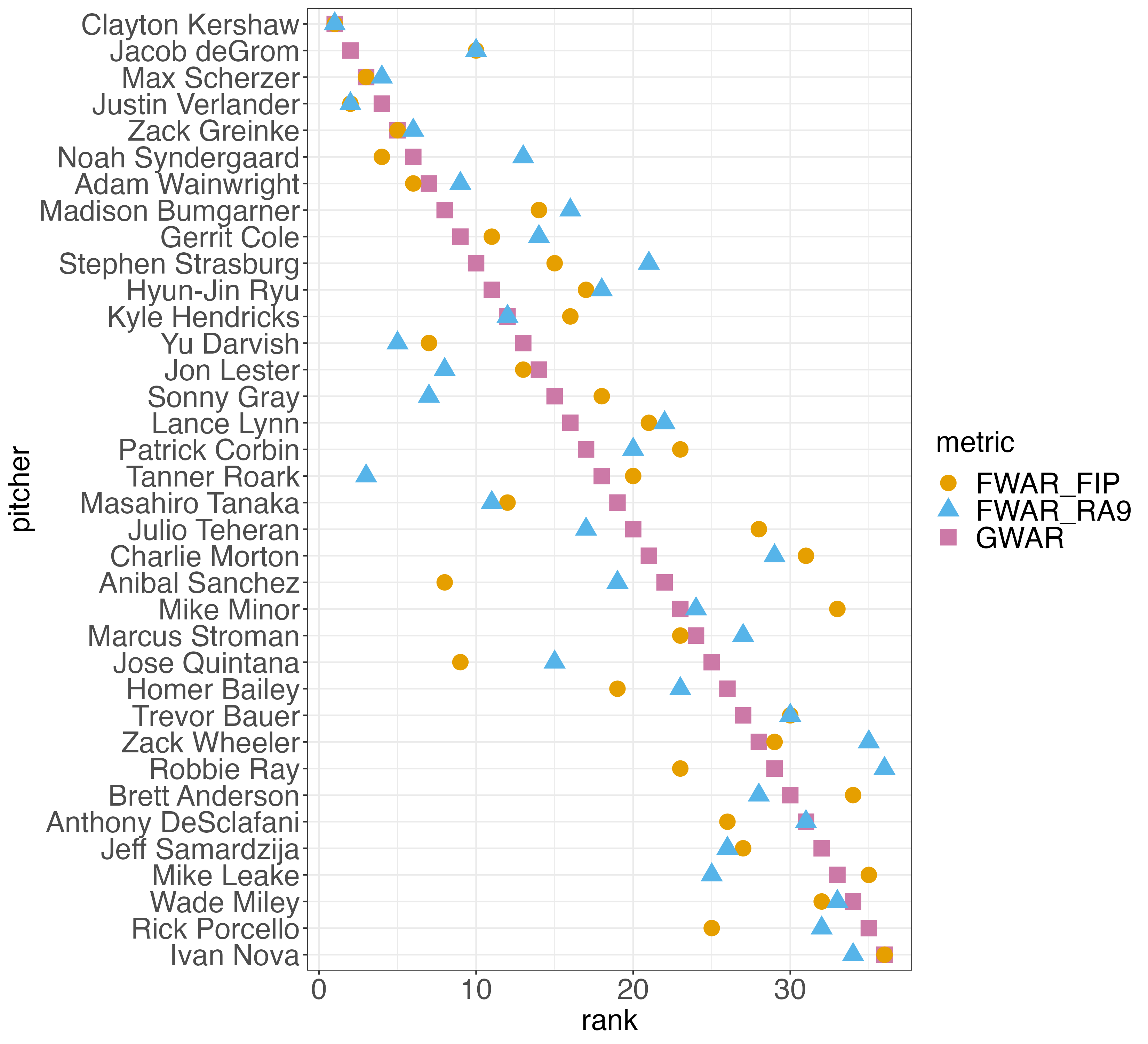}}\label{fig:plot_EB_pitRankingsRank}}%
    \caption{
    Pitcher quality estimates $\widehat\mu_p$ (Figure (a)) and their associated ranks $\widehat{R}_p$ (Figure (b)) for a set of starting pitchers prior to the 2019 season.
    } 
    \label{fig:plot_EB_pitRankings}
\end{figure}

In Figure~\ref{fig:plot_EB_pitRankings} we visualize starting pitcher rankings prior to the 2019 season according to $\widehat\mu_p$ (left) and the associated ranks $\widehat{R}_p$ (right).
Clayton Kershaw has the highest $\widehat\mu_p^{\GWAR}$ and Ivan Nova has the lowest. 

%%%%%%%%%%%%%%%%%%%%%
%%%%%%%%%%%%%%%%%%%%%

%%%%%%%%%%%%%%%%%%%%%%%%%%%%%%%%%%%%%%%%%%%%%%%%%%%%%%%%%%%%%%%%%%%%%%%%%%%%%%%%%%%%
\subsection{Empirical Bayes estimator of pitcher quality built from FanGraphs $\WAR$}

Our estimator $\widehat{\mu}_p^{\FWAR}$ of latent pitcher quality built from pitcher $p$'s previous seasons' observed FanGraphs $\WAR$ differs methodologically from our estimator built from Grid $\WAR$ in that $\FWAR$ is computed on the seasonal level and $\GWAR$ is computed on the game level.
Accordingly, we slightly modify the procedure from the previous section. %to account for this.

To begin, again index each starting pitcher by $p \in \{1,...,\bP\}$ and index pitcher $p$'s games by $g \in \{1,...,N_p\}$.
Let $X_{pg}$ denote pitcher $p$'s \textit{unobserved} FanGraphs $\WAR$ in game $g$.
Note that we observe his total $\FWAR$,
\begin{equation}
\label{eqn:sum_FGw}
    X_p = \sum_{g=1}^{N_p} X_{pg}.
\end{equation}
As before, we use model~\eqref{eqn:emp_bayes_pq}, which implies
\begin{align}
\begin{split}
\label{eqn:emp_bayes_pq_FWAR_X}
    X_{p} &\sim \N(N_p\mu_p, N_p\sigma^2_p), \\
    \mu_p  &\sim \N(\mu, \tau^2).
\end{split}
\end{align}
Therefore the posterior mean of pitcher $p$'s latent pitcher quality $\mu_p$ is
\begin{align}
\begin{split}
\label{eqn:post_mean_pq_FWAR}
    \widehat{\mu}_p := \E[\mu_p | X_{p}] = \frac{\frac{X_p}{\sigma^2_p} + \frac{\mu}{\tau^2}}{\frac{N_p}{\sigma^2_p} + \frac{1}{\tau^2}}.
\end{split}
\end{align}
This estimator is analogous to that from Equation~\eqref{eqn:post_mean_pq}, using $\FWAR$ instead of $\GWAR$.

As before, we use a parametric Empirical Bayes approach to estimate each starting pitcher's latent quality from his FanGraphs $\WAR$.
In particular, we compute maximum likelihood estimates for $\mu$, $\tau^2$, and $\{\sigma^2_p\}$ using the FanGraphs data $\{X_p\}$, which we plug in to Formula~\eqref{eqn:post_mean_pq_FWAR}.
We again begin finding the MLE by noting the marginal distribution of $X_p$ according to model~\eqref{eqn:emp_bayes_pq_FWAR_X}, 
\begin{equation}
    \label{eqn:marginal_EB_fwar}
    X_p \sim \N(N_p\cdot\mu, \ N_p\cdot(\tau^2+\sigma^2_p)).
\end{equation}
Thus the log-likelihood of the full FanGraphs dataset $\{X_p\}$ is proportional to
\begin{equation}
    \ell(X_p|\mu,\tau^2,\sigma^2_p) \propto -\frac{1}{2} \sum_{p=1}^{\bP} \log(\tau^2 + \sigma^2_p) - \frac{1}{2} \sum_{p=1}^{\bP} \frac{  (\frac{X_{p}}{N_p} - \mu)^2  }{(\tau^2 + \sigma^2_p)}.
\end{equation}
Setting the derivative of the log-likelihood with respect to $\mu$ (resp., $\tau^2$) equal to 0 and solving for $\mu$ (resp., $\tau^2$) yields the following equations,
\begin{align}
    \label{eqn:mu_mle_EB_fwar}
    & \mu = \frac{\sum_{p} (X_{p} / N_p) / (\tau^2 + \sigma^2_p)}{\sum_{p} 1 / (\tau^2 + \sigma^2_p)}.
\end{align}
and
\begin{align}
    \label{eqn:tau_sq_mle_EB_fwar}
    & \sum_{p} \frac{1}{\tau^2+\sigma^2_p} = \sum_{p} \frac{(X_{p}/N_p-\mu)^2}{(\tau^2+\sigma^2_p)^2}.
\end{align}
So, in designing an iterative algorithm analogous to Algorithm~\ref{algo:mle_EB_gwar} but for FanGraphs $\WAR$, we replace Equations \eqref{eqn:mu_mle_EB_gwar} (Step 1) and \eqref{eqn:tau_sq_mle_EB_gwar} (Step 2) with Equations \eqref{eqn:mu_mle_EB_fwar} and \eqref{eqn:tau_sq_mle_EB_fwar}.

Setting the derivative of the log-likelihood with respect to $\sigma^2_p$ equal to 0 and solving for $\sigma^2_p$, however, yields a trivial equation which doesn't include $\sigma^2_p$.
Intuitively this is because we can't glean information about $\sigma^2_p$ since we don't observe the game-level FanGraphs $\WAR$ $X_{pg}$. 
Therefore, in designing an iterative algorithm analogous to Algorithm~\ref{algo:mle_EB_gwar} but for FanGraphs $\WAR$, we eliminate Equation~\eqref{eqn:sig_sq_mle_EB_gwar} (Step 3) and replace $\sigma^2_p$ in Steps 1 and 2 with a constant hyperparameter $\sigma^2$.
We then choose the value of $\sigma^2$ which minimizes the $\rmse$ between resulting estimated pitcher quality $\widehat\mu_p$ and observed mean game $\FWAR$ in a hold-out set.
We detail the full procedure in Algorithm~\ref{algo:mle_EB_fwar}.

%%%%%%%%%%%%%%%%
\begin{algorithm}%[hbt!]
    \caption{Compute the MLE of $\mu$ and $\tau^2$ from model~\eqref{eqn:emp_bayes_pq_FWAR_X} }  %from Model~\eqref{eqn:emp_bayes_pq}}
    \label{algo:mle_EB_fwar}
    \begin{algorithmic}[1]
        \Procedure{1}{}
            \State \textbf{Input:} FanGraphs $\WAR$ $\{X_{p}: 1 \leq p \leq \bP\}$, \ $\epsilon$, \ $\sigma^2$

            \State \textbf{Initialization:}
                \State $\mu(t=0) = \frac{1}{\bP} \sum_{p} X_{p}$
                \State $\tau^2(t=0) = \var(\{X_{p}/N_p: 1\leq p \leq \bP\}) - \frac{1}{\bP} \sigma^2$    %\hspace{\algorithmicindent} 
                \State $t = 1$

            \While{TRUE}
                \State \textbf{Step 1.} Solve for $\mu$ and save the result as $\mu(t)$:
                $$\mu = \frac{\sum_{p} (X_{p} / N_p) / (\tau^2 + \sigma^2)}{\sum_{p} 1 / (\tau^2 + \sigma^2)}.$$
                \State \textbf{Step 2.} Solve for $\tau^2$ (e.g., using a root finder) and save the result as $\tau^2(t)$:
                $$\sum_{p} \frac{1}{\tau^2+\sigma^2} = \sum_{p} \frac{(X_{p}/N_p-\mu)^2}{(\tau^2+\sigma^2)^2}.$$
                \If{$|\mu(t)-\mu(t-1)| < \epsilon$ and $|\tau^2(t)-\tau^2(t-1)| < \epsilon$}
                    \State \textbf{break} the while loop
                \Else
                    \State $t = t + 1$
                \EndIf
            \EndWhile
            \State \textbf{Output:} $\widehat\mu = \mu(t)$, $\widehat\tau^2 = \tau^{2}(t)$
        \EndProcedure
        % \\ 
        \State
        \Procedure{2}{}
            \State \textbf{Input:} FanGraphs $\WAR$ $\{X_{p}: 1 \leq p \leq \bP\}$, \ $\epsilon$
            \State \textbf{Initialization:}
                \State Split $\{X_{p}\}$ into a training set $\{X^{(train)}_{p}\}$ and a validation set $\{X^{(test)}_{p}\}$
                \State In particular, $\{X^{(train)}_{p}\}$ is all $\FWAR$ from 2010-2016, $\{X^{(test)}_{p}\}$ is all $\FWAR$ from 2017-2018
                \State Sigmas = vector of smartly chosen positive values 
                \State Losses = empty vector 
            \For{$\sigma^2$ in Sigmas}
                \State \textbf{Step 1.} Run Procedure 1 with inputs $\{X^{(train)}_{p}\}$, $\epsilon$, and $\sigma^2$
                \State \textbf{Step 2.} Use $\widehat\mu$ and $\widehat\tau^2$ from Step 1 to estimate pitcher quality $\widehat\mu_p$ (Formula~\eqref{eqn:post_mean_pq_FWAR})
                \State \textbf{Step 3.} Append $\rmse(\{ \widehat\mu_p, \ X^{(test)}_{p}\} )$ to Losses
            \EndFor
            \State \textbf{Output:} $\widehat\sigma^2_p \equiv \sigma^2$, where $\sigma^2$ corresponds to the minimum value in Losses
        \EndProcedure
    \end{algorithmic}
\end{algorithm}
%%%%%%%%%%%%%%%%

\begin{figure}[htb!]
    \centering{}
    \subfloat[]{{\includegraphics[width=7cm]{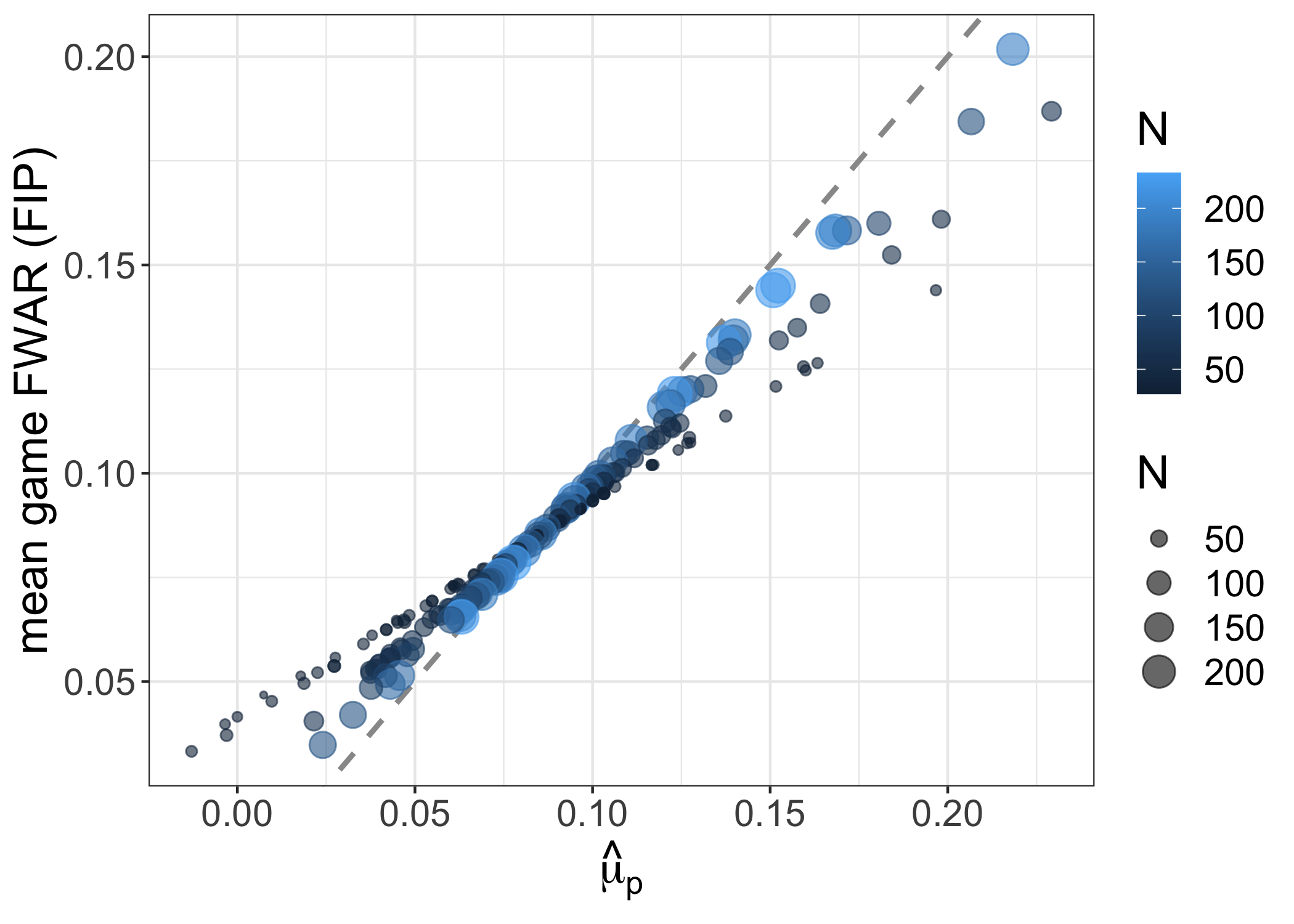}}\label{fig:plot_EB_FWAR_FIP}}%
    \qquad
    \subfloat[]{{\includegraphics[width=7cm]{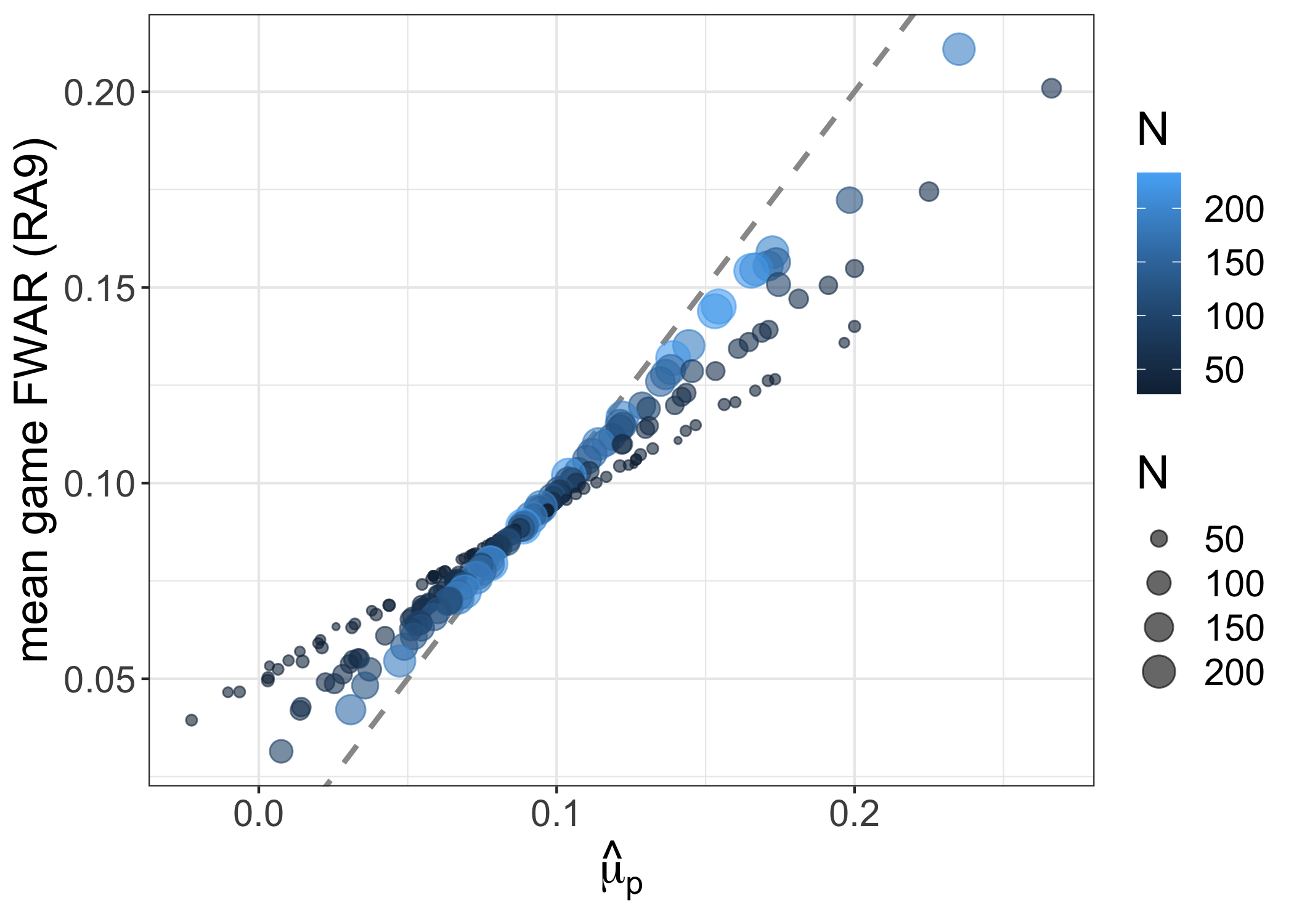}}\label{fig:plot_EB_FWAR_RA9}}%
    \caption{
    For starting pitchers $p$ from 2010 to 2018, his mean game $\FWAR$ versus his Empirical Bayes estimator $\widehat\mu_p^{\FWAR}$, built from $\FWARf$ in Figure (a) and $\FWARr$ in Figure (b).
    The dashed gray line is the line $y=x$.
    } 
    \label{fig:plot_EB_FWAR}
\end{figure}

Using the same dataset of starting pitchers from 2010 to 2018 as before, we run Algorithm~\ref{algo:mle_EB_fwar}, yielding maximum likelihood estimators of $\mu$ and $\tau^2$ and an estimate of $\sigma^2_p \equiv \sigma^2$.
With $\epsilon = 10^{-5}$, the algorithm converges again after just four iterations.
Then, we plug these estimators into Formula~\eqref{eqn:post_mean_pq_FWAR}, yielding parametric Empirical Bayes estimators of $\{\mu_p\}$.
In Figure~\ref{fig:plot_EB_FWAR} we compare these estimates $\{\widehat\mu^{FWAR}_p\}$ to each pitcher's mean game FanGraphs $\WAR$ from 2010 to 2018.
As before, for players with fewer games played (small gray dots), $\widehat\mu_p$ is shrunk towards the overall mean $\mu$. 
Also, for players with enough games played (large blue dots), $\widehat\mu_p$ is essentially pitcher $p$'s mean game $\FWAR$, lying on the line $y=x$.

%%%%%%%%%%%%%%%%%%%%%
%%%%%%%%%%%%%%%%%%%%%

A weakness of our Empirical Bayes approach is that it assumes latent pitcher quality $\mu_p$ is constant over the decade from 2010 to 2019.
Player quality is, however, non-stationary over time.
Therefore, in future work we suggest using a similar Empirical Bayes approach to estimate pitcher quality, except downweighting data further back in time (e.g., using exponential decay weighting as in \citet{darko}) in the posterior mean Formulas \eqref{eqn:post_mean_pq} and \eqref{eqn:post_mean_pq_FWAR}.

In Figure~\ref{fig:plot_EB_pitRankings} we visualize starting pitcher rankings prior to the 2019 season according to $\widehat\mu_p$ (left) and the associated ranks $\widehat{R}_p$ (right).
Clayton Kershaw has the highest $\widehat\mu_p^{\FWARf}$ and $\widehat\mu_p^{\FWARr}$,  and Ivan Nova has the lowest such values.
We see that there is a nontrivial difference between pitcher quality estimates and rankings built from Grid $\WAR$ and FanGraphs $\WAR$.

%%%%%%%%%%%%%%%%%%%%%%%%%%%%%%%%%%%%%%%%%%%%%%%%%%%%%%%%%%%%%%%%%%%%%%%%%%%%%%%%%%%%
%%%%%%%%%%%%%%%%%%%%%%%%%%%%%%%%%%%%%%%%%%%%%%%%%%%%%%%%%%%%%%%%%%%%%%%%%%%%%%%%%%%%
\section{Estimating the Park Effects $\alpha$}\label{sec:park_effects}

In this section, we detail why we use ridge regression to estimate park effects.
First, in Section~\ref{sec:existing_park_fx}, we discuss existing park factors from ESPN, FanGraphs, and Baseball Reference.
Then, in Section~\ref{sec:problems_w_existingParkFx}, we discuss problems with these existing park effects.
In Section~\ref{sec:parkFxModel}, we introduce our park effects model, designed to yield park effects which represent the expected runs scored in a half-inning at a ballpark above that of an average park, if an average offense faces an average defense. 
Then, in Sections~\ref{sec:first_sim_study} and \ref{sec:second_sim_study}, we conduct two simulation studies which show that ridge repression works better than other methods at estimating park effects.
Then, in Section~\ref{sec:compare_existing_parkFx}, we show that our ridge park effects have better out-of-sample predictive performance than existing park effects from ESPN and FanGraphs.
Finally, in Sections~\ref{sec:sec:threeYrParkFx} and \ref{sec:parkFxFinal}, we discuss our final ridge park effects, fit on data from all half-innings from 2017 to 2019.

\subsection{Existing Park Effects.}\label{sec:existing_park_fx}

FanGraphs, Baseball Reference, and ESPN each have runs-based park factors, which are all variations of a common formula: the ratio of runs per game at home to runs per game on the road. In particular, ESPN's park factors are an unadjusted version of this formula \citep{espn_pf},
\begin{equation}
\alpha_{ESPN} = \frac{(\text{home runs scored } + \text{ home runs allowed})/(\text{home games})}{(\text{road runs scored } + \text{ road runs allowed})/(\text{road games})}.
\label{eqn:espn_pf}
\end{equation}

FanGraphs modifies this formula by imposing a form of regression onto the park factors \citep{FanGraphs_pf}. In particular, the FanGraphs park factors $\alpha_{FanGraphs}$ are computed via
\begin{equation}
\begin{cases}
 \xi = ( (\text{home runs per game}) - (\text{road runs per game}) )/(\text{number of MLB teams}), \\
 PF_{raw} = (\text{home runs per game}) / ( (\text{road runs per game}) + \xi), \\
 iPF = (PF_{raw} + 1)/2, \\
 \alpha_{FanGraphs} = 1 - (1 - iPF) \cdot w,
\end{cases}
\label{eqn:FanGraphs_pf}
\end{equation}
where $w$ is a regression weight determined by the number of years in the dataset (e.g., for a three year park factor, $w = 0.8$). 

Finally, Baseball Reference's park factors are a long series of adjustments on top of ESPN's park factors, computed separately for batters and pitchers \citep{BR_pf}. In particular, Baseball Reference begins with 
\begin{equation}
\frac{(\text{home runs scored})/(\text{home games})}{(\text{road runs scored})/(\text{road games})}
\label{eqn:br_pf_1}
\end{equation}
for batters and 
\begin{equation}
\frac{(\text{home runs allowed})/(\text{home games})}{(\text{road runs allowed})/(\text{road games})}
\label{eqn:br_pf_2}
\end{equation}
for pitchers as base park factors. Then, they apply several adjustments on top of these base values. For instance, they adjust for the quality of the home team and the fact that the batter doesn't face its own pitchers. These adjustments, however, are a long series of convoluted calculations, so we do not repeat them here.

\subsection{Problems with Existing Park Effects}\label{sec:problems_w_existingParkFx}

There are several problems with these existing runs-based park effects. 
First, ESPN and FanGraphs do not adjust for offensive and defensive quality at all, and Baseball Reference adjusts for only a fraction of team quality. It is important to adjust for team quality in order to de-bias the park factors. For example, the Colorado Rockies play in the NL West, a division with good offensive teams such as the Dodgers, Giants, and Padres. So, by ignoring offensive quality in creating park factors, the Rockies' park factor may be an overestimate, since many of the runs scored at their park may be due to the offensive power of the NL West rather than the park itself. 
By ignoring team quality, the ESPN and FanGraphs park factors are biased. Baseball Reference's park factors adjust for the fact that a team doesn't face its own pitchers, albeit through a convoluted series of ad-hoc calculations. Although adjusting for not facing one's own pitchers slightly de-biases the park factors, it does not suffice as a full adjustment of the offensive and defensive quality of a team's schedule.

Second, these existing runs-based park effects do not come from a statistical model. This makes it more difficult to quantitatively measure which park factors are the ``best'', for instance via some loss function. In other words, it is more difficult to quantitatively know that Baseball Reference's park factors are \textit{actually} more accurate than FanGraphs', in some mathematical sense, besides that it claims to adjust for some biases in its derivation, although we discuss a way to do so in Section~\ref{sec:compare_existing_parkFx}.
Another benefit of a statistical model is that it will allow us to adjust for the offensive and defensive quality of a team and its opponents \textit{simultaneously}.
Finally, a statistical model will give us a firm physical interpretation of the park factors.

Hence, in this paper, we create our own park factors, which are the fitted coefficients of a statistical model that adjusts for team offensive and defensive quality.

\subsection{Our Park Effects Model}\label{sec:parkFxModel}

In this section, we introduce our park effects model, designed to yield park effects which represent the expected runs scored in a half-inning at a ballpark above that of an average park, if an average offense faces an average defense. 

We index each half-inning in our dataset by $i$, each park by $j$, and each team-season by $k$. 
We define the park matrix $\P$ so that $\P_{ij}$ is 1 if the $i^{th}$ half-inning is played in park $j$, and 0 otherwise. 
Similarly, we define the offense matrix $\OO$ so that $\OO_{ik}$ is 1 if the $k^{th}$ team-season is on offense during the $i^{th}$ half-inning, and 0 otherwise, and define the defense matrix $\D$ so that $\D_{ik}$ is 1 if the $k^{th}$ team-season is on defense during the $i^{th}$ half-inning, and 0 otherwise. 
We denote the runs scored during the $i^{th}$ half-inning by $y_i$. 
Then, we model $y_i$ using a linear model,
\begin{align}
    y_i &= \beta_0 + \sum_{j \neq \text{ANA}} \P_{ij} \beta_j^{(\text{park})} + \sum_{k \neq \text{ANA2017}} \OO_{ik} \beta_k^{(\text{off})} + \sum_{k \neq \text{ANA2017}} \D_{ik} \beta_k^{(\text{def})} + \epsilon_i,
    \label{eqn:parkfx_model_A}
\end{align}
where $\epsilon_i$ is mean-zero noise,
\begin{equation}
    \E[\epsilon_i] = 0.
\end{equation}
Succinctly, we model
\begin{equation}
    y_i = \XX_{i\ast}\beta + \epsilon_i,
    \label{eqn:parkfx_model_B}
\end{equation}
where
\begin{equation}
    \XX = 
    \begin{bmatrix}
        1, \ \ \P, \ \  \OO, \ \ \D
    \end{bmatrix}
\end{equation}
and
\begin{equation}
    \beta^\top = (\beta_0, \ \ \beta^{(\text{park})\top}, \ \  \beta^{(\text{off})\top}, \ \ \beta^{(\text{def})\top}).
\end{equation}
The coefficients are fitted relative to the first park ANA (the Anaheim Angels) and relative to the first team-season ANA2017 (the Angels in 2017). By including distinct coefficients for each offensive-team-season and each defensive-team-season, we adjust for offensive and defensive quality simultaneously in fitting our park factors.
Finally, in order to make our park effects represent the expected runs scored in a half-inning at a ballpark above that of an average park, we subtract the mean park effect from each park effect,
\begin{equation}
    \beta^{(\text{park})} \leftarrow \beta^{(\text{park})} - \frac{1}{n}\sum_{j=1}^{n} \beta^{(\text{park})}_j.
\end{equation}

%%%%%%%%%%%%%%%%%%%%%%%%%%%%%%%%%%%%%%%%%%%%%%%%%%%%%%%%%%%%%%%%%
\subsection{First Simulation Study}\label{sec:first_sim_study}

We have a park effects model, Formula~\eqref{eqn:parkfx_model_A}, but it is not immediately obvious which algorithm we should use to fit the model. In particular, due to multicollinearity in the observed data matrix $\XX$, ordinary least squares is sub-optimal. Hence we run a simulation study in order to test various methods of fitting model~\eqref{eqn:parkfx_model_A}, using the method which best recovers the ``true'' simulated park effects as the park factor algorithm to be used in computing Grid $\WAR$: ridge regression.

\textbf{Simulation setup.} In our first simulation study, we assume that the park, team offensive quality, and team defensive quality coefficients are independent. Specifically, we simulate 25 ``true'' parameter vectors $\{\beta^{[m]}\}_{m=1}^{25}$ according to 
\begin{equation}
\begin{cases}
 \beta_0 = 0.4, \\
 \beta^{(\text{park})}_j \overset{iid}{\sim} \N(0.04, 0.065), \\
 \beta^{(\text{off})}_k \overset{iid}{\sim} \N(0.02, 0.045), \\
 \beta^{(\text{def})}_k \overset{iid}{\sim} \N(0.03, 0.07).
\end{cases}
\label{eqn:beta_sim}
\end{equation}
Then, we assemble our data matrix $\XX$ to consist of every half-inning from 2017 to 2019. Then, we simulate 25 ``true'' outcome vectors $\{y^{[m]}\}_{m=1}^{25}$ according to 
\begin{equation}
    y_i = \text{Round}\big( \N_{+}(\XX_{i\ast}\beta, \ 1) \big).
\label{eqn:y_sim}
\end{equation}
We use a truncated normal distribution, denoted by $\N_{+}$, in order to make $y_i$ positive. We round $y_i$ so that it is a positive integer, since $y_i$ represents the runs scored in the $i^{th}$ inning. Although we don't directly simulate $\epsilon_i$, our simulated $y_i$ still adheres to model~\eqref{eqn:parkfx_model_B}, as it has mean $\XX_{i\ast}\beta$. 
We choose the values in Formula~\eqref{eqn:beta_sim} so that the simulated outcome vectors $\{y^{[m]}\}_{m=1}^{25}$ seem reasonable in representing the runs scored in a half-inning.
% We draw the ``true'' coefficient values according to formula \ref{eqn:beta_sim} so that $y_i$ resembles the observed vector of runs scored in each inning.

Our goal is to recover the park effects $\beta^{(\text{park})}$, so our evaluation metric of an estimator $\hat \beta$ is the average simulation error of the fitted park effects,
\begin{equation}
    \frac{1}{25} \sum_{m=1}^{25} \bigg\| \hat\beta^{(\text{park})[m]} - \beta^{(\text{park})[m]} \bigg\|_2.
\label{eqn:parkFx_loss}
\end{equation}

Note that it doesn't make sense to compare the existing ESPN and FanGraphs park factors to park effects methods based on model~\eqref{eqn:parkfx_model_A} as part of this simulation study because they are not based on model~\eqref{eqn:parkfx_model_A}. In fact, ESPN and FanGraphs park effects are not based on any statistical model. 
Rather, in Section~\ref{sec:compare_existing_parkFx}, we separately compare these existing park factors to our model-based park factors. % using RMSE.

\textbf{Method 1: OLS without adjusting for team quality.} The naive method of estimating the park factors $\beta^{(park)}$ is ordinary least squares regression while ignoring team offensive quality and team defensive quality, as done in \citet[Formula 11]{BaumerJensenMatthews+2015+69+84}. In other words, fit the park coefficients using OLS on the following model,
\begin{align}
    y_i &= \beta_0 + \sum_{j \neq \text{ANA}} \P_{ij} \beta_j^{(\text{park})} + \epsilon_i.
    \label{eqn:parkfx_simModel_1}
\end{align}
In failing to adjust for offensive and defensive quality, we expect this algorithm to perform poorly.

\textbf{Method 2: OLS.} Next, we adjust for offensive quality, defensive quality, and park simultaneously using ordinary least squares (OLS) on model~\eqref{eqn:parkfx_model_A}. 
This method is similar to that from \citet{AcharyaParkFactors}, although they compute game-level park factors and we compute half-inning-level park factors.
This yields an unbiased estimate of the park effects, and so we expect this method to perform better than the previous one. 
% This method is similar to that from \citet{AcharyaParkFactors}, which uses OLS with fixed effects to adjust for offensive and defensive quality, although they compute game-level park factors and we compute half-inning-level park factors.

\textbf{Method 3: Three-Part OLS.} Although OLS using model~\eqref{eqn:parkfx_model_A} is unbiased, the fitted coefficients have high variance due to the multicollinearity of the data matrix $\XX$. In particular, the park matrix $\P$ is correlated with the offensive team matrix $\OO$ and the defensive team matrix $\D$ because in each half-inning, either the team on offense or defense is the home team. We may visualize the collinearity in $\XX$ by denoting all of the half-innings (rows) in which the road team is batting by $road$, denoting the half-innings in which the home team is batting by $home$, and writing $\XX$ (e.g., for one season of data) as
\begin{equation}
    \XX = 
    \begin{bmatrix}
        1 \ \ \P_{road,\ast} \ \  \OO_{road,\ast} \ \ \P_{road,\ast} \\
        \ \ 1 \ \ \P_{home,\ast} \ \  \P_{home,\ast} \ \ \D_{home,\ast} \\
    \end{bmatrix}.
\label{eqn:X_collinear}
\end{equation}

To address this collinearity issue, we propose a three-part OLS algorithm. First, we estimate the offensive quality coefficients during half-innings in which the road team is batting. We do so via OLS on the following model,
\begin{align}
    y_{road} = \beta_0 1 + \P_{road,\ast} \beta^{(\text{park-def})} +\OO_{road,\ast} \beta^{(\text{off})} + \epsilon.
\end{align}
This yields a decent estimate $\hat\beta^{(\text{off})}$ of $\beta^{(\text{off})}$, in particular because for one season of innings, $\P_{road,\ast} = \D_{road,\ast}$, and for multiple seasons of innings, $\P_{road,\ast} \simeq \D_{road,\ast}$.

Second,  we estimate the defensive quality coefficients during half-innings in which the home team is batting. We do so via OLS using the following model,
\begin{align}
    y_{home} = \beta_0 1 + \P_{home,\ast} \beta^{(\text{park-off})} +\D_{home,\ast} \beta^{(\text{def})} + \epsilon.
\end{align}
This yields a decent estimate $\hat\beta^{(\text{def})}$ of $\beta^{(\text{def})}$, in particular because $\P_{home,\ast} \simeq \OO_{home,\ast}$.

Third, we use the fitted team quality coefficients $\hat\beta^{(\text{off})}$ and $\hat\beta^{(\text{def})}$ on all half-innings to obtain the park effects. 
Specifically, we run OLS on the following model,
\begin{align}
    y_i = \beta_0 + \beta_1 \cdot \sum_{k \neq \text{ANA2017}} \OO_{ik} \hat\beta_k^{(\text{off})} + \beta_2 \cdot \sum_{k \neq \text{ANA2017}} \D_{ik} \hat\beta_k^{(\text{def})} + \sum_{j \neq \text{ANA}} \P_{ij} \beta_j^{(\text{park})} + \epsilon_i,
    \label{eqn:parkfx_simModel_2}
\end{align}
yielding fitted park coefficients $\hat\beta^{(\text{park})}$.

\textbf{Method 4: Ridge.} Finally, we use ridge regression to fit model~\eqref{eqn:parkfx_model_A}. In the presence of multicollinearity, ridge regression coefficient estimates may improve upon OLS estimates by introducing a small amount of bias in order to reduce the variance of the estimates \citep{ridge}. 
We tune the ridge parameter $\lambda$ using cross validation.
% We tune the ridge parameter $\lambda = 0.25$ on an out-of-sample validation set consisting of all innings from 2014 to 2016. 

\textbf{Simulation Results.} Recall from the start of this section that we simulate 25 sets of ``true'' parameters $\{\beta^{[m]}\}_{m=1}^{25}$ and 25 ``true'' outcome vectors $\{y^{[m]}\}_{m=1}^{25}$. Then, using the observed data matrix $\XX$, which consists of all half-innings from 2017 to 2019, we run each of our five methods, yielding parameter estimates, and evaluate them using the average simulation error from Formula~\eqref{eqn:parkFx_loss}. We report the results in Table~\ref{table:sim1_results}. 

%%%%%%%%%%%%%%%%%
\begin{table}[hbt!]
\centering
\begin{tabular}{ll} \hline
method & simulation error \\ \hline
Ridge & \textbf{0.0244} \\
% OLS with biased adjustments of team quality & \textbf{0.0266} \\
OLS & 0.0326 \\
Three-part OLS & 0.0330 \\
OLS without adjusting for team quality & 0.0610 \\
\hline
\end{tabular}
\caption{
Results of our first simulation study. %Ridge regression works best.
}
\label{table:sim1_results}
\end{table}

% OLS with biased adjustments of team quality performs the worst by far. This is because adjusting for the biased measures of team quality leads to biased park effects. OLS without adjusting for team quality performs second worst. Ignoring team quality still leads to a biased estimate of the park effects, but since the team quality coefficients are small in magnitude, ignoring them is better than using wrong estimates. OLS and three-part OLS, which include proper adjustments for team quality, perform similarly and are second best. Three-part OLS turns out not to be an improvement over OLS because despite the multicollinearity, there is enough linear independence between the batting-road and batting-home half-innings to obtain reasonably accurate team quality estimates. Additionally, three-part OLS uses half as much data to estimate team quality, which is significant because the outcome variable inning runs is so noisy. Lastly, as expected, ridge regression performs the best, and is significantly better than OLS. In Figure \ref{fig:sim1_compare}, we visualize one of the 25 simulations by plotting the ``true'' park effects against the ridge estimates and the OLS estimates. We see that ridge regression. OLS appears to be biased, probably from overfitting to the noise, whereas ridge appears to be unbiased, lying evenly around the line $y=x$. 

\begin{figure}[hbt!]
\centering
\includegraphics[width=10cm]{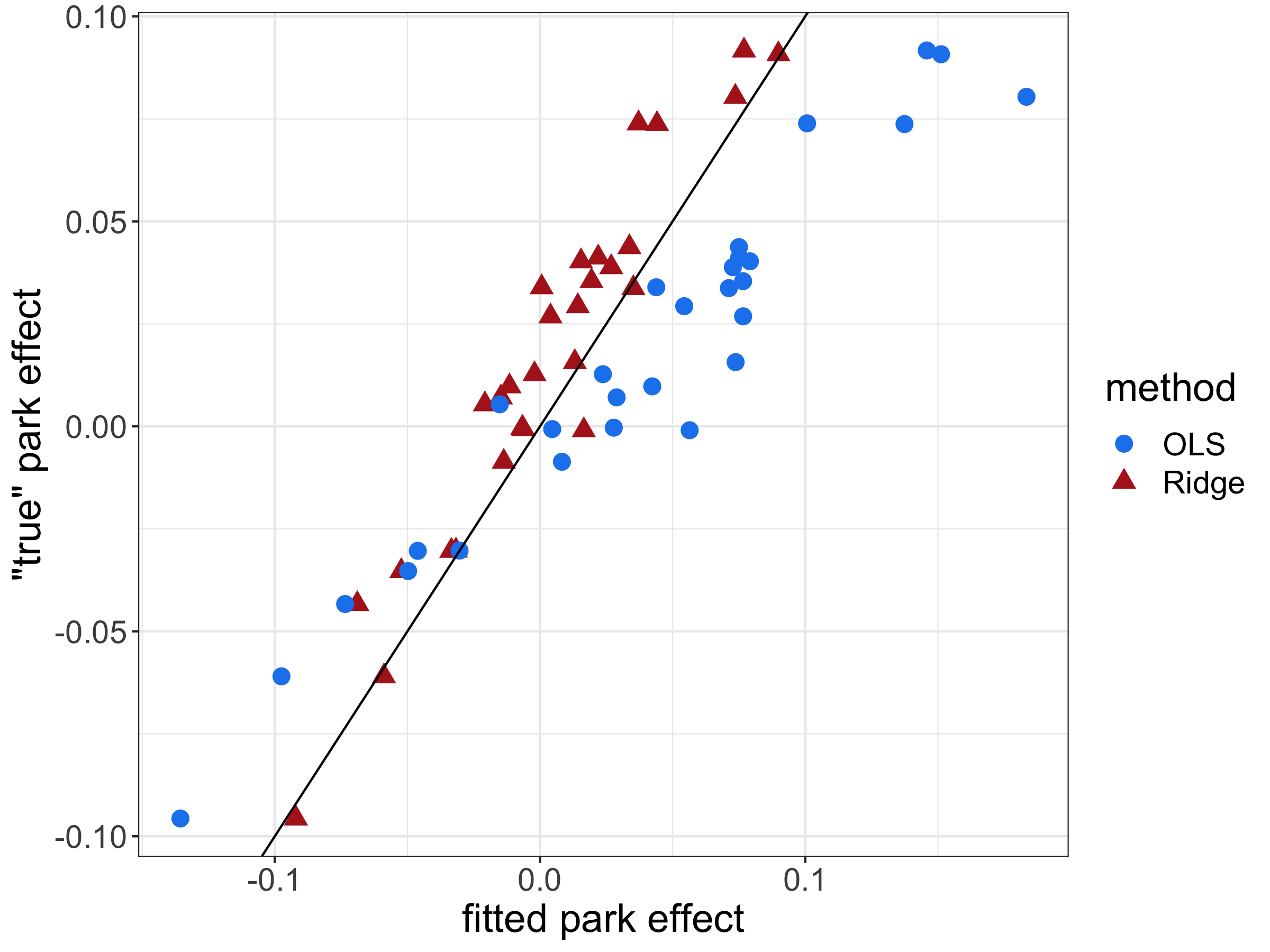}
\caption{For one of 25 simulations from our first simulation study from Section~\ref{sec:first_sim_study}, plot the ``true'' park effects against the ridge estimates and the OLS estimates. The line $y=x$, shown in black, represents a perfect fit between the ``true'' and fitted park effects. The OLS estimates are biased, whereas the ridge estimates lie more evenly around the line $y=x$.} 
\label{fig:sim1_compare}
\end{figure}

The OLS estimator without adjusting for team quality performs worst, as ignoring team quality leads to a biased estimate of the park effects. OLS and three-part OLS, which include proper adjustments for team quality, perform similarly and are second best. Three-part OLS turns out not to be an improvement over OLS because despite the multicollinearity, there is enough linear independence between the batting-road and batting-home half-innings to obtain reasonably accurate team quality estimates. Also, if $\XX$ contains multiple years worth of data, three-part OLS leads to slightly biased estimates of $\beta^{\text{(off)}}$ and $\beta^{\text{(def)}}$ since steps one and two only adjust for park. Additionally, three-part OLS uses half as much data to estimate team quality, which is significant because the outcome variable inning runs is so noisy. Lastly, ridge regression performs the best, and is significantly better than OLS. In Figure~\ref{fig:sim1_compare}, we visualize one of the 25 simulations by plotting the ``true'' park effects against the ridge estimates and the OLS estimates. We see that OLS is biased, whereas ridge lies more evenly around the line $y=x$. 

%%%%%%%%%%%%%%%%%
%%%%%%%%%%%%%%%%%

%%%%%%%%%%%%%%%%%%%%%%%%%%%%%
\subsection{Second Simulation Study}\label{sec:second_sim_study}

A primary criticism of the first simulation study from Section~\ref{sec:first_sim_study} is that in actual baseball, the offensive and defensive quality coefficients are not independent. Rather, often times offensive and defensive qualities are correlated within MLB divisions. For instance, in 2021, the Rays, Red Sox, Yankees, and Blue Jays of the AL East each had at least 91 wins, and so were all good offensive teams. Correlated offensive and defensive qualities within divisions introduces additional collinearity into the data matrix $\XX$, since teams play other teams within their division at a disproportionately high rate.

Another criticism of the first simulation study is that it treats Colorado's park effect as a draw from the same distribution as the other park effects, whereas in real life we know Colorado's park effect is an outlier as a result of the high altitude. 

So, in this section, we conduct a second simulation study which incorporates intra-divisional collinearity and forces Colorado's park effect to be an outlier. Specifically, we simulate 25 ``true'' parameter vectors $\{\beta^{[m]}\}_{m=1}^{25}$ according to 
\begin{equation}
\begin{cases}
 \beta_0 = 0.15, \\
 \beta^{(\text{park $\neq$ DEN)}}_j \overset{iid}{\sim}  \N(0.04, 0.065), \\
 \beta^{(\text{park = DEN)}} = 0.32, \\
 \beta^{(\text{div, off)}} \overset{iid}{\sim}  \N(0.02, 0.05), \\
 \beta^{(\text{off)}}_k \overset{iid}{\sim} \N(\beta^{(\text{div, off})}, 0.02), \\
 \beta^{(\text{div, def)}} \overset{iid}{\sim}  \N(0.03, 0.0), \\
 \beta^{(\text{def)}}_k \overset{iid}{\sim}  \N(\beta^{(\text{div, def})}, 0.033),
\end{cases}
\label{eqn:beta_sim_2}
\end{equation}
where DEN refers to Coors Field in Denver, Colorado. In other words, each division has its own offensive and defensive quality means, coercing the offensive and defensive qualities within each division to be correlated. We choose the values in Formula~\eqref{eqn:beta_sim_2} so that the simulated outcome vectors $\{y^{[m]}\}_{m=1}^{25}$ seem reasonable in representing the runs scored in a half-inning. 

Because Colorado is an outlier in this simulation study, we judge our methods based on two loss functions: the Colorado park effect average simulation error,
\begin{equation}
    \frac{1}{25} \sum_{m=1}^{25} \bigg| \hat\beta^{(\text{park})[m]}_{j = \text{DEN}} - \beta^{(\text{park})[m]}_{j = \text{DEN}} \bigg|,
\label{eqn:parkFx_loss_2Col}
\end{equation}
and the non-Colorado park effect average simulation error,
\begin{equation}
    \frac{1}{25} \sum_{m=1}^{25} \bigg\| \hat\beta^{(\text{park})[m]}_{j \neq \text{DEN}} - \beta^{(\text{park})[m]}_{j \neq \text{DEN}} \bigg\|_2.
\label{eqn:parkFx_loss_2nonCol}
\end{equation}
The remainder of the second simulation study proceeds identically to the first simulation study from the previous section. 

\begin{figure}[hbt!]
\centering
\includegraphics[width=10cm]{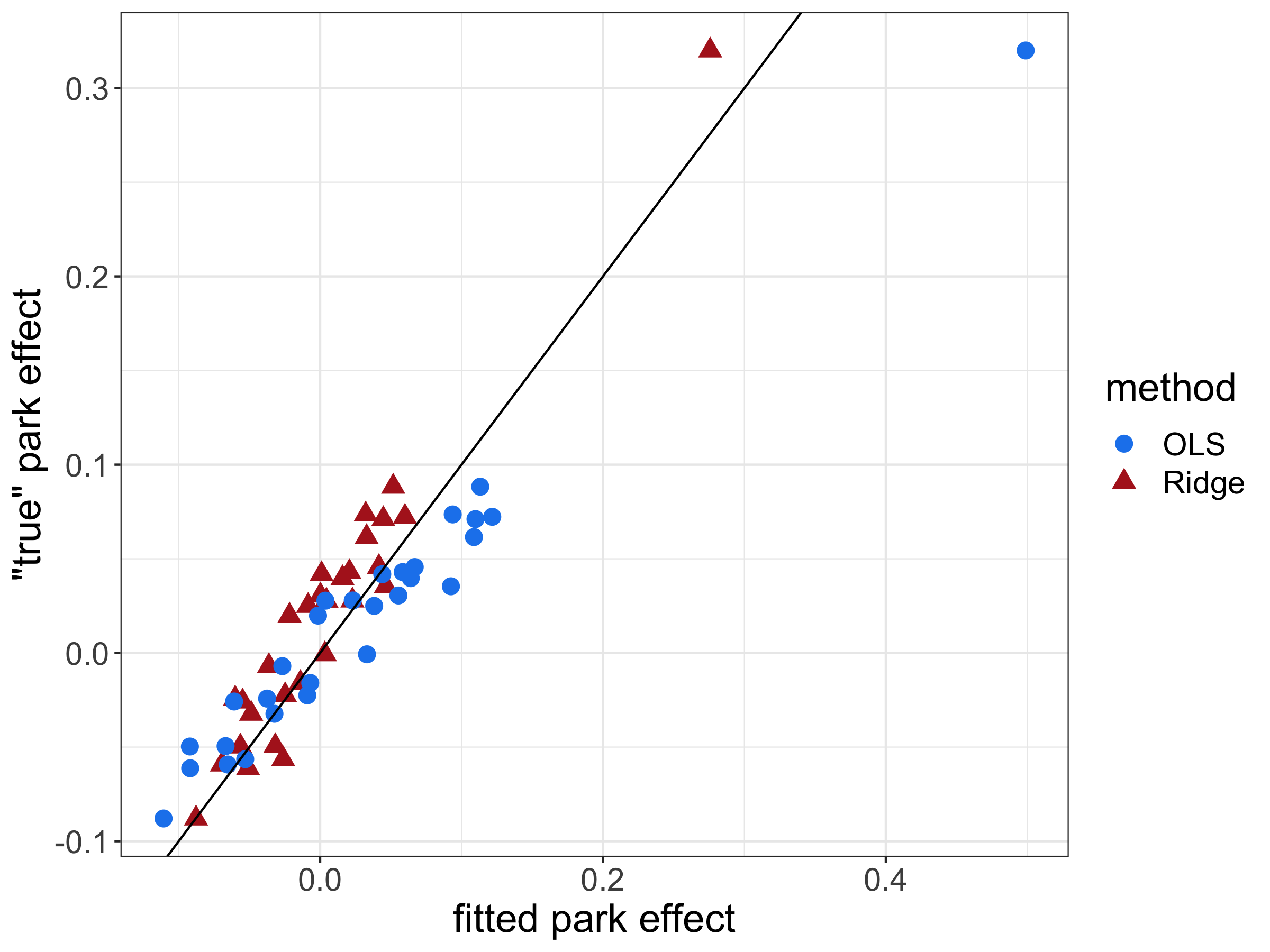}
\caption{For one of the 25 simulations from our second simulation study from Section~\ref{sec:second_sim_study}, plot the ``true'' park effects against the ridge estimates and the OLS estimates. 
The line $y=x$, shown in black, represents a perfect fit between the ``true'' and fitted park effects. The OLS estimates are biased, whereas the ridge estimates lie more evenly around the line $y=x$. In particular, ridge regression much better captures the park effect of the outlier, Denver.} 
\label{fig:sim2_compare}
\end{figure}

We report the results of our second simulation study in Table~\ref{table:sim2_results}. Again, ridge regression performs the best. In particular, ridge performs significantly better than the other methods on the outlier Colorado, and better than the other methods on the other parks. In Figure~\ref{fig:sim2_compare}, we visualize one of the 25 simulations by plotting the ``true'' park effects against the ridge estimates and the OLS estimates. We see that ridge regression successfully fits the Colorado park effect, whereas OLS significantly overestimates Colorado. Colorado as an outlier in OLS exerts high leverage over the rest of the park effects, swaying their estimates upwards. One might suggest removing the outlier Colorado from the dataset and estimating it separately, but doing weakens the estimates of the other teams in its division as a result of removing too many games from the set schedule which determines the data matrix $\XX$. 
So, in both the first simulation study from Section~\ref{sec:first_sim_study} and the second simulation study from this section, ridge regression most successfully estimates the ``true'' simulated park effects.

%%%%%%%%%%%%%%%%%
\begin{table}[hbt!]
\centering
\begin{tabular}{ lll } \hline
method & non-Colorado simulation error & Colorado error \\ \hline
Ridge & \textbf{0.0442} & \textbf{0.0502} \\
% OLS with biased adjustments of team quality & \textbf{0.0283} & \textbf{0.0267} \\
OLS & 0.0483 & 0.1791 \\
Three-part OLS & 0.0490 & 0.1807 \\
OLS without adjusting for team quality & 0.0751 & 0.1893 \\
\hline
\end{tabular}
\caption{Results of our second simulation study.}
\label{table:sim2_results}
\end{table}

%%%%%%%%%%%%%%%%%

%%%%%%%%%%%%%%%%%%%%%%%%%%%%%%%%%%%%%%%%%
\subsection{Comparing Existing Park Effects and Ridge Park Effects}\label{sec:compare_existing_parkFx}

Now, in this section, we compare our ridge park factors, which perform best in our simulation studies from Sections~\ref{sec:first_sim_study} and~\ref{sec:second_sim_study}, to existing park effects from ESPN and FanGraphs.

\textbf{Transforming ESPN and FanGraphs park factors to an ``additive'' scale.} Our ridge and OLS park effects of a ballpark, based on model~\eqref{eqn:parkfx_model_A}, are ``additive'' in the sense that they represent the expected runs scored in a half-inning at that park \textit{above} that of an average park, if an average offense faces an average defense. On the other hand, ESPN and FanGraphs park factors, defined in Formulas~\eqref{eqn:espn_pf} and~\eqref{eqn:FanGraphs_pf}, are ``multiplicative'' in the sense that they represent the \textit{ratio} of runs created at home to runs created on the road. Therefore, in order to compare these existing park factors to our park factors, we need to put them on the same scale. In particular, we transform the ESPN and FanGraphs park effects into ``additive'' park effects. To do so, we take the mean runs scored in a half-inning,
\begin{equation}
    \overline{y} = 0.5227,
    \label{eqn:ybar}
\end{equation}
and multiply it by a ``multiplicative'' park factor subtracted by 1. For example, if the ESPN Colorado park factor $\alpha$ in 2019 is 1.34, representing that teams score $34\%$ more runs in Colorado than in other parks, then the transformed ``additive'' ESPN park factor is
\begin{equation}
    (\alpha - 1) \cdot \overline{y} = (0.34) \cdot (0.5227) = 0.178.
\end{equation}
After this transformation, the ESPN and FanGraphs park factors also represent the expected runs scored in a half-inning at that park above that of an average park.

\begin{figure}[hbt!]
\centering
\includegraphics[width=15cm]{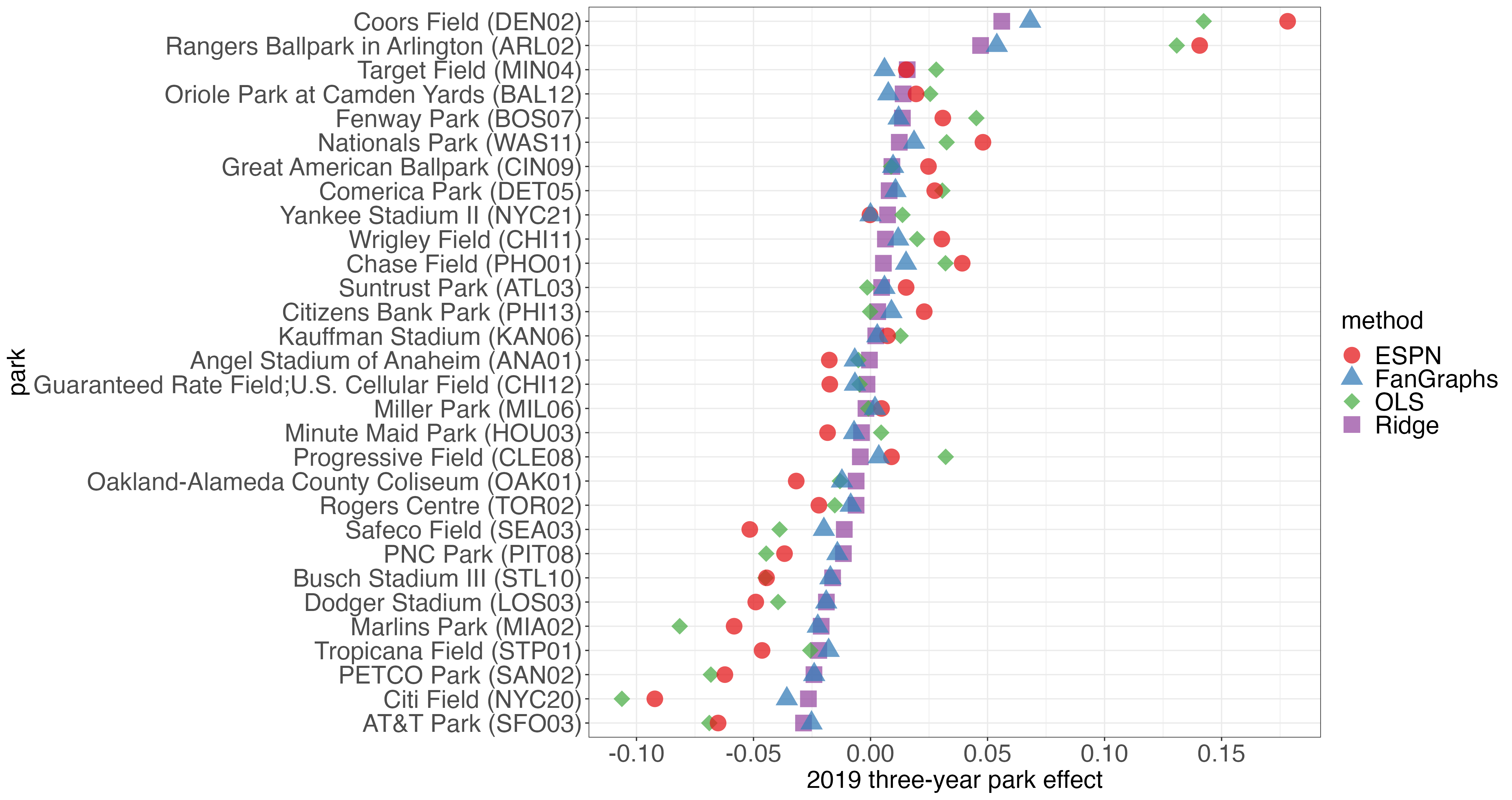}
\caption{Ridge, OLS, ESPN, and FanGraphs 2019 three-year park effects (where the latter two are transformed to an ``additive'' scale).
The abbreviations are the Retrosheet ballpark codes.
 % NYC20 refers to Citi Field, NYC21 refers to Yankee Stadium II, CHI11 refers to Wrigley field, and CHI12 refers to the White Sox's park
 } 
\label{fig:park_fx_1719}
\end{figure}

\textbf{Visualizing these park effect methods.}  In Figure~\ref{fig:park_fx_1719} we visualize the ridge, OLS, ESPN, and FanGraphs park effects (where the latter two are transformed to an ``additive'' scale), fit on all half-innings from 2017 to 2019. We use the park abbreviations from Retrosheet, our data source, as discussed in Section~\ref{sec:our_data}. As expected, we see that ridge park factors are a shrunk version of OLS park factors, and FanGraphs park factors are a shrunk version of ESPN park factors. 
The FanGraphs and ridge park factors are remarkably similar.
Also, as expected, Coors Field (DEN02) has the largest park effect for all four methods. The Texas Ranger's ballpark in Arlington (ARL02) has the second highest park effect for all four methods. These two parks have significantly larger park effects than all the other ones. Additionally, the Mets' ballpark (NYC20) has the lowest park effect for all four methods. 

%%%%%%%%%%%%%%%%%
%%%%%%%%%%%%%%%%%

\begin{figure}[hbt!]
\centering
\includegraphics[width=15cm]{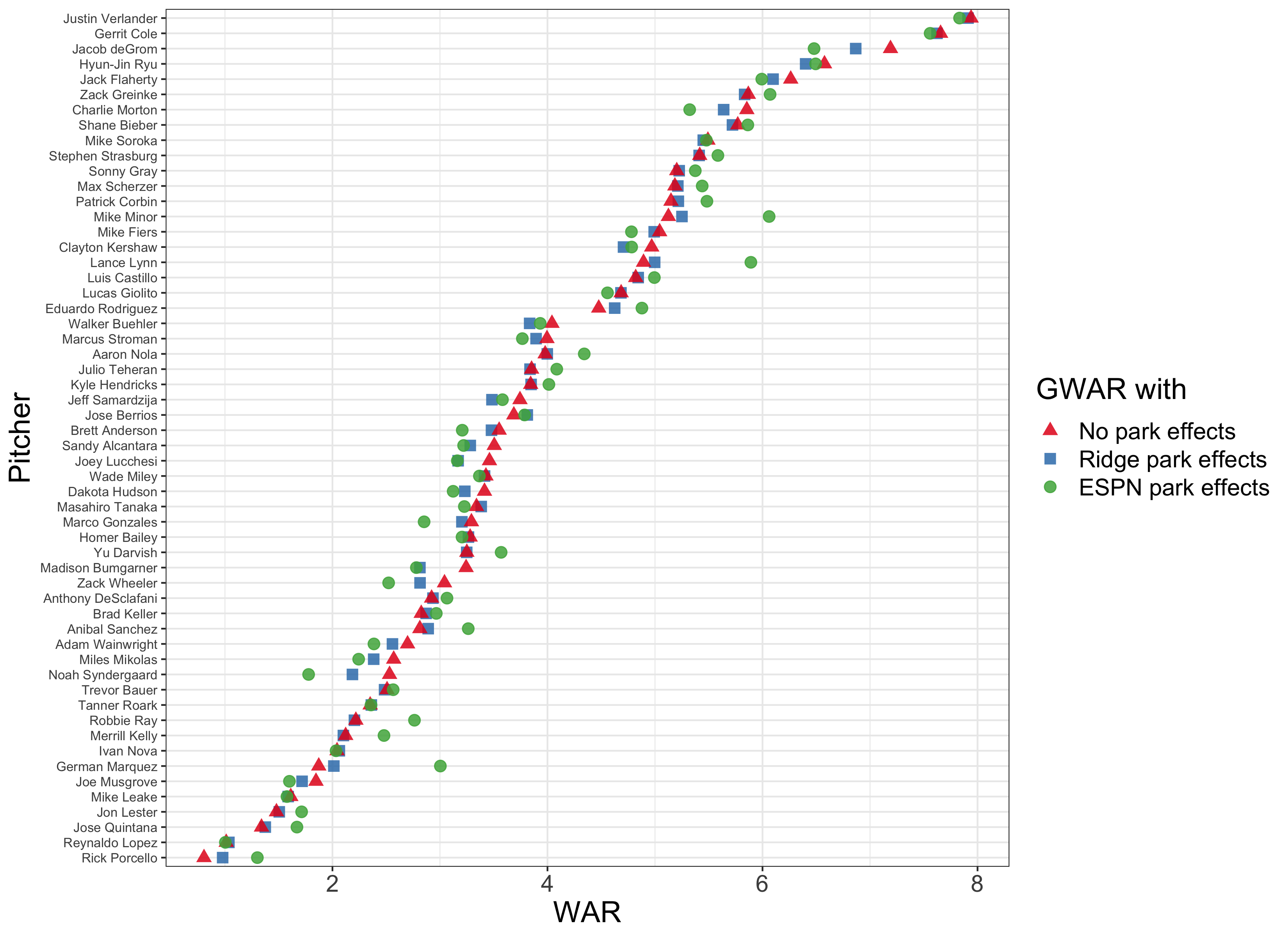}
\caption{Grid $\WAR$ on a set of pitchers from 2019 without park effects (red triangles), with ridge park effects (blue squares), and with ESPN park effects (green circles).} 
\label{fig:gwar_parkFx_comp}
\end{figure}

Additionally, we visualize how these various park effects impact the Grid $\WAR$ of various starting pitchers. In Figure~\ref{fig:gwar_parkFx_comp}, we show the 2019 seasonal Grid $\WAR$ for a set of starters without park effects, with ridge park effects, and with (transformed) ESPN park effects. 
For most pitchers, the impact of including a park adjustment is small.
For some pitchers, the impact of an ESPN park adjustment is massive.
For instance, the $\GWAR$ of Mike Minor and Lance Lynn, who pitched for the Rangers in 2019, each increases by a staggering one whole $\WAR$. 
Ridge park factors have a much more muted impact than ESPN park factors.
This makes sense, as ridge regression shrinks the park coefficients closer to 0.
For a few pitchers, however, even the ridge park effects make a nontrivial impact on their $\GWAR$. 
This also makes sense, as some park effects, such as those of the Mets, Rockies, and Rangers, are far enough from zero. 
For instance, the $\GWAR$ of Noah Syndergaard and Jacob deGrom, who pitched for the Mets, each decreases by about one-quarter of a $\WAR$.

%%%%%%%%%%%%%%%%%
%%%%%%%%%%%%%%%%%

\textbf{Comparing these park effects quantitatively.} In deciding which park effects to use in our final Grid $\WAR$ calculations, we quantitatively compare the ridge, OLS, ESPN, and FanGraphs park factors. In particular, we compare the out-of-sample predictive performance of these park effects. 

We begin by fitting each park factor method using data from all half-innings from 2014 to 2016. Note that the OLS and ridge park factors adjust for team offensive and defensive quality, whereas the ESPN and FanGraphs park factors don't. So, in order to fairly quantitatively compare which of these park factor methods is ``best'', we adjust for team quality on all of these methods. Specifically, using the fitted park factors $\hat\alpha$ from 2014-2016 for a given method, we regress out team offense and team defense indicators via OLS on the following model, where $\P$, $\OO$, and $\D$ are the data matrices consisting of all half-innings from 2017 to 2019, 
\begin{align}
    y_i = \beta_0 + \beta_1 \sum_{j \neq \text{ANA}} \P_{ij} \hat\alpha_j + \cdot \sum_{k \neq \text{ANA2017}} \OO_{ik} \beta_k^{(\text{off})} + \cdot \sum_{k \neq \text{ANA2017}} \D_{ik} \beta_k^{(\text{def})} + \epsilon_i.
    \label{eqn:regress_out_teamQuality}
\end{align}
Then, using these adjusted models based on Formula~\eqref{eqn:regress_out_teamQuality}, we predict the expected runs scored $\hat y_i$ in each half-inning $i$ from 2017 to 2019, which are out-of-sample predictions relative to the park effects $\hat\alpha$ which were estimated on data from 2014-2016.
Finally, we compute the out-of-sample RMSE. 

Each of the four methods has the same out-of-sample RMSE, 1.504. For reference, the RMSE of the overall mean is 1.508, so park factors do improve prediction, albeit slightly. But, because the runs scored in a half-inning is so noisy, and because the differences across parks are so slight, out-of-sample RMSE isn't sensitive enough to quantitatively show which park factors have the best predictive performance.

To more clearly understand the differences in methods, we calculate the \textit{ecological RMSE} to quantitatively compare the various park factor methods. This is done by first fixing a ballpark $p$. Then for each park factor method, we take the mean of the vector of predicted runs scored in each half-inning at the given park $p$, yielding $\hat{\overline{y}}_p$. Then we find the mean of the vector of observed runs scored in each half-inning at that park $p$, yielding $\overline{y}_p$. Finally, we compute the RMSE of the regression of   $(\overline{y}_p)$ on $(\hat{\overline{y}}_p)$ (each vectors of length $30$), yielding the out-of-sample ecological  RMSE. In Table~\ref{table:ecological_rmse} we show the out-of-sample ecological RMSE for several park factor methods. The ridge park effects perform best, outperforming the FanGraphs and ESPN park factors, mainly since Ridge adjusts for offensive and defensive quality. The ridge park effects outperform the OLS park effects for the same reasons discussed in our simulation studies from Sections~\ref{sec:first_sim_study} and~\ref{sec:second_sim_study}.

%%%%%%%%%%%%%%%%%
\begin{table}[hbt!]
\centering
\begin{tabular}{ ll } \hline
Park Effect & Ecological RMSE \\ \hline
Ridge & \textbf{0.01516} \\
FanGraphs & 0.01578 \\
ESPN & 0.01579 \\
OLS & 0.01672\\
Overall mean $\overline{y}$ & 0.04658 \\ 
\hline
\end{tabular}
\caption{Out-of-sample ecological RMSE of predicting the runs scored in a half-inning using various park effect methods.}
\label{table:ecological_rmse}
\end{table}

Hence the ridge regression park effects lead to better out-of-sample predictions of the runs scored in a half-inning than existing park factor methods from ESPN and FanGraphs.

%%%%%%%%%%%%%%%%%%%%%%%%%%%%%%%%%%%%%%%%%
\subsection{2019 Three-Year Park Effects}\label{sec:sec:threeYrParkFx}

As discussed in Sections~\ref{sec:first_sim_study} and~\ref{sec:second_sim_study}, the ridge park effects outperform other park effect methods in two simulation studies. Further, as discussed in Section~\ref{sec:compare_existing_parkFx}, ridge park effects outperform existing park effect methods from ESPN and FanGraphs. Therefore, we use ridge park effects in computing Grid $\WAR$. 
In particular, we use ridge regression on our observed dataset $\{y, \XX\}$ consisting of all half-innings from 2017 to 2019, tuning the ridge parameter $\lambda$ using cross validation, to fit our park effects, shown in Figure~\ref{fig:final_parkFx} in Section~\ref{sec:parkFxFinal}.
% with $\lambda = 0.25$ 

%%%%%%%%%%%%%%%%%%%%%%%%%%%%%%%%%%%%%%%%%%%%%%%%%%%%%%%%%%%%%%%%%
%%%%%%%%%%%%%%%%%%%%%%%%%%%%%%%%%%%%%%%%%%%%%%%%%%%%%%%%%%%%%%%%%
%%%%%%%%%%%%%%%%%%%%%%%%%%%%%%%%%%%%%%%%%%%%%%%%%%%%%%%%%%%%%%%%%
\section{Grid $\WAR$ across modern baseball history}\label{sec:GWARhistory}

\begin{figure}[p]
    \centering{}
    \subfloat[]{
        \includegraphics[width=0.45\textwidth]{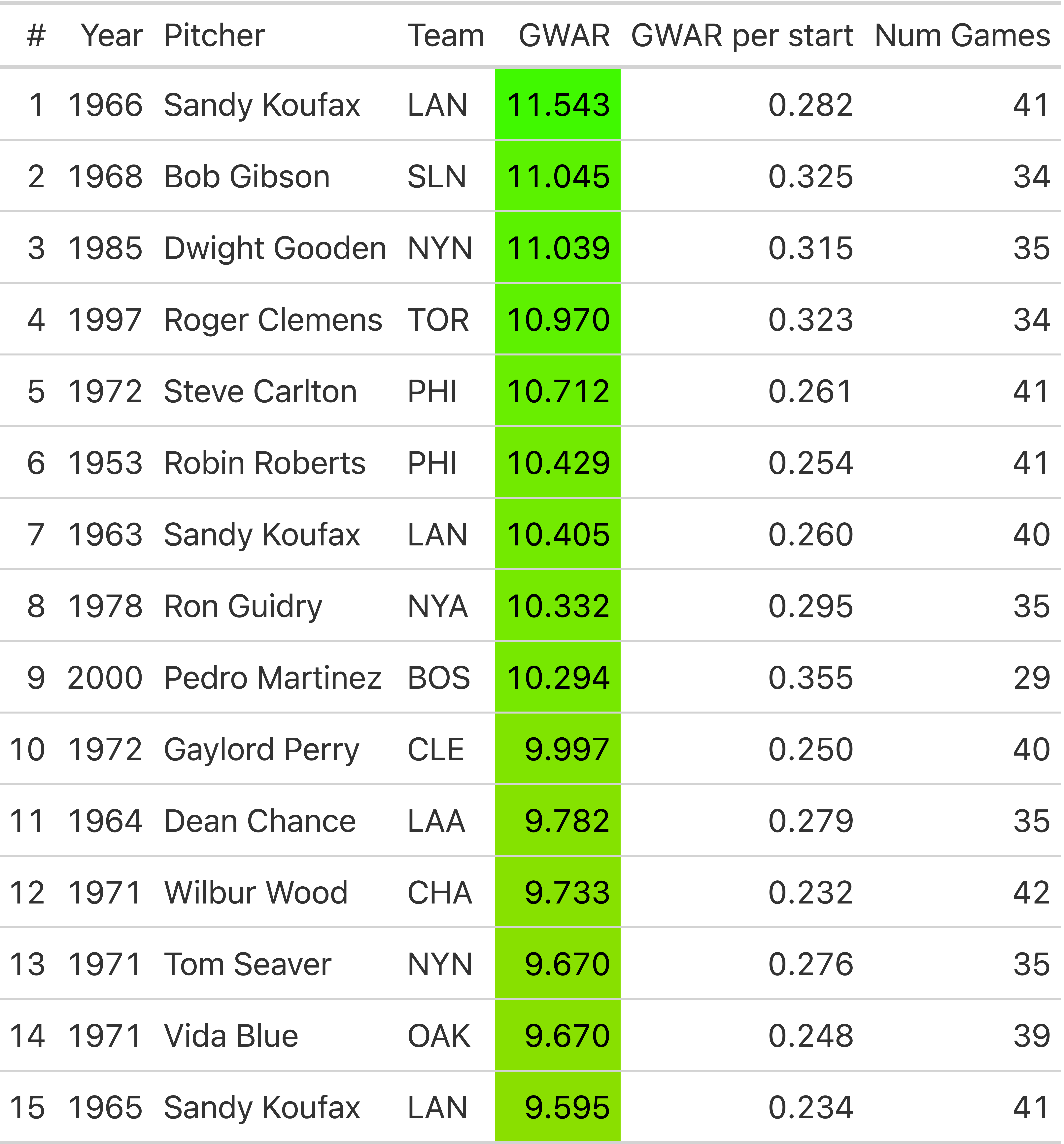}
        \label{fig:gwar_allTime_szn_total}
    } 
    \qquad
    \subfloat[]{
        \includegraphics[width=0.45\textwidth]{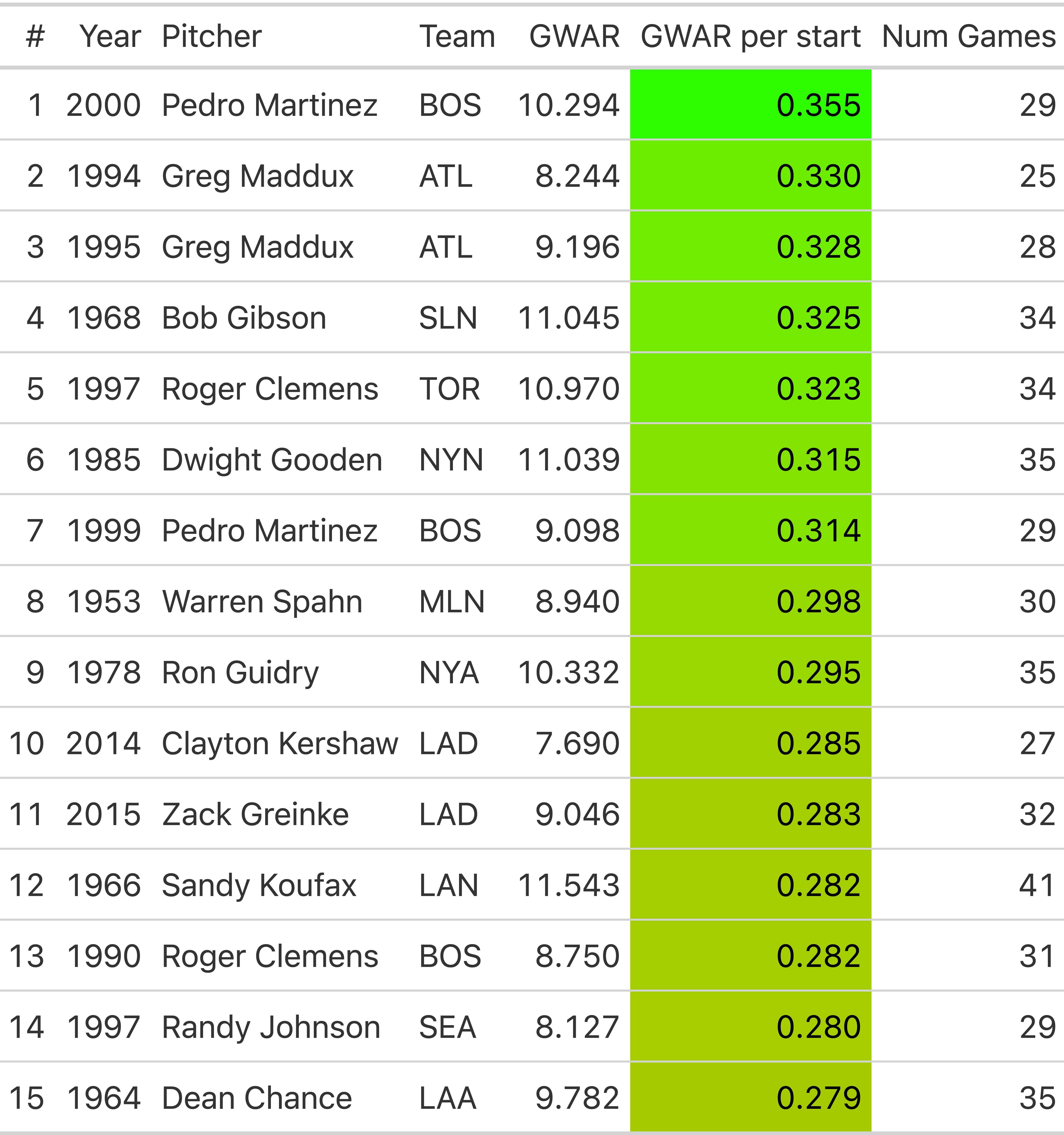}
        \label{fig:gwar_allTime_szn_eff}
    }
    \caption{
        The top 15 starting pitcher-seasons post 1951 by total Grid $\WAR$ (Figure (a)) and Grid $\WAR$ per game (Figure (b)).
    } 
    \label{fig:gwar_allTime_szn}
\end{figure}

We visualize the top 15 starting pitcher-seasons of all time by total Grid $\WAR$ in Figure~\ref{fig:gwar_allTime_szn_total}.
The pitcher with the highest total Grid $\WAR$ of all time in a single season is Sandy Koufax and it is not even close. In 1966, he accumulated 11.54 $\GWAR$ over 41 games in his final season which is half a game more $\GWAR$ than the second best season (Bob Gibson had 11.05 Grid $\WAR$ over 34 games in 1968) and the third best (Dwight Gooden had 11.04 Grid $\WAR$ over 35 games in 1985). Koufax's 1966 season is an illuminating example of the value of Grid $\WAR$ compared to standard formulations. While his 1966 is the standout season of all time in terms of Grid WAR,  it is just the sixth highest seasonal $\FWARr$ and the $20^{th}$ highest seasonal $\FWARf$. The other methods incorrectly overweight his three outlying blow-up games (i.e. less than -0.1 $\GWAR$). This is an excellent example of why it is a philosophical mistake to ignore variance and convexity. In 1966, there were two versions of Koufax: the ``left arm of God'' or worse than replacement. The ``left arm of God'' threw eight complete game shutouts and 9 one-run complete games. Grid WAR properly accounts for this variation while the standard metrics do not. Among the 15 all time best seasons, no other pitcher appears more than once, while Koufax appears on the list three times (1963, 1965, 1966) with his 1964 season only falling short because Koufax lost a quarter of the season due to injury. Koufax's ``duality'' is not just chance variation, it is a systematic attribute and a significant contributor to his early retirement after the 1966 season.

%%%%%%%%%%%%%%%%%%%%%
%%%%%%%%%%%%%%%%%%%%%

%%%%%%%%%%%%%%%%%%%%%
\begin{figure}[p]
    \centering{}
    \subfloat[]{
        \includegraphics[width=0.45\textwidth]{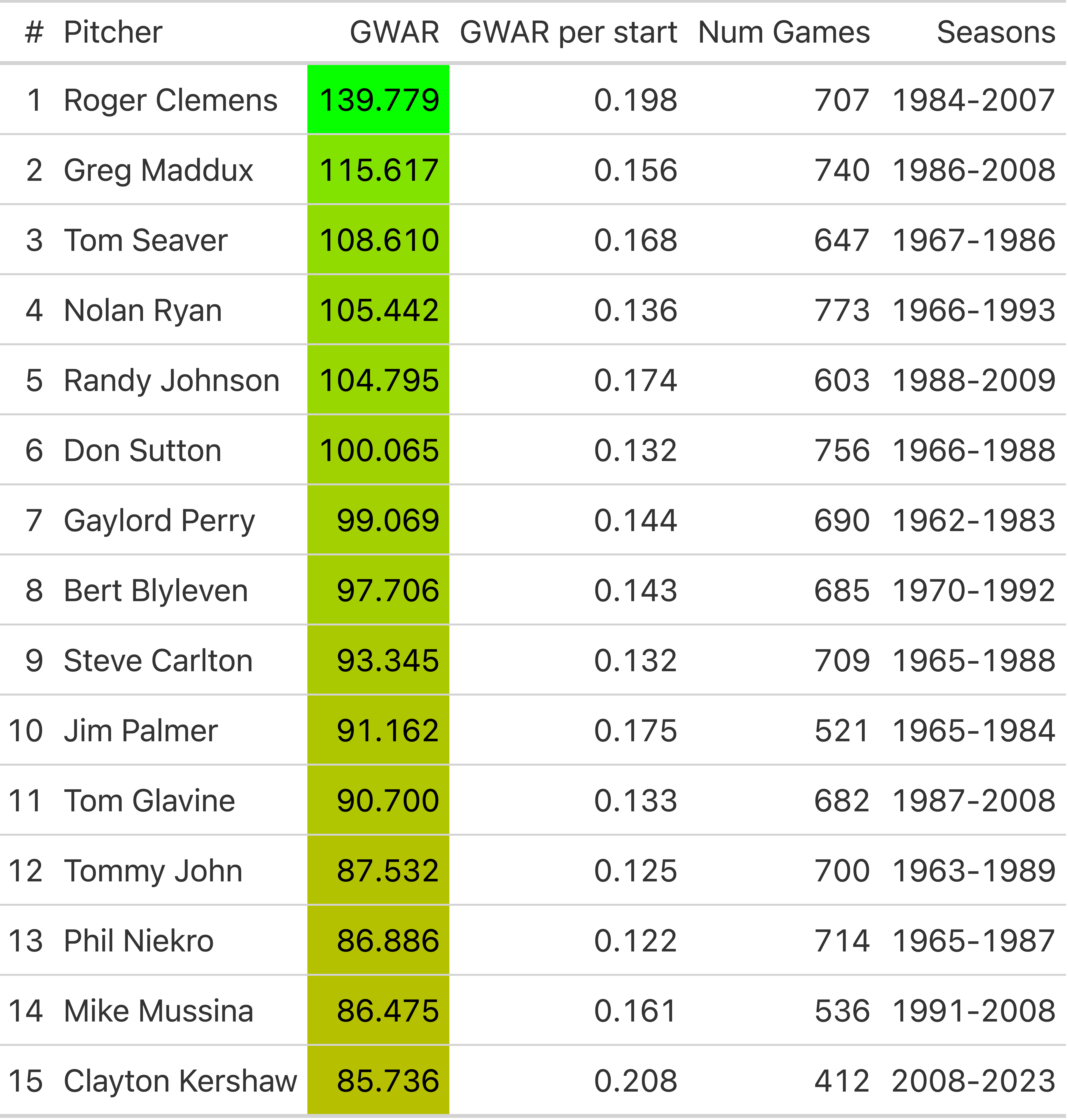}
        \label{fig:gwar_allTime_career_total}
    }
    \qquad
    \subfloat[]{
        \includegraphics[width=0.45\textwidth]{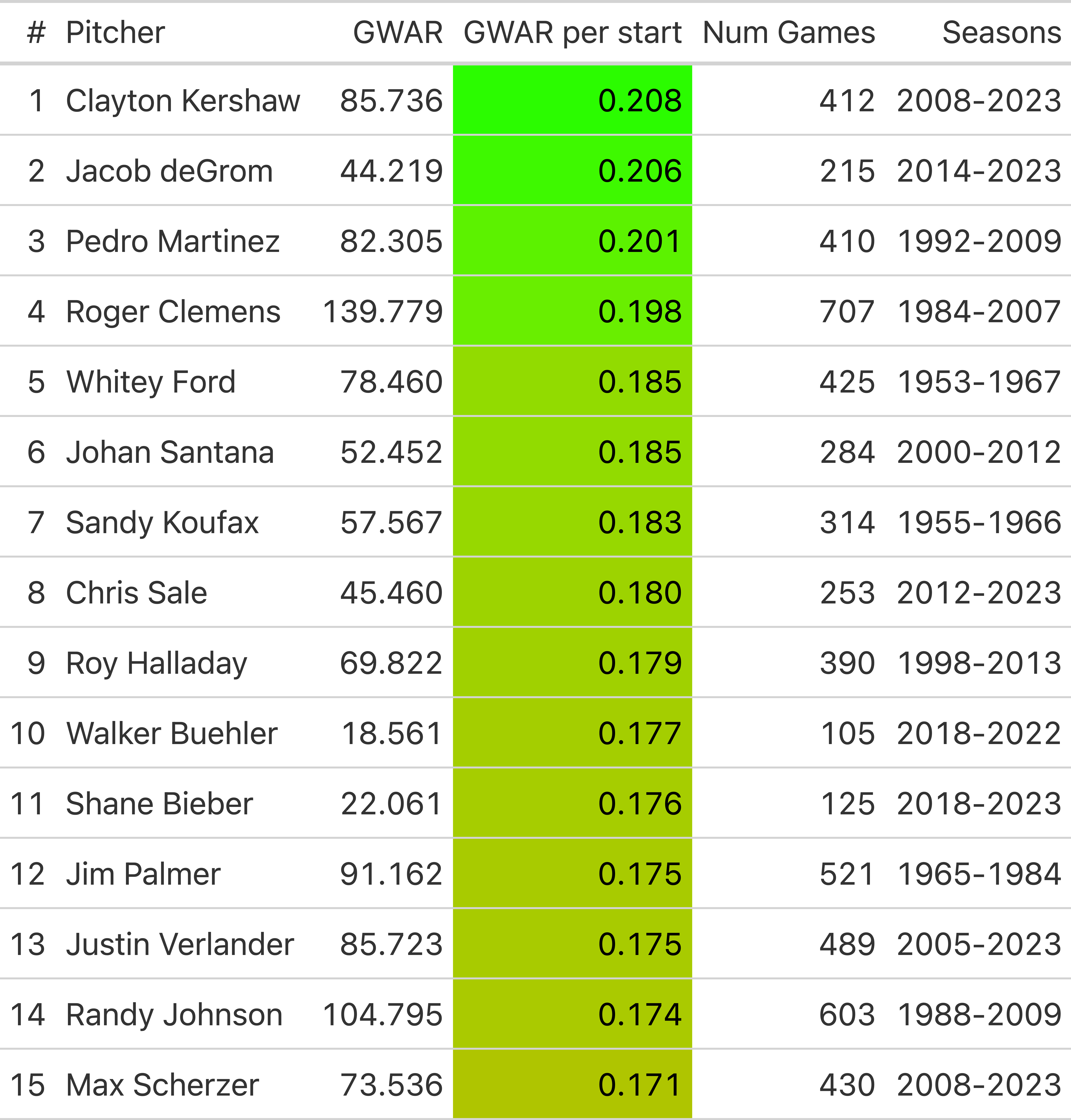}
        \label{fig:gwar_allTime_career_eff}
    }
    \caption{
        The top 15 starting pitchers post 1951 by career Grid $\WAR$ (Figure (a)) and career Grid $\WAR$ per game (Figure (b)).
    } 
    \label{fig:gwar_allTime_career}
\end{figure}
%%%%%%%%%%%%%%%%%%%%%

Grid $\WAR$ certainly favors pitchers that start more often and pitch deeper into the game. 
Consequently, all of the top  pitcher-seasons of all time, according to cumulative Grid $\WAR$, occurred prior to 2001.
Moreover, no pitcher-season since 2001 has eclipsed 9.1 $\GWAR$ (Greinke in 2009 has the highest at 9.068). Just five pitcher seasons since 2009 have eclipsed 8.5 $\GWAR$ (Weaver in 2011, deGrom in 2018, Arrieta in 2015, and Greinke in 2009 and 2015),
and only two pitcher-seasons since 2016 have eclipsed eight $\GWAR$ (deGrom in 2018 and Verlander in 2019). 
Starting pitchers are fundamentally less valuable today because top starters used to pitch many more games per season and  more innings per game.
Koufax started every fourth game in 1966, whereas high-end pitchers today start in every fifth game. Furthermore, the complete game is a thing of the past. 
Today, starters are typically removed prior to the seventh inning as a reaction to the time through the order penalty \citep{BrillDeshpandeWyner+2023}.

We visualize the top 15 starting pitcher-seasons of all time by Grid $\WAR$ per game in Figure~\ref{fig:gwar_allTime_szn_eff}.
Today's pitchers fare substantially better on a per-start basis. 
The most efficient individual pitcher-season of all time, subject to starting at least 25 games, is Pedro Martinez who in 2000 finished with 0.355 Grid $\WAR$ over 29 games. Greg Maddux's 1994 season is second all-time,  with 0.330 Grid $\WAR$ over 25 games. Greg Maddux has third place too,  with 0.328 Grid $\WAR$ over 28 games in 1995. These seasons were remarkable in terms of consistency. 
Pedro had zero blow-up games in 2000. All but two of his games resulted in positive $\GWAR$, and his worst game featured just -0.068 $\GWAR$.
Similarly, across 1994 and 1995, Maddux had just five negative $\GWAR$ games in 53 starts, with at worst $-0.13$ $\GWAR$ in a game.

We visualize the top 15 starting pitchers of all time by career Grid $\WAR$ per game in Figure~\ref{fig:gwar_allTime_career_eff}.
The most efficient starter of all time over his career (minimum of 100 games) is Clayton Kershaw who currently averages  0.208 Grid $\WAR$ per start over 410 games. Jacob deGrom is a close second with 0.206 over his 215 games. Pedro Martinez is third with an average of  0.201 over 410 games. 
Interestingly, only two of the top 10 pitchers of all time by career efficiency played their entire careers prior to 2000 (the exceptions are Whitey Ford and Sandy Koufax). 

We visualize the top 15 starting pitchers of all time by total career Grid $\WAR$ in Figure~\ref{fig:gwar_allTime_career_total}.
The pitcher with the most career Grid $\WAR$ is Roger Clemens, who accumulated 139.8 $\GWAR$ across 24 seasons from 1984 to 2007.
The starter with the second highest career Grid $\WAR$, Greg Maddux, doesn't come close to Clemens, as he accumulated 115.6 $\GWAR$ over 23 seasons from 1986 to 2008.
Three of the top ten starters according to career $\GWAR$ (Clemens, Maddux, and Randy Johnson) played in the 1980s, 90s, and 00s, and the remaining seven (Tom Seaver, Nolan Ryan, Don Sutton, Gaylord Perry, Bert Blyleven, Steve Carlton, and Jim Palmer) played in the 1960s, 70s, and 80s.
Just five of the top 25 starters by career Grid $\WAR$, on the other hand, began their career after 2000.
The best such starters are Justin Verlander (85.7 $\GWAR$ over 19 seasons from 2005 to 2003) and Clayton Kershaw (85.1 $\GWAR$ over 16 seasons from 2008 to 2023).
These are the $15^{th}$ and $16^{th}$ ranked pitchers, and they don't come anywhere particularly close to Clemens or even Maddux, although their careers aren't over.
As discussed before, the pitchers with the highest career Grid $\WAR$ come from the previous millennium because top starters back then pitched more games per season and pitched more innings per game. 

Finally, we measure a starters peak performance by the maximum total Grid $\WAR$ across all his contiguous four year stretches.
We combine a starter's peak ranking with his ranking by total career Grid $\WAR$ using the geometric mean, $\textsf{GeomMean}(a,b) = \sqrt{a \cdot b}$.
We use the geometric mean as opposed to the arithmetic mean so as to value pitchers much more the closer they are to being the best (rank one).
In Figure~\ref{fig:gwar_allTime_geom} we show the peak $\GWAR$ rank and geometric mean rank of the top 30 starting pitchers post 1951.
Greg Maddux has the best geometric mean rank, as he has the second highest career $\GWAR$ and the fourth highest peak $\GWAR$.
We also include whether the pitcher was elected to the Hall of Fame by the Baseball Writers' Association of America (BBWAA) or the veteran's committee (Veterans), and hence call Figure~\ref{fig:gwar_allTime_geom} the hall-of-fame chart. 
Each of these pitchers who aren't still playing (Kershaw, Verlander, Scherzer), recently retired (Sabathia), or mired by controversy (Clemens, Schilling) made the Hall of Fame, except notably Kevin Brown and Dave Steib ($\# 10$ and $\# 29$ ranked geometric mean, respectively). 
These latter two have without a doubt been snubbed and should be in Cooperstown.
There are many articles online corroborating this view.\footnote{
    \url{https://blogs.fangraphs.com/should-kevin-brown-be-in-the-hall-of-fame/}, \url{https://www.fishstripes.com/22914710/kevin-brown-hall-of-fame-case}
}

%%%%%%%%%%%%%%%%%%%%%
\begin{figure}[p]
    \centering{}
    \includegraphics[width=13cm]{writeup_plots/plot_best_pit_4_yr_peak_HOF_Adi}
    \caption{
    The hall-of-fame chart, or the top 30 starting pitchers post 1951 by the geometric mean of career Grid $\WAR$ and peak four-year Grid $\WAR$.
    } 
    \label{fig:gwar_allTime_geom}
\end{figure}
%%%%%%%%%%%%%%%%%%%%%

% \end{document}

% \newpage
% \begin{center}
% {\LARGE \textbf{Appendix}}
% \end{center}
% % {\LARGE \textbf{Supplementary Materials}}
%% \clearpage
%% \appendix
% \input{appendix}

\end{document}